\documentclass[%
 reprint,
 amsmath,amssymb,
 aps,
]{revtex4-2}

\usepackage{graphicx}
\usepackage{dcolumn}
\usepackage{bm}
\usepackage{float}
\usepackage{xcolor}
\usepackage{braket}

\begin{document}

\preprint{APS/123-QED}

\title{A DFT+U type functional derived to explicitly address the flat plane condition}

\author{Andrew Burgess$^1$}

\author{Edward Linscott$^2$}%

 \author{David D. O'Regan$^{1}$}%
 \email{david.o.regan@tcd.ie}
\affiliation{%
 $^1$School of Physics, Trinity College Dublin, The University of Dublin, Ireland \\
 $^2$Theory and Simulation of Materials (THEOS), Faculté des Sciences et Techniques de l’Ingénieur, École Polytechnique Fédérale de Lausanne, CH-1015 Lausanne, Switzerland
}%

\date{\today}

\begin{abstract}
A new DFT+$U$ type corrective functional is derived from first principles to enforce the flat plane condition on localized subspaces, thus dispensing with the need for an ad hoc derivation from the Hubbard model. The newly derived functional as given by equation \ref{eqn:BLOR_trace_version}, yields relative errors below $0.6 \%$ in the total energy of the dissociated s-block dimers as well as the dissociated H$_5^+$ ring system. In comparison bare PBE and PBE+$U$ (using Dudarev's 1998 Hubbard functional) yields relative energetic errors as high as $8.0 \%$ and $20.5\%$ respectively. 
\end{abstract}

\maketitle


Since the inception of the Hohenberg Kohn Theorems \cite{HK_theorems}, practitioners of density functional theory (DFT) have sought more accurate, reliable and efficient density functional approximations (DFAs) to the exchange correlation functional $E_{xc}^{\rm approx}$ \cite{LSDA,PBE,PBEsol,becke88,LYP,becke93,HSE,SCAN,TPSS,wB97M-V,CAM-QTP-02,TTwPBEh}. Despite these DFAs' remarkable success in predicting mechanical properties \cite{big_screening_of_mechanical_props} and crystallographic structures \cite{pickard}, they still exhibit significant failures in the prediction of molecular bond dissociation \cite{perdew_bond_dissocation,dutoi_bond_dissocation,partitionDFT}, band gaps in solids \cite{perdew_band_gap_problem,bandgap_benchmark,yang_bandgap} and polymorph energy differences in transition metal oxides \cite{TiO2,MnO,carter}. Many of these failures can be attributed to the breaking of certain exact physical constraints, namely (a) the piecewise linearity condition with respect to electron count \cite{pwl} and (b) the constancy condition with respect to magnetisation \cite{pwc_part1,pwc_part2}. The breaking of these two exact conditions is referred to as many-electron self interaction error (MSIE) \cite{perdew_bond_dissocation,MSIE} and static correlation error (SCE) \cite{pwc_part2} respectively. The generalisation of these two conditions is referred to as the ``flat plane condition" \cite{flat_plane_condition}. 

\begin{figure}[H]
\centering
\includegraphics[scale=0.5]{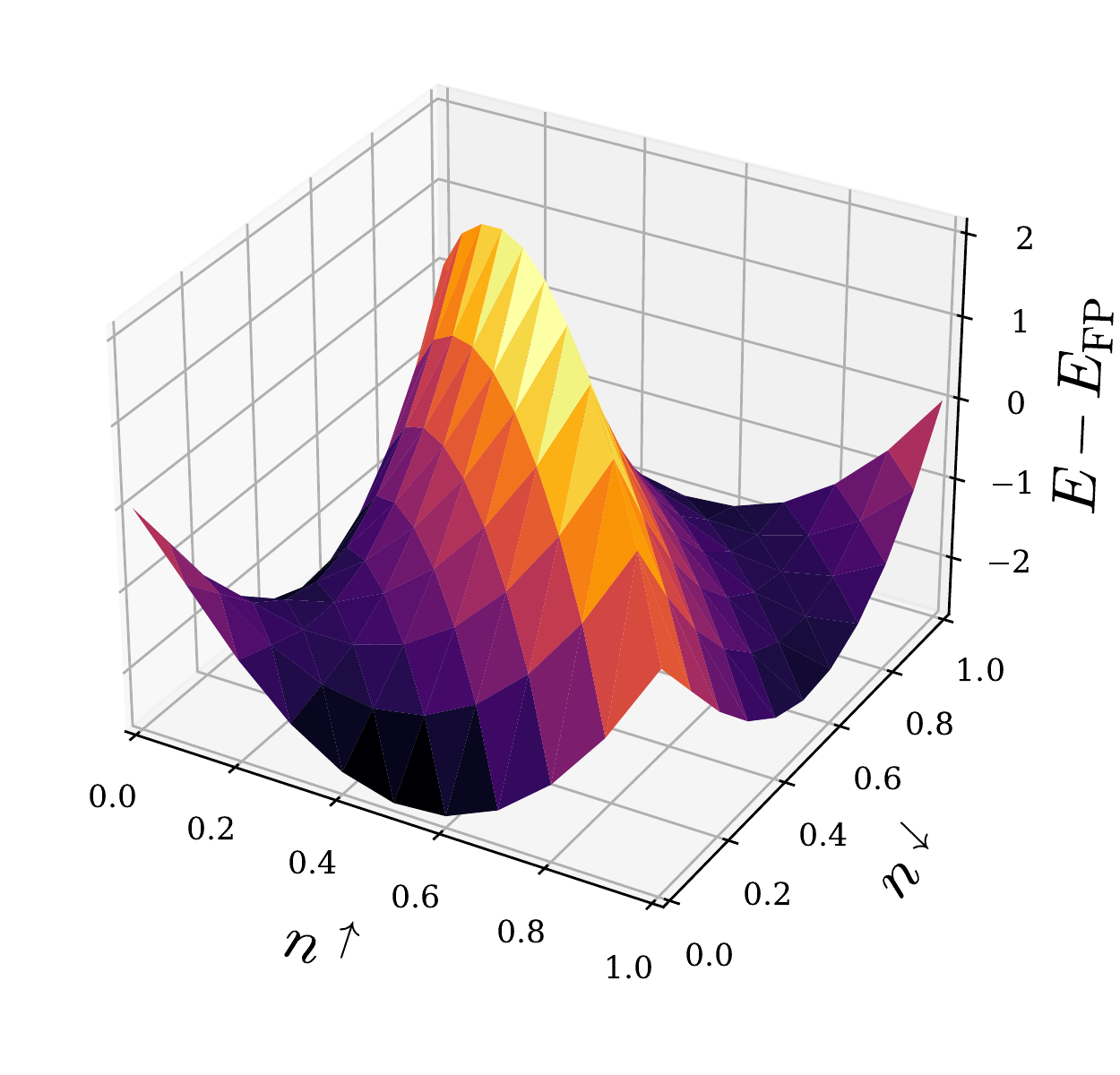}
\caption{Deviation of the total energy $E$ of the He atom/ion for different values of spin up ($n^{\upharpoonright}$) and spin down ($n^{\downharpoonright}$) occupancy using the PBE exchange-correlation functional \cite{PBE}. The PBE total energy is assumed to be exact at integer values of $n^{\upharpoonright}$ and $n^{\downharpoonright}$ for the He species. The exact energy from the flat plane condition is denoted as $E_{FP}$.}
\label{fig:deviation_from_flat_plane}
\end{figure}

For a two-electron system it is known that the total energy with respect to electron count and magnetisation, $E_{\rm tot}[N_{\rm tot},M_{\rm tot}]$ will typically be composed of two flat planes which meet with a derivative discontinuity along the $N_{\rm tot}=1$ line. This is referred to as a `Type 1' flat plane \cite{two_types_of_flat_planes} and will occur when the convexity condition is met:
\begin{equation}
2E_{\rm tot}[N_{\rm tot}]<E_{\rm tot}[N_{\rm tot}+1]+E_{\rm tot}[N_{\rm tot}-1],
\end{equation}
for $N_{\rm tot}=1$. This particular two-electron flat plane structure will be referred to as the ``diamond" for brevity. An equivalent flat plane will also form for the individual components of the total energy such as the Hartree-exchange-correlation Energy, $E_{\rm Hxc}$. Total electronic energies for systems with certain integer numbers of spin-up and spin-down electrons are well-approximated by currently available DFAs. However, large deviations from the exact total energies have been reported for systems with non-integer values, as shown in figure \ref{fig:deviation_from_flat_plane}.

There are similar conditions for many-atom systems. For a system of $N_{\rm atom}$ isolated atomic sites with a total of $N_{\rm tot}$ electrons, a piecewise linearity condition with respect to electron count should occur separately at each atomic site where $N_{\rm atom}, N_{\rm tot}\in \mathbb{N}$ but $N_{\rm tot}/N_{\rm atom} \notin \mathbb{N}$. The total energy of the system with $N_{\rm tot}/N_{\rm atom}$ electrons at each atomic site should be equal in energy to the system with $k$ sites with $N+1$ electrons and $N_{\rm atom}-k$
sites with $N$ electrons, where $N_{\rm tot}=NN_{\rm atom}+k$ and $N=\lfloor N_{\rm tot}/N_{\rm atom} \rfloor$. However, current DFAs yield incorrect energies for systems with fractional occupancies at the atomic sites \cite{mori-sanchez}. We refer to this error as local-MSIE as in this case one varies the local occupancy at the atomic site $N$, as opposed to the global electron count $N_{\rm tot}$ \cite{globalversuslocal}. Analogously, there exists local-SCE and a local analogue of the flat plane condition. Local-MSIE and local-SCE will lead to erroneous total energies for systems with integer global electron counts $N_{\rm tot}$.

Assuming local-MSIE is predominantly quadratic in nature (as has been reported for global-MSIE \cite{MSIE-quadratic}), the local-MSIE at an atomic site can be alleviated with an energetic correction of the form:
\begin{equation}
E_{\rm u}=\frac{U_{\rm eff}}{2}\bigg[(N-N_0)-(N-N_0)^2\bigg],
\label{eqn:localMSIEcorrection}
\end{equation}
where $N$ is the local occupancy at the atomic site, $\left \lfloor{N}\right \rfloor =N_0$ and $U_{\rm eff}$ is a corrective parameter. 

DFT$+U$ functionals \cite{anisimov,anisimov1993,Liechtenstein,dudarev} have often been employed as a correction to local-MSIE. Much like equation~\ref{eqn:localMSIEcorrection}, DFT$+U$-like functionals comprise of linear- and quadratic-occupancy-dependent energy corrections. For example, Dudarev's 1998 Hubbard corrective functional \cite{dudarev} is given by
\begin{equation}
E_{\rm u}^{\rm Dudarev}=\frac{U_{\rm eff}}{2}\sum_{\sigma m}n^{I\sigma}_{mm}-(n^{I\sigma}_{mm})^2.
\label{eqn:dudarev_dftu}
\end{equation}
Unlike equation \ref{eqn:localMSIEcorrection}, here the corrections are given in terms of subspace occupancy matrix elements:
\begin{equation}
n^{I\sigma}_{mm'}=\braket{\phi_{m'}^I|\hat{\rho}^{\sigma}|\phi_{m}^I},
\end{equation}
where $\hat{\rho}^{\sigma}$ is the spin-$\sigma$ Kohn Sham density operator and $\{\phi_m\}$ are the set of atomically localized orbitals at atom $I$ (the atomic site index is often suppressed for clarity). Equation~\ref{eqn:dudarev_dftu} was written in the basis of localized orbitals which diagonalize this subspace occupancy matrix. 

In the case where (a) the fractional occupancy at the atomic site is limited to the $s$-spin channel of one orbital $\phi_m$, i.e. $n^{s}_{mm}=N-N_0$,
and (b) all other orbitals $\phi_{m'}$ are fully occupied or unoccupied, Dudarev's 1998 functional provides a perfect correction for local-MSIE. 

Despite DFT+$U$'s success in alleviating local-MSIE in this limiting case, here we stress two points. Firstly, the DFT$+U$ method was originally derived from the Hubbard model and it is merely fortuitous that it acts as a correction to local-MSIE. Secondly, the DFT$+U$ method does not correct static correlation error, and will therefore not satisfy the local flat plane condition.

In this letter we instead derive a new DFT$+U$ type functional, disregarding entirely its connection with the Hubbard model and instead motivating its form entirely on the local analogue of the flat plane condition. Such a functional should, for a single orbital subspace, satisfy the following four key conditions: 
\begin{enumerate}
\item be a continuous function of the subspace electron count $N$ and subspace magnetisation $M$. 
\item yield no correction at integer values of $N$ and $M$. This is desirable because (semi-)local functionals are expected to yield accurate total energies in this case.
\item have a constant curvature of $-U^{\sigma}$ with respect to $n^{\sigma}$. This is desirable because (semi-)local functionals are expected to have a spurious curvature with respect to $n^{\sigma}$, due to their deviation from the local-flat plane condition.
\item have a constant curvature of $J$ with respect to $M$. This is desirable because (semi-)local functionals are expected to have a spurious curvature with respect to $M$, again due to their deviation from the local-flat plane condition.

\end{enumerate}
The functional which satisfies these four key conditions is BLOR:
\begin{widetext}
\begin{align}
\label{eqn:BLOR_trace_version}
E_{\rm BLOR}= \left\{
\begin{array}{*6{>{\displaystyle}c}}
\frac{U^{\upharpoonright}+U^{\downharpoonright}}{4}{\rm Tr}[\hat{N}-\hat{N}^2]
&+&
\frac{J}{2}{\rm Tr}[\hat{M}^2-\hat{N}^2]
&+&
\frac{U^{\upharpoonright}-U^{\downharpoonright}}{4}{\rm Tr}[\hat{M}-\hat{N}\hat{M}]
& , \
{\rm Tr}[\hat{N}] \leq {\rm Tr}[\hat{P}]
 \\
\underbrace{\frac{U^{\upharpoonright}+U^{\downharpoonright}}{4}{\rm Tr}[(\hat{N}-\hat{P})-(\hat{N}-\hat{P})^2]}_{\text{Symmetric-MSIE term}}
&+&
\underbrace{\frac{J}{2}{\rm Tr}[\hat{M}^2-(\hat{N}-2\hat{P})^2]}_{\text{SCE term}}
&+&
\underbrace{\frac{U^{\upharpoonright}-U^{\downharpoonright}}{4}{\rm Tr}[\hat{M}-\hat{N}\hat{M}]}_{\text{Asymmetric-MSIE term}}
& , \
{\rm Tr}[\hat{N}] > {\rm Tr}[\hat{P}]
\end{array}
\right.
\end{align}
\end{widetext}
where $\hat{P}$ is the subspace projection operator: $\hat{P}=\sum_m\ket{\phi_m}\bra{\phi_m}$. The subspace occupancy and magnetisation operators can be expressed in terms of the spin resolved subspace occupancy operators: $
\hat{N}=\hat{n}^{\upharpoonright}+\hat{n}^{\downharpoonright}$ and $ \hat{M}=\hat{n}^{\upharpoonright}-\hat{n}^{\downharpoonright}$, where $\hat{n}^{\sigma}=\hat{P}\hat{\rho}^{\sigma}\hat{P}$. The magnitude of the correction is controlled by three scalars: $U^\upharpoonright$, $U^\downharpoonright$, and $J$, which correspond respectively to the curvature with respect to $n^\upharpoonright$, $n^\downharpoonright$, and $M$. A full derivation of BLOR is given in  \ref{sec:BLOR-derivation}. One can show that conditions (1)-(4) are uniquely satisfied by BLOR (see \ref{sec:BLOR-uniqueness}). The lower and upper versions of the functional have a similar form (the lower version of BLOR is the case where ${\rm Tr}[\hat{N}] \leq {\rm Tr}[\hat{P}]$).

\begin{figure*}
  \includegraphics[width=\textwidth]{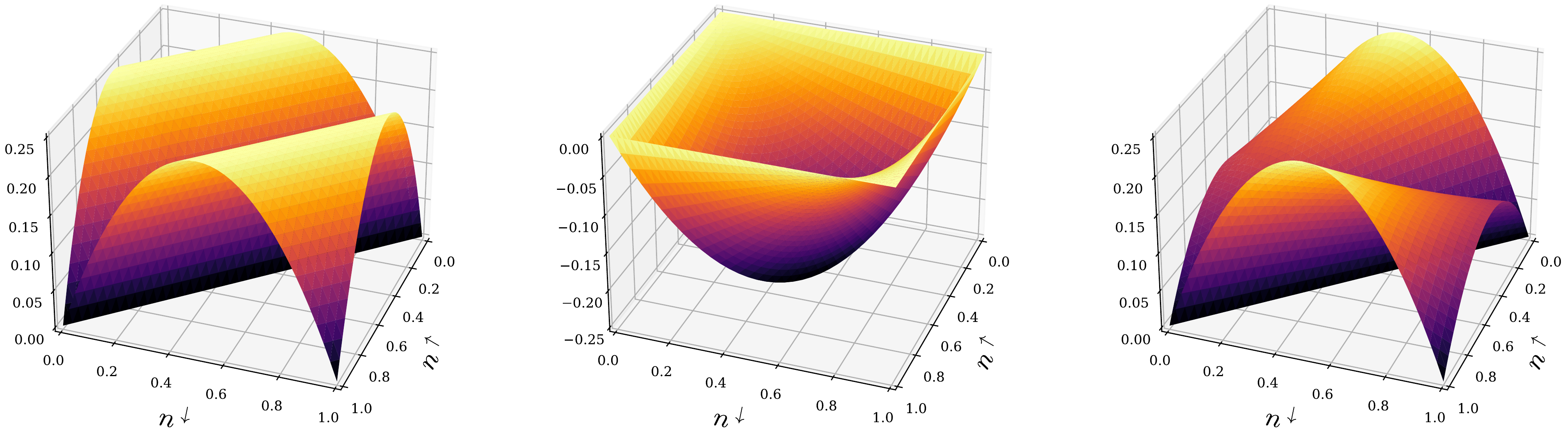}
  \caption{The left panel presents the symmetric-MSIE term for an s-orbital subspace as a function of spin up ($n^{\upharpoonright}$) and spin down ($n^{\downharpoonright}$) subspace occupancy. The centre panel presents the SCE term as a function of $n^{\upharpoonright}$ and $n^{\downharpoonright}$. The right panel presents the sum of the symmetric-MSIE and asymmetric-MSIE terms as a function of $n^{\upharpoonright}$ and $n^{\downharpoonright}$.}
  \label{fig:BLOR_msie_correction}
\end{figure*}

The first term is referred to as the symmetric-MSIE term because for a single orbital subspace it yields zero correction at integer values of $N$ and yields its maximum correction at $N=\frac{1}{2}, \frac{3}{2}$ as shown in the left panel of figure~\ref{fig:BLOR_msie_correction}. 

The second term is labelled as the SCE-term because for a single orbital subspace it yields zero correction when the subspace is maximally spin polarised and yields its maximum correction at $M=0$ for a given value of $N$ as shown in the middle panel of figure~\ref{fig:BLOR_msie_correction}. 

The asymmetric-MSIE term will contribute to $E_{\rm BLOR}$ when an effective magnetic field acts on the subspace. In this case, we cannot assume that the curvatures $U^{\upharpoonright}$ and $U^{\downharpoonright}$ are equal in magnitude. This effective magnetic field may be caused by an external magnetic field acting on the isolated atomic site. More notably, in practical calculations the target subspace will not be entirely isolated from its surrounding environment, such as the 3d subspace of face-centered cubic nickel. The 3d atomic subspace will experience an internal exchange-correlation magnetic field from the surrounding nickel atoms and hence we expect that $U^{\upharpoonright} \neq U^{\downharpoonright}$ for this system. The difference in magnitude is accounted for in the asymmetric-MSIE term. The combination of the symmetric- and asymmetric-MSIE terms is depicted in the right panel of figure~\ref{fig:BLOR_msie_correction}, which unlike the left panel, shows a different curvature along the maximally spin up polarised line compared to the maximally spin down polarised line.  

BLOR can also be expressed in terms of subspace occupancy matrix elements as:
\begin{widetext}
\begin{align}
\label{eqn:BLOR_sum_version}
E_{\rm BLOR}= \left\{
\begin{array}{*8{>{\displaystyle}c}}
\sum_{\sigma m m'}\frac{U^{\sigma}}{2}n^{\sigma}_{mm'}\delta_{mm'}-\frac{U^{\sigma}}{2}n^{\sigma}_{mm'}n^{\sigma}_{m'm}
&-&
\frac{U^{\sigma}+2J}{2}n^{\sigma}_{mm'}n^{\bar{\sigma}}_{m'm}
&\bf{}&
\bf{}
& , \
N \leq 2l+1
 \\
\sum_{\sigma m m'} \bigg(U^{\sigma}+\frac{U^{\bar{\sigma}}}{2}+2J\bigg)n^{\sigma}_{mm'}\delta_{mm'}-\frac{U^{\sigma}}{2}n^{\sigma}_{mm'}n^{\sigma}_{m'm}
&-&
\frac{U^{\sigma}+2J}{2}n^{\sigma}_{mm'}n^{\bar{\sigma}}_{m'm}
& - &
\frac{U^{\sigma}+2J}{2(2l+1)}
& , \
N > 2l+1
\end{array}
\right.
\end{align}
\end{widetext}

BLOR has many similarities with existing functionals. For example, Himmetoglu's \cite{coco} DFT$+U+J$ functional was recently modified by Bajaj et al \cite{bajaj2017,bajaj2019} to obtain jmDFT, a functional designed to correct for deviations from the global flat plane condition. However, jmDFT fails to satisfy conditions 3 and 4. Meanwhile, setting $U^{\sigma}=U_{\rm eff}$, the first two terms of BLOR in the lower-half plane are equal to Dudarev's 1998 Hubbard functional. Furthermore, for non-spin polarised systems we have that $U^{\upharpoonright}=U^{\downharpoonright}=U-J$ and the BLOR functional in the lower half plane simplifies to Moynihan et al's DFT+$U$+$J$ method with self consistent formulae for the $U$ and $J$ parameters \cite{moynihan_thesis}.

Before BLOR is applied to test systems, the corrective parameters $U^{\sigma}$ and $J$ must first be carefully chosen. Our aim is to use BLOR to explicitly enforce the $E_{Hxc}$ flat plane condition on localized states embedded within a material environment. To achieve this, one can define the local curvature with respect to the spin resolved subspace occupancy $n^{\sigma}$ as:
\begin{equation}
\label{eqn:Udefine}
U^{\sigma}=\bigg(\frac{\partial^2 E_{Hxc}^{\rm approx}[\rho_{\rm loc}(\bf{r})]}{\partial (n^{\sigma})^2}\bigg)_{n^{-\sigma}}, 
\end{equation}
and with respect to the subspace magnetisation as:
\begin{equation}
\label{eqn:Jdefine}
J=-\bigg(\frac{\partial^2 E_{Hxc}^{\rm approx}[\rho_{\rm loc}(\bf{r})]}{\partial (M)^2}\bigg)_{N},
\end{equation}
where $N$ \& $M$ are the subspace electron count and magnetisation, and $\rho_{\rm loc}(\bf{r})$ is the electron density associated with the localized electrons. By explicitly enforcing the $E_{Hxc}$ flat plane condition on localized states we have implicitly assumed that all local curvature is spurious \cite{globalversuslocal}. This is true for an ensemble of isolated atomic/molecular species but in most practical cases this is an approximation.

In this work, the corrective parameters $U^{\sigma}$ and $J$ were not calculated directly as second-order partial derivatives as defined by equations \ref{eqn:Udefine} \& \ref{eqn:Jdefine}. We chose instead to compute the corrective parameters from the Hxc potential. This can be achieved using Linscott et al's minimum tracking linear response methodology \cite{linscott,moynihan}, which defines the spin-resolved Hxc kernel as:
\begin{equation}
f^{\sigma \sigma '}=\frac{\partial}{\partial n^{\sigma '}}\bigg(\frac{{\rm Tr}[\hat{P}^I \frac{\delta E_{\rm Hxc}}{\delta {\hat{\rho}^{\sigma}}}]}{{\rm Tr}[\hat{P}^I]}\bigg)_{n^{\bar{\sigma}'}},
\end{equation}
where $\frac{\delta E_{\rm Hxc}}{\delta {\hat{\rho}^{\sigma}}}$ is the Hxc potential operator. Within this formalism the spin resolved Hubbard parameters for BLOR can be set as the diagonal elements of the Hxc kernel:
\begin{equation}
U^{\sigma}=f^{\sigma \sigma }.
\end{equation}
In all other Hubbard functionals, the spin-agnostic $U$ parameter was evaluated by Linscott et al's simple $2\times2$ prescription:
\begin{equation}
\label{eqn:Usimple}
U=\frac{1}{4}(f^{\upharpoonright \upharpoonright}+f^{\upharpoonright \downharpoonright}+f^{\downharpoonright \upharpoonright}+f^{\upharpoonright \downharpoonright}), 
\end{equation}
Finally, the Hund's $J$ parameter can be computed as:
\begin{equation}
\label{eqn:Jsimple}
J=-\frac{1}{4}(f^{\upharpoonright \upharpoonright}-f^{\upharpoonright \downharpoonright}-f^{\downharpoonright \upharpoonright}+f^{\upharpoonright \downharpoonright}). 
\end{equation}

We note that by constructing $J$ from the elements of the spin-resolved Hxc kernel, we side-step the need to perform a constrained DFT calculation (as required by equation~\ref{eqn:Jdefine}): the above equation for $J$ obtains the same result via two unconstrained linear-response calculations \cite{linscott}.

As shown in \ref{sec:hubbard-param-tests}, the Hubbard functionals were also tested using a variety of prescriptions for the $U$ and $J$ parameters. All further computational details can be found in \ref{sec:comp-details}, while \ref{sec:param-table} gives a practical scheme for implementing the various corrective functionals.

In equations \ref{eqn:BLOR_trace_version} and \ref{eqn:BLOR_sum_version} BLOR is expressed in a generalized from which readily allows its implementation for $s$, $p$, $d$ or $f$ valence orbitals. However, in this letter we explore BLOR's application solely to $s$-valence species, in which case there is no ambiguity as to whether the local flat plane condition should be enforced on the localized subspace as a whole or on each localized orbital in the subspace separately. The later of these two options has been used to give BLOR in its current form however, analysis of this choice through bench marking with $p$ and $d$ valence species will be left to future work.

\begin{figure}[H]
\centering
\includegraphics[scale=0.2]{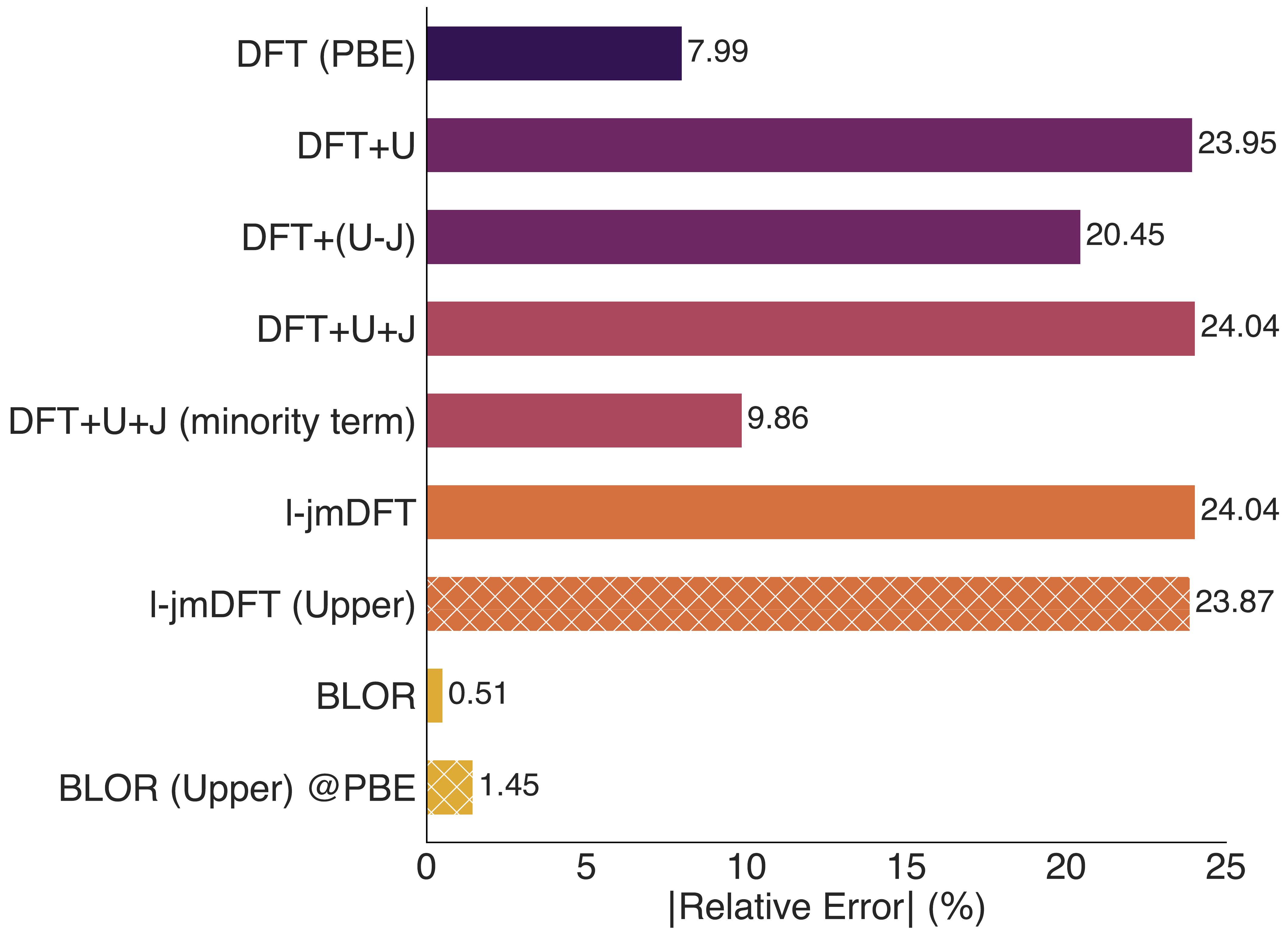}
\caption{ Bar chart of the relative errors in the total energies of H$_2$ at a bond length of 9 bohr radii using different corrective functionals \cite{PBE,dudarev,coco,bajaj2017,bajaj2019}. The raw DFT calculations were performed with the PBE exchange correlation functional \cite{PBE}. The DFT+$U$ and DFT$+(U-J)$ relative errors were computed using Dudarev et al's 1998 functional with the effective Hubbard parameter ($U_{\rm eff}$) set as $U$ and $U-J$ respectively. Both the lower and upper versions of l-jmDFT and BLOR are included in the bar chart. In the dissociated limit of H$_2$ each H atom will be singly occupied and hence the lower versions of BLOR and l-jmDFT offers the correct prescription for dissociated H$_2$ despite the computed occupancy being greater than one at the large but finite bond length of 9 bohr radii. The BLOR (Upper) corrective functional was evaluated on the PBE density as discussed in the main text.}
\label{fig:h2_bar_chart}
\end{figure}

The first set of $s$-valence systems BLOR was tested on were $s$-block dimers, (namely H$_2$, He$_2^+$, Li$_2$ and Be$_2^+$) with large internuclear separation lengths. It is assumed that at these elongated bond lengths the energy of the X$_2$ dimer:
\begin{equation}
E[{\rm X}_2]=2E[{\rm X}].
\end{equation}
The subspace occupancies of the atomic species will be located at the vertices of the diamond and hence the bare Perdew-Burke-Ernzerhof  approximation \cite{PBE} (PBE) is expected to be reasonably accurate for $E[{\rm X}]$. We thus assume that $2E_{\rm PBE}[{\rm X}]$ yields the exact total energy of our stretched ${\rm X}_2$ species. This approximation avoids discrepancies in the total energy caused by using a pseudopotential. The atomic subspaces of dissociated H$_2$ and Li$_2$ are approximately located along the $N=1$ line of the diamond (the fold) and are thus dominated by local-SCE. The atomic subspaces of dissociated He$_2^+$ and Be$_2^+$ are approximately located along the edges of the diamond and are thus dominated by local-MSIE. These errors will result in the computed $E[X_2] \neq 2E[X]$ for the stretched $X_2$ species. In figure~\ref{fig:h2_bar_chart} we present the relative errors in the total energies for H$_2$ at a bond length of 9 bohr radii using different corrective functionals. 

PBE yields a significant relative error of $7.99 \%$, however most of the corrective functionals significantly worsen the PBE result, yielding errors up to $24.04 \%$. Use of BLOR in the lower half-plane yields a very low error of $ 0.510 \%$. As shown in \ref{sec:dimer-bar-charts}, BLOR yields even lower relative errors for dissociated He$_2^+$, Li$_2$ and Be$_2^+$. 

In the bar chart jmDFT is denoted as l-jmDFT (localised-jmDFT). The jmDFT functional was designed to correct for deviations from the global flat plane condition and was the main inspiration for the development of BLOR, which instead focuses on the local flat plane condition. In this paper the jmDFT functional is implemented to correct for deviations from the local flat plane condition as opposed to the global equivalent. Furthermore, we use the simple $2 \times 2$ to compute the $U$ and $J$ parameters for the jmDFT functional, which is not how the functional was intended to be applied. The poor performance of l-jmDFT is thus unsurprising. 

For the DFT+$U+J$ method, inclusion of the minority spin term was found to lower the relative errors in the total energies across all five test cases. In this letter we present the DFT+$U$+$J$ results including the minority spin term, but results omitting the minority spin term can be found in the SI.

Excluding the minority spin term, it is possible to reformulate the DFT+$U$ and DFT+$U$+$J$ functionals in terms of an MSIE-term: ${\rm Tr}[\hat{N}-\hat{N}^2]$ and a SCE-term: ${\rm Tr}[\hat{M}^2-\hat{N}^2]$, with different linear combinations of $U$ and $J$ as prefactors. In the case of stretched H$_2$, the MSIE-term is negligable because the atomic occupancy is equal to one in the fully dissociated limit. Thus, the failure of the DFT+$U$ and DFT+$U$+$J$ functionals to predict the correct total energy can be attributed to the incorrect SCE-term prefactor. Indeed, computing the total energy of H$_2$ at a $9a_0$ bond length with both MSIE and magnetic-term prefactors equal to zero and the correct SCE prefactor of $J/2$ yields a relative error of $0.81\%$. 

Several of the corrective functionals (including BLOR) were found to yield the incorrect ordering of the KS orbitals upon self-consistent application of the corrective functional. Whenever this occurred the corrective functional was applied non-self consistently, i.e. the total energy was evaluated on the PBE density, hence we have BLOR@PBE etc. For all corrective functionals where no KS orbital re-ordering occurs, the total energy was evaluated both self-consistently and non-self-consistently and the difference between the two was found to be negligable. This demonstrates that BLOR yields correct total energies but fails to properly correct the KS potential, rectifying this issue will be left to future work. 

\begin{figure}[H]
\includegraphics[scale=0.2]{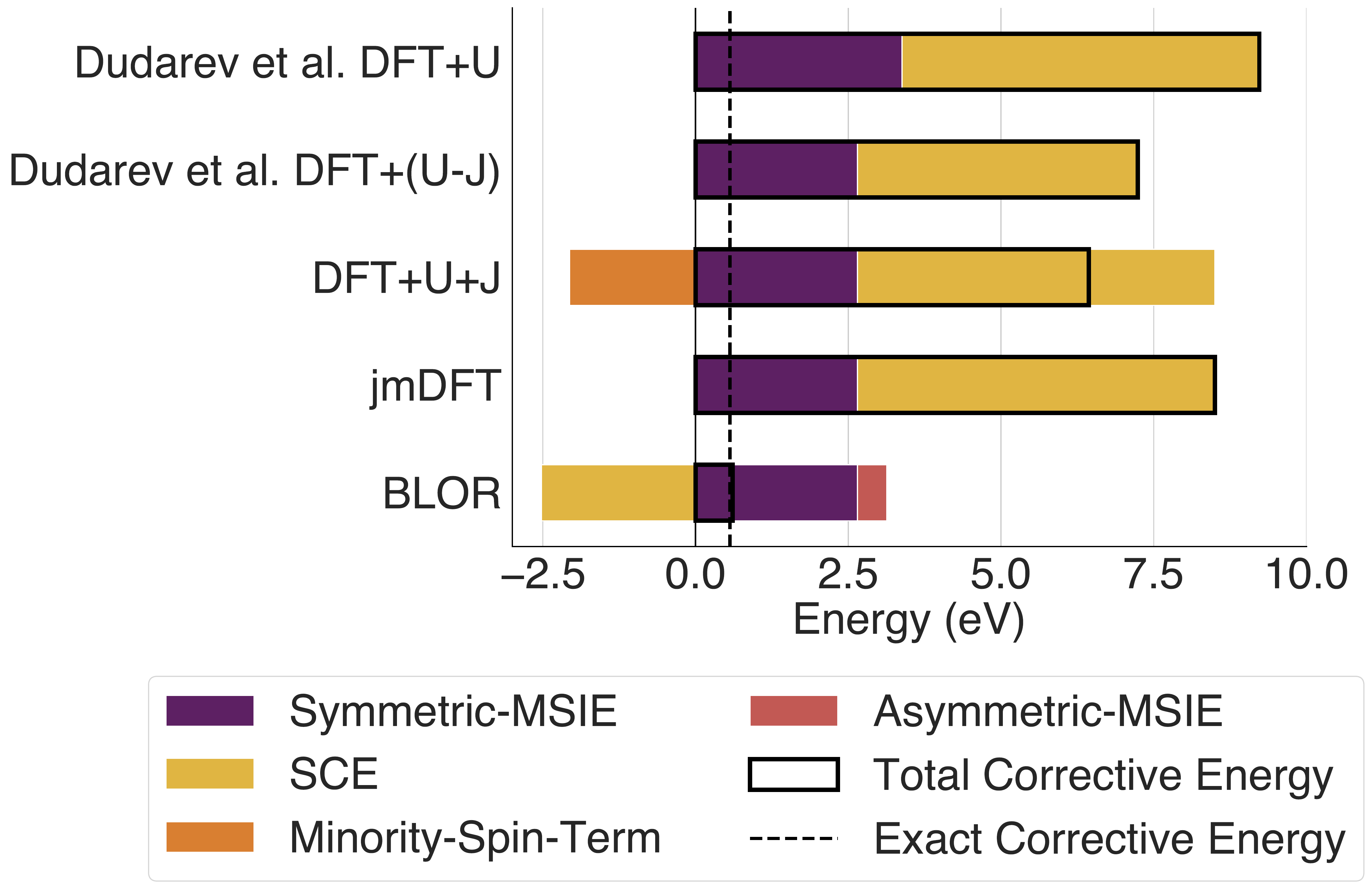}
\caption{The decomposition of the total corrective energy associated with several functionals \cite{dudarev,dudarev_2019,coco,bajaj2017,bajaj2019}, into a symmetric-MSIE term, a SCE term, a minority-spin term and an asymmetric-MSIE term.}
\label{fig:energy_decomposition}
\end{figure}

The second system we tested BLOR on was a dissociated hydrogen ring system, which suffers from both local-MSIE and local-SCE (in a system where both local-MSIE and local-SCE are present, error cancellation may occur). Dissociated H$_5^+$ is the smallest hydrogen ring system where: (1) the subspaces are not located along the edge or fold of the diamond and (2) the system does not suffer from KS orbital degeneracy problems (where a degenerate pair of KS orbitals is occupied by a single KS particle). All corrective functionals as well as bare PBE were found to give the incorrect KS orbital ordering for this system. To stabilise the correct  KS orbital ordering a potential of the form:
\begin{equation}
\label{eqn:stabilising_potential2}
\hat{v}^{\sigma}=G\hat{n}^{-\sigma}
\end{equation}
was applied to the atomic subspaces. The total PBE energy and spin resolved subspace occupancies were then evaluated as a function of $G$ and extrapolated to $G=0$ to get the correct PBE energy and spin resolved subspace occupancies. These occupancies were then used to obtain the total energy of the H$_5^+$ system evaluated on the PBE density for different corrective functionals.
\vspace{-0.6mm}
\begin{figure}[H]
\centering
\includegraphics[scale=0.35]{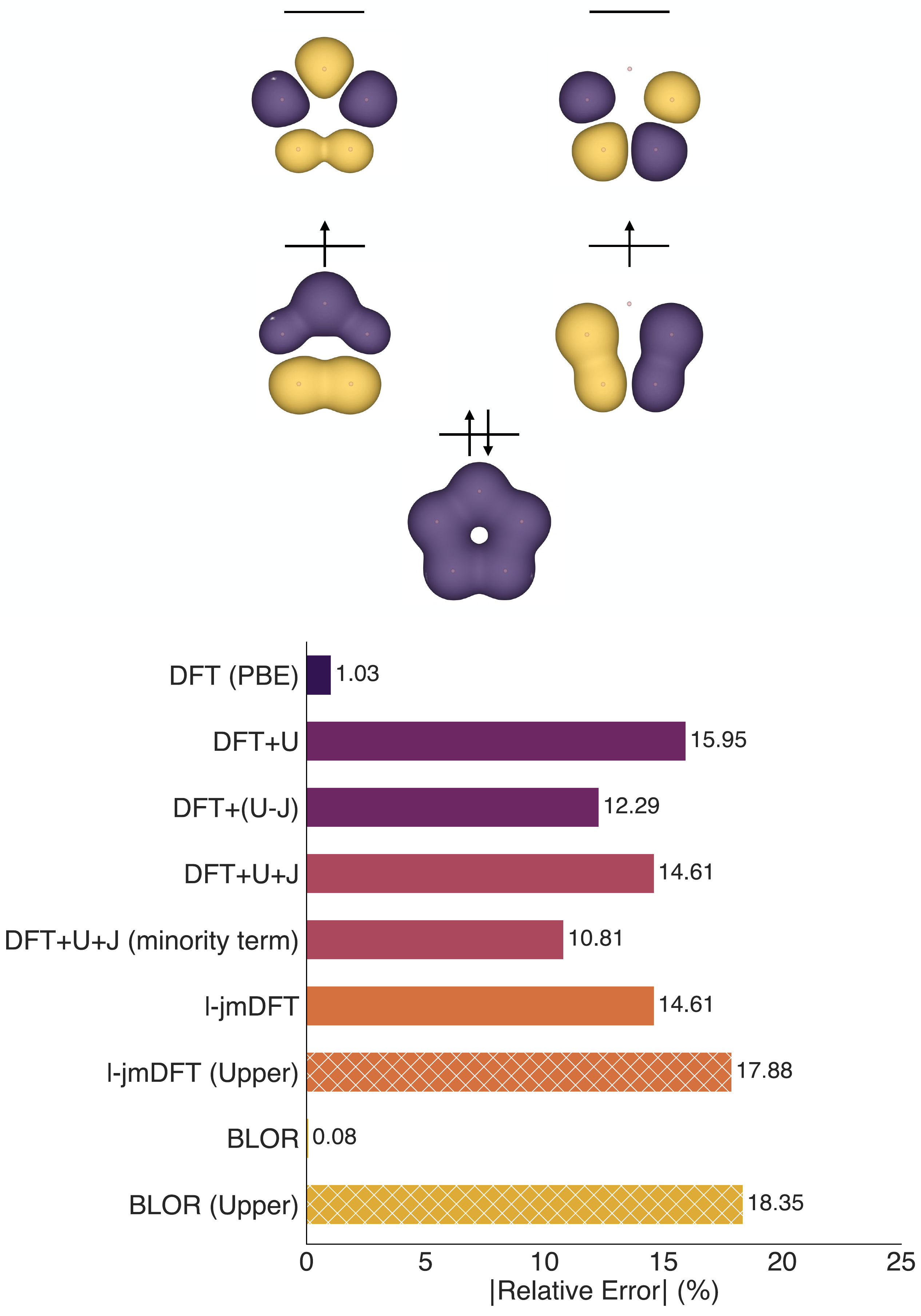}
\caption{The top panel displays the five lowest spin up Kohn Sham orbitals of the dissociated H$_5^+$ ring at an isosurface value of 0.003. The bottom planel displays a bar chart of the relative errors in the total energy of dissociated H$_5^+$ at an internuclear separation of 8 bohr radii using different corrective functionals \cite{PBE,kopernik,dudarev,dudarev_2019,moynihan_thesis,coco,shishkin2017,shishkin2019,seo2007,bajaj2017,bajaj2019}, which have been applied non-self consistently on the PBE density. The atomic subspace occupancy is significantly less than one and will be equal to 0.8 at the dissociated limit, thus the lower versions of BLOR and l-jmDFT offers the correct prescription for this system.}
\label{fig:h5_bar_chart}
\end{figure}
As shown in figure~\ref{fig:h5_bar_chart}, bare PBE yields a very low relative error of $1.03 \%$ for the dissociated H$_5^+$ system. Application of any functional is found to worsen the bare PBE result, with the exception of BLOR, which yields a relative error of $0.08\%$. This extremely low error is investigated further in figure \ref{fig:energy_decomposition}, where the total energy associated with several corrective functionals is decomposed into a symmetric-MSIE term, a SCE term and an asymmetric-MSIE term. All corrective functionals shown yield similar positive energies for the symmetric-MSIE term. However, the corrective functionals all yield large positive SCE terms with the exception of BLOR, which in turn leads to a significant overestimation of the total corrective energy. 
\vspace{3cm}
In conclusion, our newly derived corrective functional BLOR yielded relative energetic errors below $0.6\%$ across all five dissociated $s$-block species. This performance was unmatched by any of the other DFT$+U$ type functionals tested. But despite yielding highly accurate energetic results, the BLOR functional was found to worsen the KS potential. This problem was bypassed by evaluating the total energy  with the PBE density (BLOR@PBE) and a non-self consistent energy correction scheme was proposed for future practical use of the new functional. Most notably, our DFT+$U$ type corrective functional has been derived entirely from first principles and alleviates the need to rely on an ad hoc derivation from the Hubbard model. 

The research conducted in this publication was funded by the Irish Research Council under grant number GOIPG/2020/1454. All calculations were performed on the Boyle cluster maintained by the Trinity Centre for High Performance Computing. This cluster was funded through grants from the European Research Council and Science Foundation Ireland.

\bibliography{main}

\providecommand{\noopsort}[1]{}\providecommand{\singleletter}[1]{#1}%
\begin{thebibliography}{52}%
\makeatletter
\providecommand \@ifxundefined [1]{%
 \@ifx{#1\undefined}
}%
\providecommand \@ifnum [1]{%
 \ifnum #1\expandafter \@firstoftwo
 \else \expandafter \@secondoftwo
 \fi
}%
\providecommand \@ifx [1]{%
 \ifx #1\expandafter \@firstoftwo
 \else \expandafter \@secondoftwo
 \fi
}%
\providecommand \natexlab [1]{#1}%
\providecommand \enquote  [1]{``#1''}%
\providecommand \bibnamefont  [1]{#1}%
\providecommand \bibfnamefont [1]{#1}%
\providecommand \citenamefont [1]{#1}%
\providecommand \href@noop [0]{\@secondoftwo}%
\providecommand \href [0]{\begingroup \@sanitize@url \@href}%
\providecommand \@href[1]{\@@startlink{#1}\@@href}%
\providecommand \@@href[1]{\endgroup#1\@@endlink}%
\providecommand \@sanitize@url [0]{\catcode `\\12\catcode `\$12\catcode
  `\&12\catcode `\#12\catcode `\^12\catcode `\_12\catcode `\%12\relax}%
\providecommand \@@startlink[1]{}%
\providecommand \@@endlink[0]{}%
\providecommand \url  [0]{\begingroup\@sanitize@url \@url }%
\providecommand \@url [1]{\endgroup\@href {#1}{\urlprefix }}%
\providecommand \urlprefix  [0]{URL }%
\providecommand \Eprint [0]{\href }%
\providecommand \doibase [0]{https://doi.org/}%
\providecommand \selectlanguage [0]{\@gobble}%
\providecommand \bibinfo  [0]{\@secondoftwo}%
\providecommand \bibfield  [0]{\@secondoftwo}%
\providecommand \translation [1]{[#1]}%
\providecommand \BibitemOpen [0]{}%
\providecommand \bibitemStop [0]{}%
\providecommand \bibitemNoStop [0]{.\EOS\space}%
\providecommand \EOS [0]{\spacefactor3000\relax}%
\providecommand \BibitemShut  [1]{\csname bibitem#1\endcsname}%
\let\auto@bib@innerbib\@empty
\bibitem [{\citenamefont {Hohenberg}\ and\ \citenamefont
  {Kohn}(1964)}]{HK_theorems}%
  \BibitemOpen
  \bibfield  {author} {\bibinfo {author} {\bibfnamefont {P.}~\bibnamefont
  {Hohenberg}}\ and\ \bibinfo {author} {\bibfnamefont {W.}~\bibnamefont
  {Kohn}},\ }\bibfield  {title} {\bibinfo {title} {Inhomogeneous {{Electron
  Gas}}},\ }\href {https://doi.org/10.1103/PhysRev.136.B864} {\bibfield
  {journal} {\bibinfo  {journal} {Phys. Rev.}\ }\textbf {\bibinfo {volume}
  {136}},\ \bibinfo {pages} {B864} (\bibinfo {year} {1964})}\BibitemShut
  {NoStop}%
\bibitem [{\citenamefont {Vosko}\ \emph {et~al.}(1980)\citenamefont {Vosko},
  \citenamefont {Wilk},\ and\ \citenamefont {Nusair}}]{LSDA}%
  \BibitemOpen
  \bibfield  {author} {\bibinfo {author} {\bibfnamefont {S.~H.}\ \bibnamefont
  {Vosko}}, \bibinfo {author} {\bibfnamefont {L.}~\bibnamefont {Wilk}},\ and\
  \bibinfo {author} {\bibfnamefont {M.}~\bibnamefont {Nusair}},\ }\bibfield
  {title} {\bibinfo {title} {Accurate spin-dependent electron liquid
  correlation energies for local spin density calculations: A critical
  analysis},\ }\href {https://doi.org/10.1139/p80-159} {\bibfield  {journal}
  {\bibinfo  {journal} {Can. J. Phys.}\ }\textbf {\bibinfo {volume} {58}},\
  \bibinfo {pages} {1200} (\bibinfo {year} {1980})}\BibitemShut {NoStop}%
\bibitem [{\citenamefont {Perdew}\ \emph {et~al.}(1996)\citenamefont {Perdew},
  \citenamefont {Burke},\ and\ \citenamefont {Ernzerhof}}]{PBE}%
  \BibitemOpen
  \bibfield  {author} {\bibinfo {author} {\bibfnamefont {J.~P.}\ \bibnamefont
  {Perdew}}, \bibinfo {author} {\bibfnamefont {K.}~\bibnamefont {Burke}},\ and\
  \bibinfo {author} {\bibfnamefont {M.}~\bibnamefont {Ernzerhof}},\ }\bibfield
  {title} {\bibinfo {title} {Generalized {{Gradient Approximation Made
  Simple}}},\ }\href {https://doi.org/10.1103/PhysRevLett.77.3865} {\bibfield
  {journal} {\bibinfo  {journal} {Phys. Rev. Lett.}\ }\textbf {\bibinfo
  {volume} {77}},\ \bibinfo {pages} {3865} (\bibinfo {year}
  {1996})}\BibitemShut {NoStop}%
\bibitem [{\citenamefont {Perdew}\ \emph {et~al.}(2008)\citenamefont {Perdew},
  \citenamefont {Ruzsinszky}, \citenamefont {Csonka}, \citenamefont {Vydrov},
  \citenamefont {Scuseria}, \citenamefont {Constantin}, \citenamefont {Zhou},\
  and\ \citenamefont {Burke}}]{PBEsol}%
  \BibitemOpen
  \bibfield  {author} {\bibinfo {author} {\bibfnamefont {J.~P.}\ \bibnamefont
  {Perdew}}, \bibinfo {author} {\bibfnamefont {A.}~\bibnamefont {Ruzsinszky}},
  \bibinfo {author} {\bibfnamefont {G.~I.}\ \bibnamefont {Csonka}}, \bibinfo
  {author} {\bibfnamefont {O.~A.}\ \bibnamefont {Vydrov}}, \bibinfo {author}
  {\bibfnamefont {G.~E.}\ \bibnamefont {Scuseria}}, \bibinfo {author}
  {\bibfnamefont {L.~A.}\ \bibnamefont {Constantin}}, \bibinfo {author}
  {\bibfnamefont {X.}~\bibnamefont {Zhou}},\ and\ \bibinfo {author}
  {\bibfnamefont {K.}~\bibnamefont {Burke}},\ }\bibfield  {title} {\bibinfo
  {title} {Restoring the {{Density-Gradient Expansion}} for {{Exchange}} in
  {{Solids}} and {{Surfaces}}},\ }\href
  {https://doi.org/10.1103/PhysRevLett.100.136406} {\bibfield  {journal}
  {\bibinfo  {journal} {Phys. Rev. Lett.}\ }\textbf {\bibinfo {volume} {100}},\
  \bibinfo {pages} {136406} (\bibinfo {year} {2008})}\BibitemShut {NoStop}%
\bibitem [{\citenamefont {Becke}(1988)}]{becke88}%
  \BibitemOpen
  \bibfield  {author} {\bibinfo {author} {\bibfnamefont {A.~D.}\ \bibnamefont
  {Becke}},\ }\bibfield  {title} {\bibinfo {title} {Density-functional
  exchange-energy approximation with correct asymptotic behavior},\ }\href
  {https://doi.org/10.1103/PhysRevA.38.3098} {\bibfield  {journal} {\bibinfo
  {journal} {Phys. Rev. A}\ }\textbf {\bibinfo {volume} {38}},\ \bibinfo
  {pages} {3098} (\bibinfo {year} {1988})}\BibitemShut {NoStop}%
\bibitem [{\citenamefont {Lee}\ \emph {et~al.}(1988)\citenamefont {Lee},
  \citenamefont {Yang},\ and\ \citenamefont {Parr}}]{LYP}%
  \BibitemOpen
  \bibfield  {author} {\bibinfo {author} {\bibfnamefont {C.}~\bibnamefont
  {Lee}}, \bibinfo {author} {\bibfnamefont {W.}~\bibnamefont {Yang}},\ and\
  \bibinfo {author} {\bibfnamefont {R.~G.}\ \bibnamefont {Parr}},\ }\bibfield
  {title} {\bibinfo {title} {Development of the {{Colle-Salvetti}}
  correlation-energy formula into a functional of the electron density},\
  }\href {https://doi.org/10.1103/PhysRevB.37.785} {\bibfield  {journal}
  {\bibinfo  {journal} {Phys. Rev. B}\ }\textbf {\bibinfo {volume} {37}},\
  \bibinfo {pages} {785} (\bibinfo {year} {1988})}\BibitemShut {NoStop}%
\bibitem [{\citenamefont {Becke}(1993)}]{becke93}%
  \BibitemOpen
  \bibfield  {author} {\bibinfo {author} {\bibfnamefont {A.~D.}\ \bibnamefont
  {Becke}},\ }\bibfield  {title} {\bibinfo {title} {Density-functional
  thermochemistry. {{III}}. {{The}} role of exact exchange},\ }\href
  {https://doi.org/10.1063/1.464913} {\bibfield  {journal} {\bibinfo  {journal}
  {The Journal of Chemical Physics}\ }\textbf {\bibinfo {volume} {98}},\
  \bibinfo {pages} {5648} (\bibinfo {year} {1993})}\BibitemShut {NoStop}%
\bibitem [{\citenamefont {Heyd}\ \emph {et~al.}(2003)\citenamefont {Heyd},
  \citenamefont {Scuseria},\ and\ \citenamefont {Ernzerhof}}]{HSE}%
  \BibitemOpen
  \bibfield  {author} {\bibinfo {author} {\bibfnamefont {J.}~\bibnamefont
  {Heyd}}, \bibinfo {author} {\bibfnamefont {G.~E.}\ \bibnamefont {Scuseria}},\
  and\ \bibinfo {author} {\bibfnamefont {M.}~\bibnamefont {Ernzerhof}},\
  }\bibfield  {title} {\bibinfo {title} {Hybrid functionals based on a screened
  {{Coulomb}} potential},\ }\href {https://doi.org/10.1063/1.1564060}
  {\bibfield  {journal} {\bibinfo  {journal} {J. Chem. Phys.}\ }\textbf
  {\bibinfo {volume} {118}},\ \bibinfo {pages} {8207} (\bibinfo {year}
  {2003})}\BibitemShut {NoStop}%
\bibitem [{\citenamefont {Sun}\ \emph {et~al.}(2015)\citenamefont {Sun},
  \citenamefont {Ruzsinszky},\ and\ \citenamefont {Perdew}}]{SCAN}%
  \BibitemOpen
  \bibfield  {author} {\bibinfo {author} {\bibfnamefont {J.}~\bibnamefont
  {Sun}}, \bibinfo {author} {\bibfnamefont {A.}~\bibnamefont {Ruzsinszky}},\
  and\ \bibinfo {author} {\bibfnamefont {J.~P.}\ \bibnamefont {Perdew}},\
  }\bibfield  {title} {\bibinfo {title} {Strongly {{Constrained}} and
  {{Appropriately Normed Semilocal Density Functional}}},\ }\href
  {https://doi.org/10.1103/PhysRevLett.115.036402} {\bibfield  {journal}
  {\bibinfo  {journal} {Phys. Rev. Lett.}\ }\textbf {\bibinfo {volume} {115}},\
  \bibinfo {pages} {036402} (\bibinfo {year} {2015})}\BibitemShut {NoStop}%
\bibitem [{\citenamefont {Tao}\ \emph {et~al.}(2003)\citenamefont {Tao},
  \citenamefont {Perdew}, \citenamefont {Staroverov},\ and\ \citenamefont
  {Scuseria}}]{TPSS}%
  \BibitemOpen
  \bibfield  {author} {\bibinfo {author} {\bibfnamefont {J.}~\bibnamefont
  {Tao}}, \bibinfo {author} {\bibfnamefont {J.~P.}\ \bibnamefont {Perdew}},
  \bibinfo {author} {\bibfnamefont {V.~N.}\ \bibnamefont {Staroverov}},\ and\
  \bibinfo {author} {\bibfnamefont {G.~E.}\ \bibnamefont {Scuseria}},\
  }\bibfield  {title} {\bibinfo {title} {Climbing the {{Density Functional
  Ladder}}: {{Nonempirical Meta--Generalized Gradient Approximation Designed}}
  for {{Molecules}} and {{Solids}}},\ }\href
  {https://doi.org/10.1103/PhysRevLett.91.146401} {\bibfield  {journal}
  {\bibinfo  {journal} {Phys. Rev. Lett.}\ }\textbf {\bibinfo {volume} {91}},\
  \bibinfo {pages} {146401} (\bibinfo {year} {2003})}\BibitemShut {NoStop}%
\bibitem [{\citenamefont {Mardirossian}\ and\ \citenamefont
  {{Head-Gordon}}(2016)}]{wB97M-V}%
  \BibitemOpen
  \bibfield  {author} {\bibinfo {author} {\bibfnamefont {N.}~\bibnamefont
  {Mardirossian}}\ and\ \bibinfo {author} {\bibfnamefont {M.}~\bibnamefont
  {{Head-Gordon}}},\ }\bibfield  {title} {\bibinfo {title} {{{$\omega$B97M-V}}:
  {{A}} combinatorially optimized, range-separated hybrid, meta-{{GGA}} density
  functional with {{VV10}} nonlocal correlation},\ }\href
  {https://doi.org/10.1063/1.4952647} {\bibfield  {journal} {\bibinfo
  {journal} {J. Chem. Phys.}\ }\textbf {\bibinfo {volume} {144}},\ \bibinfo
  {pages} {214110} (\bibinfo {year} {2016})}\BibitemShut {NoStop}%
\bibitem [{\citenamefont {Haiduke}\ and\ \citenamefont
  {Bartlett}(2018)}]{CAM-QTP-02}%
  \BibitemOpen
  \bibfield  {author} {\bibinfo {author} {\bibfnamefont {R.~L.~A.}\
  \bibnamefont {Haiduke}}\ and\ \bibinfo {author} {\bibfnamefont {R.~J.}\
  \bibnamefont {Bartlett}},\ }\bibfield  {title} {\bibinfo {title}
  {Non-empirical exchange-correlation parameterizations based on exact
  conditions from correlated orbital theory},\ }\href
  {https://doi.org/10.1063/1.5025723} {\bibfield  {journal} {\bibinfo
  {journal} {J. Chem. Phys.}\ }\textbf {\bibinfo {volume} {148}},\ \bibinfo
  {pages} {184106} (\bibinfo {year} {2018})}\BibitemShut {NoStop}%
\bibitem [{\citenamefont {Lin}\ and\ \citenamefont
  {Van~Voorhis}(2019)}]{TTwPBEh}%
  \BibitemOpen
  \bibfield  {author} {\bibinfo {author} {\bibfnamefont {Z.}~\bibnamefont
  {Lin}}\ and\ \bibinfo {author} {\bibfnamefont {T.}~\bibnamefont
  {Van~Voorhis}},\ }\bibfield  {title} {\bibinfo {title} {Triplet tuning: A
  novel family of non-empirical exchange-correlation functionals},\ }\href
  {https://doi.org/10.1021/acs.jctc.8b00853} {\bibfield  {journal} {\bibinfo
  {journal} {J. Chem. Theory Comput.}\ }\textbf {\bibinfo {volume} {15}},\
  \bibinfo {pages} {1226} (\bibinfo {year} {2019})}\BibitemShut {NoStop}%
\bibitem [{\citenamefont {{de Jong}}\ \emph {et~al.}(2015)\citenamefont {{de
  Jong}}, \citenamefont {Chen}, \citenamefont {Angsten}, \citenamefont {Jain},
  \citenamefont {Notestine}, \citenamefont {Gamst}, \citenamefont {Sluiter},
  \citenamefont {Krishna~Ande}, \citenamefont {{van der Zwaag}}, \citenamefont
  {Plata}, \citenamefont {Toher}, \citenamefont {Curtarolo}, \citenamefont
  {Ceder}, \citenamefont {Persson},\ and\ \citenamefont
  {Asta}}]{big_screening_of_mechanical_props}%
  \BibitemOpen
  \bibfield  {author} {\bibinfo {author} {\bibfnamefont {M.}~\bibnamefont {{de
  Jong}}}, \bibinfo {author} {\bibfnamefont {W.}~\bibnamefont {Chen}}, \bibinfo
  {author} {\bibfnamefont {T.}~\bibnamefont {Angsten}}, \bibinfo {author}
  {\bibfnamefont {A.}~\bibnamefont {Jain}}, \bibinfo {author} {\bibfnamefont
  {R.}~\bibnamefont {Notestine}}, \bibinfo {author} {\bibfnamefont
  {A.}~\bibnamefont {Gamst}}, \bibinfo {author} {\bibfnamefont
  {M.}~\bibnamefont {Sluiter}}, \bibinfo {author} {\bibfnamefont
  {C.}~\bibnamefont {Krishna~Ande}}, \bibinfo {author} {\bibfnamefont
  {S.}~\bibnamefont {{van der Zwaag}}}, \bibinfo {author} {\bibfnamefont
  {J.~J.}\ \bibnamefont {Plata}}, \bibinfo {author} {\bibfnamefont
  {C.}~\bibnamefont {Toher}}, \bibinfo {author} {\bibfnamefont
  {S.}~\bibnamefont {Curtarolo}}, \bibinfo {author} {\bibfnamefont
  {G.}~\bibnamefont {Ceder}}, \bibinfo {author} {\bibfnamefont {K.~A.}\
  \bibnamefont {Persson}},\ and\ \bibinfo {author} {\bibfnamefont
  {M.}~\bibnamefont {Asta}},\ }\bibfield  {title} {\bibinfo {title} {Charting
  the complete elastic properties of inorganic crystalline compounds},\ }\href
  {https://doi.org/10.1038/sdata.2015.9} {\bibfield  {journal} {\bibinfo
  {journal} {Sci Data}\ }\textbf {\bibinfo {volume} {2}},\ \bibinfo {pages}
  {150009} (\bibinfo {year} {2015})}\BibitemShut {NoStop}%
\bibitem [{\citenamefont {Zilka}\ \emph {et~al.}(2017)\citenamefont {Zilka},
  \citenamefont {Dudenko}, \citenamefont {Hughes}, \citenamefont {Williams},
  \citenamefont {Sturniolo}, \citenamefont {Franks}, \citenamefont {Pickard},
  \citenamefont {Yates}, \citenamefont {Harris},\ and\ \citenamefont
  {Brown}}]{pickard}%
  \BibitemOpen
  \bibfield  {author} {\bibinfo {author} {\bibfnamefont {M.}~\bibnamefont
  {Zilka}}, \bibinfo {author} {\bibfnamefont {D.~V.}\ \bibnamefont {Dudenko}},
  \bibinfo {author} {\bibfnamefont {C.~E.}\ \bibnamefont {Hughes}}, \bibinfo
  {author} {\bibfnamefont {P.~A.}\ \bibnamefont {Williams}}, \bibinfo {author}
  {\bibfnamefont {S.}~\bibnamefont {Sturniolo}}, \bibinfo {author}
  {\bibfnamefont {W.~T.}\ \bibnamefont {Franks}}, \bibinfo {author}
  {\bibfnamefont {C.~J.}\ \bibnamefont {Pickard}}, \bibinfo {author}
  {\bibfnamefont {J.~R.}\ \bibnamefont {Yates}}, \bibinfo {author}
  {\bibfnamefont {K.~D.~M.}\ \bibnamefont {Harris}},\ and\ \bibinfo {author}
  {\bibfnamefont {S.~P.}\ \bibnamefont {Brown}},\ }\bibfield  {title} {\bibinfo
  {title} {Ab initio random structure searching of organic molecular solids:
  Assessment and validation against experimental data},\ }\href
  {https://doi.org/10.1039/C7CP04186A} {\bibfield  {journal} {\bibinfo
  {journal} {Phys. Chem. Chem. Phys.}\ }\textbf {\bibinfo {volume} {19}},\
  \bibinfo {pages} {25949} (\bibinfo {year} {2017})}\BibitemShut {NoStop}%
\bibitem [{\citenamefont {Ruzsinszky}\ \emph {et~al.}(2006)\citenamefont
  {Ruzsinszky}, \citenamefont {Perdew}, \citenamefont {Csonka}, \citenamefont
  {Vydrov},\ and\ \citenamefont {Scuseria}}]{perdew_bond_dissocation}%
  \BibitemOpen
  \bibfield  {author} {\bibinfo {author} {\bibfnamefont {A.}~\bibnamefont
  {Ruzsinszky}}, \bibinfo {author} {\bibfnamefont {J.~P.}\ \bibnamefont
  {Perdew}}, \bibinfo {author} {\bibfnamefont {G.~I.}\ \bibnamefont {Csonka}},
  \bibinfo {author} {\bibfnamefont {O.~A.}\ \bibnamefont {Vydrov}},\ and\
  \bibinfo {author} {\bibfnamefont {G.~E.}\ \bibnamefont {Scuseria}},\
  }\bibfield  {title} {\bibinfo {title} {Spurious fractional charge on
  dissociated atoms: {{Pervasive}} and resilient self-interaction error of
  common density functionals},\ }\href {https://doi.org/10.1063/1.2387954}
  {\bibfield  {journal} {\bibinfo  {journal} {J. Chem. Phys.}\ }\textbf
  {\bibinfo {volume} {125}},\ \bibinfo {pages} {194112} (\bibinfo {year}
  {2006})}\BibitemShut {NoStop}%
\bibitem [{\citenamefont {Dutoi}\ and\ \citenamefont
  {{Head-Gordon}}(2006)}]{dutoi_bond_dissocation}%
  \BibitemOpen
  \bibfield  {author} {\bibinfo {author} {\bibfnamefont {A.~D.}\ \bibnamefont
  {Dutoi}}\ and\ \bibinfo {author} {\bibfnamefont {M.}~\bibnamefont
  {{Head-Gordon}}},\ }\bibfield  {title} {\bibinfo {title} {Self-interaction
  error of local density functionals for alkali\textendash halide
  dissociation},\ }\href {https://doi.org/10.1016/j.cplett.2006.02.025}
  {\bibfield  {journal} {\bibinfo  {journal} {Chemical Physics Letters}\
  }\textbf {\bibinfo {volume} {422}},\ \bibinfo {pages} {230} (\bibinfo {year}
  {2006})}\BibitemShut {NoStop}%
\bibitem [{\citenamefont {Nafziger}\ and\ \citenamefont
  {Wasserman}(2015)}]{partitionDFT}%
  \BibitemOpen
  \bibfield  {author} {\bibinfo {author} {\bibfnamefont {J.}~\bibnamefont
  {Nafziger}}\ and\ \bibinfo {author} {\bibfnamefont {A.}~\bibnamefont
  {Wasserman}},\ }\bibfield  {title} {\bibinfo {title} {Fragment-based
  treatment of delocalization and static correlation errors in
  density-functional theory},\ }\href {https://doi.org/10.1063/1.4937771}
  {\bibfield  {journal} {\bibinfo  {journal} {J. Chem. Phys.}\ }\textbf
  {\bibinfo {volume} {143}},\ \bibinfo {pages} {234105} (\bibinfo {year}
  {2015})}\BibitemShut {NoStop}%
\bibitem [{\citenamefont {Perdew}(1985)}]{perdew_band_gap_problem}%
  \BibitemOpen
  \bibfield  {author} {\bibinfo {author} {\bibfnamefont {J.~P.}\ \bibnamefont
  {Perdew}},\ }\bibfield  {title} {\bibinfo {title} {Density functional theory
  and the band gap problem},\ }\href {https://doi.org/10.1002/qua.560280846}
  {\bibfield  {journal} {\bibinfo  {journal} {International Journal of Quantum
  Chemistry}\ }\textbf {\bibinfo {volume} {28}},\ \bibinfo {pages} {497}
  (\bibinfo {year} {1985})}\BibitemShut {NoStop}%
\bibitem [{\citenamefont {Borlido}\ \emph {et~al.}(2019)\citenamefont
  {Borlido}, \citenamefont {Aull}, \citenamefont {Huran}, \citenamefont {Tran},
  \citenamefont {Marques},\ and\ \citenamefont {Botti}}]{bandgap_benchmark}%
  \BibitemOpen
  \bibfield  {author} {\bibinfo {author} {\bibfnamefont {P.}~\bibnamefont
  {Borlido}}, \bibinfo {author} {\bibfnamefont {T.}~\bibnamefont {Aull}},
  \bibinfo {author} {\bibfnamefont {A.~W.}\ \bibnamefont {Huran}}, \bibinfo
  {author} {\bibfnamefont {F.}~\bibnamefont {Tran}}, \bibinfo {author}
  {\bibfnamefont {M.~A.~L.}\ \bibnamefont {Marques}},\ and\ \bibinfo {author}
  {\bibfnamefont {S.}~\bibnamefont {Botti}},\ }\bibfield  {title} {\bibinfo
  {title} {Large-{{Scale Benchmark}} of {{Exchange}}\textendash{{Correlation
  Functionals}} for the {{Determination}} of {{Electronic Band Gaps}} of
  {{Solids}}},\ }\href {https://doi.org/10.1021/acs.jctc.9b00322} {\bibfield
  {journal} {\bibinfo  {journal} {J. Chem. Theory Comput.}\ }\textbf {\bibinfo
  {volume} {15}},\ \bibinfo {pages} {5069} (\bibinfo {year}
  {2019})}\BibitemShut {NoStop}%
\bibitem [{\citenamefont {Cohen}\ \emph
  {et~al.}(2008{\natexlab{a}})\citenamefont {Cohen}, \citenamefont
  {{Mori-S{\'a}nchez}},\ and\ \citenamefont {Yang}}]{yang_bandgap}%
  \BibitemOpen
  \bibfield  {author} {\bibinfo {author} {\bibfnamefont {A.~J.}\ \bibnamefont
  {Cohen}}, \bibinfo {author} {\bibfnamefont {P.}~\bibnamefont
  {{Mori-S{\'a}nchez}}},\ and\ \bibinfo {author} {\bibfnamefont
  {W.}~\bibnamefont {Yang}},\ }\bibfield  {title} {\bibinfo {title} {Fractional
  charge perspective on the band gap in density-functional theory},\ }\href
  {https://doi.org/10.1103/PhysRevB.77.115123} {\bibfield  {journal} {\bibinfo
  {journal} {Phys. Rev. B}\ }\textbf {\bibinfo {volume} {77}},\ \bibinfo
  {pages} {115123} (\bibinfo {year} {2008}{\natexlab{a}})}\BibitemShut
  {NoStop}%
\bibitem [{\citenamefont {Zhu}\ and\ \citenamefont {Gao}(2014)}]{TiO2}%
  \BibitemOpen
  \bibfield  {author} {\bibinfo {author} {\bibfnamefont {T.}~\bibnamefont
  {Zhu}}\ and\ \bibinfo {author} {\bibfnamefont {S.-P.}\ \bibnamefont {Gao}},\
  }\bibfield  {title} {\bibinfo {title} {The {{Stability}}, {{Electronic
  Structure}}, and {{Optical Property}} of {{{TiO}$_2$ Polymorphs}}},\ }\href
  {https://doi.org/10.1021/jp412462m} {\bibfield  {journal} {\bibinfo
  {journal} {J. Phys. Chem. C}\ }\textbf {\bibinfo {volume} {118}},\ \bibinfo
  {pages} {11385} (\bibinfo {year} {2014})}\BibitemShut {NoStop}%
\bibitem [{\citenamefont {Schr{\"o}n}\ \emph {et~al.}(2010)\citenamefont
  {Schr{\"o}n}, \citenamefont {R{\"o}dl},\ and\ \citenamefont
  {Bechstedt}}]{MnO}%
  \BibitemOpen
  \bibfield  {author} {\bibinfo {author} {\bibfnamefont {A.}~\bibnamefont
  {Schr{\"o}n}}, \bibinfo {author} {\bibfnamefont {C.}~\bibnamefont
  {R{\"o}dl}},\ and\ \bibinfo {author} {\bibfnamefont {F.}~\bibnamefont
  {Bechstedt}},\ }\bibfield  {title} {\bibinfo {title} {Energetic stability and
  magnetic properties of {{MnO}} in the rocksalt, wurtzite, and zinc-blende
  structures: {{Influence}} of exchange and correlation},\ }\href
  {https://doi.org/10.1103/PhysRevB.82.165109} {\bibfield  {journal} {\bibinfo
  {journal} {Phys. Rev. B}\ }\textbf {\bibinfo {volume} {82}},\ \bibinfo
  {pages} {165109} (\bibinfo {year} {2010})}\BibitemShut {NoStop}%
\bibitem [{\citenamefont {Sai~Gautam}\ and\ \citenamefont
  {Carter}(2018)}]{carter}%
  \BibitemOpen
  \bibfield  {author} {\bibinfo {author} {\bibfnamefont {G.}~\bibnamefont
  {Sai~Gautam}}\ and\ \bibinfo {author} {\bibfnamefont {E.~A.}\ \bibnamefont
  {Carter}},\ }\bibfield  {title} {\bibinfo {title} {Evaluating transition
  metal oxides within {{DFT-SCAN}} and {{SCAN+U}} frameworks for solar
  thermochemical applications},\ }\href
  {https://doi.org/10.1103/PhysRevMaterials.2.095401} {\bibfield  {journal}
  {\bibinfo  {journal} {Phys. Rev. Materials}\ }\textbf {\bibinfo {volume}
  {2}},\ \bibinfo {pages} {095401} (\bibinfo {year} {2018})}\BibitemShut
  {NoStop}%
\bibitem [{\citenamefont {Perdew}\ \emph {et~al.}(1982)\citenamefont {Perdew},
  \citenamefont {Parr}, \citenamefont {Levy},\ and\ \citenamefont
  {Balduz}}]{pwl}%
  \BibitemOpen
  \bibfield  {author} {\bibinfo {author} {\bibfnamefont {J.~P.}\ \bibnamefont
  {Perdew}}, \bibinfo {author} {\bibfnamefont {R.~G.}\ \bibnamefont {Parr}},
  \bibinfo {author} {\bibfnamefont {M.}~\bibnamefont {Levy}},\ and\ \bibinfo
  {author} {\bibfnamefont {J.~L.}\ \bibnamefont {Balduz}},\ }\bibfield  {title}
  {\bibinfo {title} {Density-{{Functional Theory}} for {{Fractional Particle
  Number}}: {{Derivative Discontinuities}} of the {{Energy}}},\ }\href
  {https://doi.org/10.1103/PhysRevLett.49.1691} {\bibfield  {journal} {\bibinfo
   {journal} {Phys. Rev. Lett.}\ }\textbf {\bibinfo {volume} {49}},\ \bibinfo
  {pages} {1691} (\bibinfo {year} {1982})}\BibitemShut {NoStop}%
\bibitem [{\citenamefont {Yang}\ \emph {et~al.}(2000)\citenamefont {Yang},
  \citenamefont {Zhang},\ and\ \citenamefont {Ayers}}]{pwc_part1}%
  \BibitemOpen
  \bibfield  {author} {\bibinfo {author} {\bibfnamefont {W.}~\bibnamefont
  {Yang}}, \bibinfo {author} {\bibfnamefont {Y.}~\bibnamefont {Zhang}},\ and\
  \bibinfo {author} {\bibfnamefont {P.~W.}\ \bibnamefont {Ayers}},\ }\bibfield
  {title} {\bibinfo {title} {Degenerate {{Ground States}} and a {{Fractional
  Number}} of {{Electrons}} in {{Density}} and {{Reduced Density Matrix
  Functional Theory}}},\ }\href {https://doi.org/10.1103/PhysRevLett.84.5172}
  {\bibfield  {journal} {\bibinfo  {journal} {Phys. Rev. Lett.}\ }\textbf
  {\bibinfo {volume} {84}},\ \bibinfo {pages} {5172} (\bibinfo {year}
  {2000})}\BibitemShut {NoStop}%
\bibitem [{\citenamefont {Cohen}\ \emph
  {et~al.}(2008{\natexlab{b}})\citenamefont {Cohen}, \citenamefont
  {{Mori-S{\'a}nchez}},\ and\ \citenamefont {Yang}}]{pwc_part2}%
  \BibitemOpen
  \bibfield  {author} {\bibinfo {author} {\bibfnamefont {A.~J.}\ \bibnamefont
  {Cohen}}, \bibinfo {author} {\bibfnamefont {P.}~\bibnamefont
  {{Mori-S{\'a}nchez}}},\ and\ \bibinfo {author} {\bibfnamefont
  {W.}~\bibnamefont {Yang}},\ }\bibfield  {title} {\bibinfo {title} {Fractional
  spins and static correlation error in density functional theory},\ }\href
  {https://doi.org/10.1063/1.2987202} {\bibfield  {journal} {\bibinfo
  {journal} {J. Chem. Phys.}\ }\textbf {\bibinfo {volume} {129}},\ \bibinfo
  {pages} {121104} (\bibinfo {year} {2008}{\natexlab{b}})}\BibitemShut
  {NoStop}%
\bibitem [{\citenamefont {{Mori-S{\'a}nchez}}\ \emph
  {et~al.}(2006)\citenamefont {{Mori-S{\'a}nchez}}, \citenamefont {Cohen},\
  and\ \citenamefont {Yang}}]{MSIE}%
  \BibitemOpen
  \bibfield  {author} {\bibinfo {author} {\bibfnamefont {P.}~\bibnamefont
  {{Mori-S{\'a}nchez}}}, \bibinfo {author} {\bibfnamefont {A.~J.}\ \bibnamefont
  {Cohen}},\ and\ \bibinfo {author} {\bibfnamefont {W.}~\bibnamefont {Yang}},\
  }\bibfield  {title} {\bibinfo {title} {Many-electron self-interaction error
  in approximate density functionals},\ }\href
  {https://doi.org/10.1063/1.2403848} {\bibfield  {journal} {\bibinfo
  {journal} {J. Chem. Phys.}\ }\textbf {\bibinfo {volume} {125}},\ \bibinfo
  {pages} {201102} (\bibinfo {year} {2006})}\BibitemShut {NoStop}%
\bibitem [{\citenamefont {{Mori-S{\'a}nchez}}\ \emph
  {et~al.}(2009)\citenamefont {{Mori-S{\'a}nchez}}, \citenamefont {Cohen},\
  and\ \citenamefont {Yang}}]{flat_plane_condition}%
  \BibitemOpen
  \bibfield  {author} {\bibinfo {author} {\bibfnamefont {P.}~\bibnamefont
  {{Mori-S{\'a}nchez}}}, \bibinfo {author} {\bibfnamefont {A.~J.}\ \bibnamefont
  {Cohen}},\ and\ \bibinfo {author} {\bibfnamefont {W.}~\bibnamefont {Yang}},\
  }\bibfield  {title} {\bibinfo {title} {Discontinuous {{Nature}} of the
  {{Exchange-Correlation Functional}} in {{Strongly Correlated Systems}}},\
  }\href {https://doi.org/10.1103/PhysRevLett.102.066403} {\bibfield  {journal}
  {\bibinfo  {journal} {Phys. Rev. Lett.}\ }\textbf {\bibinfo {volume} {102}},\
  \bibinfo {pages} {066403} (\bibinfo {year} {2009})}\BibitemShut {NoStop}%
\bibitem [{\citenamefont {Yang}\ \emph {et~al.}(2016)\citenamefont {Yang},
  \citenamefont {Patel}, \citenamefont {{Miranda-Quintana}}, \citenamefont
  {{Heidar-Zadeh}}, \citenamefont {{Gonz{\'a}lez-Espinoza}},\ and\
  \citenamefont {Ayers}}]{two_types_of_flat_planes}%
  \BibitemOpen
  \bibfield  {author} {\bibinfo {author} {\bibfnamefont {X.~D.}\ \bibnamefont
  {Yang}}, \bibinfo {author} {\bibfnamefont {A.~H.~G.}\ \bibnamefont {Patel}},
  \bibinfo {author} {\bibfnamefont {R.~A.}\ \bibnamefont {{Miranda-Quintana}}},
  \bibinfo {author} {\bibfnamefont {F.}~\bibnamefont {{Heidar-Zadeh}}},
  \bibinfo {author} {\bibfnamefont {C.~E.}\ \bibnamefont
  {{Gonz{\'a}lez-Espinoza}}},\ and\ \bibinfo {author} {\bibfnamefont {P.~W.}\
  \bibnamefont {Ayers}},\ }\bibfield  {title} {\bibinfo {title} {Communication:
  {{Two}} types of flat-planes conditions in density functional theory},\
  }\href {https://doi.org/10.1063/1.4958636} {\bibfield  {journal} {\bibinfo
  {journal} {J. Chem. Phys.}\ }\textbf {\bibinfo {volume} {145}},\ \bibinfo
  {pages} {031102} (\bibinfo {year} {2016})}\BibitemShut {NoStop}%
\bibitem [{\citenamefont {{Mori-S{\'a}nchez}}\ and\ \citenamefont
  {J.~Cohen}(2014)}]{mori-sanchez}%
  \BibitemOpen
  \bibfield  {author} {\bibinfo {author} {\bibfnamefont {P.}~\bibnamefont
  {{Mori-S{\'a}nchez}}}\ and\ \bibinfo {author} {\bibfnamefont
  {A.}~\bibnamefont {J.~Cohen}},\ }\bibfield  {title} {\bibinfo {title} {The
  derivative discontinuity of the exchange\textendash correlation functional},\
  }\href {https://doi.org/10.1039/C4CP01170H} {\bibfield  {journal} {\bibinfo
  {journal} {Physical Chemistry Chemical Physics}\ }\textbf {\bibinfo {volume}
  {16}},\ \bibinfo {pages} {14378} (\bibinfo {year} {2014})}\BibitemShut
  {NoStop}%
\bibitem [{\citenamefont {Zhao}\ \emph {et~al.}(2016)\citenamefont {Zhao},
  \citenamefont {Ioannidis},\ and\ \citenamefont {Kulik}}]{globalversuslocal}%
  \BibitemOpen
  \bibfield  {author} {\bibinfo {author} {\bibfnamefont {Q.}~\bibnamefont
  {Zhao}}, \bibinfo {author} {\bibfnamefont {E.~I.}\ \bibnamefont
  {Ioannidis}},\ and\ \bibinfo {author} {\bibfnamefont {H.~J.}\ \bibnamefont
  {Kulik}},\ }\bibfield  {title} {\bibinfo {title} {Global and local curvature
  in density functional theory},\ }\href {https://doi.org/10.1063/1.4959882}
  {\bibfield  {journal} {\bibinfo  {journal} {J. Chem. Phys.}\ }\textbf
  {\bibinfo {volume} {145}},\ \bibinfo {pages} {054109} (\bibinfo {year}
  {2016})}\BibitemShut {NoStop}%
\bibitem [{\citenamefont {Hait}\ and\ \citenamefont
  {{Head-Gordon}}(2018)}]{MSIE-quadratic}%
  \BibitemOpen
  \bibfield  {author} {\bibinfo {author} {\bibfnamefont {D.}~\bibnamefont
  {Hait}}\ and\ \bibinfo {author} {\bibfnamefont {M.}~\bibnamefont
  {{Head-Gordon}}},\ }\bibfield  {title} {\bibinfo {title} {Delocalization
  {{Errors}} in {{Density Functional Theory Are Essentially Quadratic}} in
  {{Fractional Occupation Number}}},\ }\href
  {https://doi.org/10.1021/acs.jpclett.8b02417} {\bibfield  {journal} {\bibinfo
   {journal} {J. Phys. Chem. Lett.}\ }\textbf {\bibinfo {volume} {9}},\
  \bibinfo {pages} {6280} (\bibinfo {year} {2018})}\BibitemShut {NoStop}%
\bibitem [{\citenamefont {Anisimov}\ \emph {et~al.}(1991)\citenamefont
  {Anisimov}, \citenamefont {Zaanen},\ and\ \citenamefont
  {Andersen}}]{anisimov}%
  \BibitemOpen
  \bibfield  {author} {\bibinfo {author} {\bibfnamefont {V.~I.}\ \bibnamefont
  {Anisimov}}, \bibinfo {author} {\bibfnamefont {J.}~\bibnamefont {Zaanen}},\
  and\ \bibinfo {author} {\bibfnamefont {O.~K.}\ \bibnamefont {Andersen}},\
  }\bibfield  {title} {\bibinfo {title} {Band theory and {{Mott}} insulators:
  {{Hubbard U}} instead of {{Stoner I}}},\ }\href
  {https://doi.org/10.1103/PhysRevB.44.943} {\bibfield  {journal} {\bibinfo
  {journal} {Phys. Rev. B}\ }\textbf {\bibinfo {volume} {44}},\ \bibinfo
  {pages} {943} (\bibinfo {year} {1991})}\BibitemShut {NoStop}%
\bibitem [{\citenamefont {Anisimov}\ \emph {et~al.}(1993)\citenamefont
  {Anisimov}, \citenamefont {Solovyev}, \citenamefont {Korotin}, \citenamefont
  {Czy{\.z}yk},\ and\ \citenamefont {Sawatzky}}]{anisimov1993}%
  \BibitemOpen
  \bibfield  {author} {\bibinfo {author} {\bibfnamefont {V.~I.}\ \bibnamefont
  {Anisimov}}, \bibinfo {author} {\bibfnamefont {I.~V.}\ \bibnamefont
  {Solovyev}}, \bibinfo {author} {\bibfnamefont {M.~A.}\ \bibnamefont
  {Korotin}}, \bibinfo {author} {\bibfnamefont {M.~T.}\ \bibnamefont
  {Czy{\.z}yk}},\ and\ \bibinfo {author} {\bibfnamefont {G.~A.}\ \bibnamefont
  {Sawatzky}},\ }\bibfield  {title} {\bibinfo {title} {Density-functional
  theory and {{NiO}} photoemission spectra},\ }\href
  {https://doi.org/10.1103/PhysRevB.48.16929} {\bibfield  {journal} {\bibinfo
  {journal} {Phys. Rev. B}\ }\textbf {\bibinfo {volume} {48}},\ \bibinfo
  {pages} {16929} (\bibinfo {year} {1993})}\BibitemShut {NoStop}%
\bibitem [{\citenamefont {Liechtenstein}\ \emph {et~al.}(1995)\citenamefont
  {Liechtenstein}, \citenamefont {Anisimov},\ and\ \citenamefont
  {Zaanen}}]{Liechtenstein}%
  \BibitemOpen
  \bibfield  {author} {\bibinfo {author} {\bibfnamefont {A.~I.}\ \bibnamefont
  {Liechtenstein}}, \bibinfo {author} {\bibfnamefont {V.~I.}\ \bibnamefont
  {Anisimov}},\ and\ \bibinfo {author} {\bibfnamefont {J.}~\bibnamefont
  {Zaanen}},\ }\bibfield  {title} {\bibinfo {title} {Density-functional theory
  and strong interactions: {{Orbital}} ordering in {{Mott-Hubbard}}
  insulators},\ }\href {https://doi.org/10.1103/PhysRevB.52.R5467} {\bibfield
  {journal} {\bibinfo  {journal} {Phys. Rev. B}\ }\textbf {\bibinfo {volume}
  {52}},\ \bibinfo {pages} {R5467} (\bibinfo {year} {1995})}\BibitemShut
  {NoStop}%
\bibitem [{\citenamefont {Dudarev}\ \emph {et~al.}(1998)\citenamefont
  {Dudarev}, \citenamefont {Botton}, \citenamefont {Savrasov}, \citenamefont
  {Humphreys},\ and\ \citenamefont {Sutton}}]{dudarev}%
  \BibitemOpen
  \bibfield  {author} {\bibinfo {author} {\bibfnamefont {S.~L.}\ \bibnamefont
  {Dudarev}}, \bibinfo {author} {\bibfnamefont {G.~A.}\ \bibnamefont {Botton}},
  \bibinfo {author} {\bibfnamefont {S.~Y.}\ \bibnamefont {Savrasov}}, \bibinfo
  {author} {\bibfnamefont {C.~J.}\ \bibnamefont {Humphreys}},\ and\ \bibinfo
  {author} {\bibfnamefont {A.~P.}\ \bibnamefont {Sutton}},\ }\bibfield  {title}
  {\bibinfo {title} {Electron-energy-loss spectra and the structural stability
  of nickel oxide: {{An LSDA}}+{{U}} study},\ }\href
  {https://doi.org/10.1103/PhysRevB.57.1505} {\bibfield  {journal} {\bibinfo
  {journal} {Phys. Rev. B}\ }\textbf {\bibinfo {volume} {57}},\ \bibinfo
  {pages} {1505} (\bibinfo {year} {1998})}\BibitemShut {NoStop}%
\bibitem [{\citenamefont {Himmetoglu}\ \emph {et~al.}(2011)\citenamefont
  {Himmetoglu}, \citenamefont {Wentzcovitch},\ and\ \citenamefont
  {Cococcioni}}]{coco}%
  \BibitemOpen
  \bibfield  {author} {\bibinfo {author} {\bibfnamefont {B.}~\bibnamefont
  {Himmetoglu}}, \bibinfo {author} {\bibfnamefont {R.~M.}\ \bibnamefont
  {Wentzcovitch}},\ and\ \bibinfo {author} {\bibfnamefont {M.}~\bibnamefont
  {Cococcioni}},\ }\bibfield  {title} {\bibinfo {title} {First-principles study
  of electronic and structural properties of {{CuO}}},\ }\href
  {https://doi.org/10.1103/PhysRevB.84.115108} {\bibfield  {journal} {\bibinfo
  {journal} {Phys. Rev. B}\ }\textbf {\bibinfo {volume} {84}},\ \bibinfo
  {pages} {115108} (\bibinfo {year} {2011})}\BibitemShut {NoStop}%
\bibitem [{\citenamefont {Bajaj}\ \emph {et~al.}(2017)\citenamefont {Bajaj},
  \citenamefont {Janet},\ and\ \citenamefont {Kulik}}]{bajaj2017}%
  \BibitemOpen
  \bibfield  {author} {\bibinfo {author} {\bibfnamefont {A.}~\bibnamefont
  {Bajaj}}, \bibinfo {author} {\bibfnamefont {J.~P.}\ \bibnamefont {Janet}},\
  and\ \bibinfo {author} {\bibfnamefont {H.~J.}\ \bibnamefont {Kulik}},\
  }\bibfield  {title} {\bibinfo {title} {Communication: {{Recovering}} the
  flat-plane condition in electronic structure theory at semi-local {{DFT}}
  cost},\ }\href {https://doi.org/10.1063/1.5008981} {\bibfield  {journal}
  {\bibinfo  {journal} {J. Chem. Phys.}\ }\textbf {\bibinfo {volume} {147}},\
  \bibinfo {pages} {191101} (\bibinfo {year} {2017})}\BibitemShut {NoStop}%
\bibitem [{\citenamefont {Bajaj}\ \emph {et~al.}(2019)\citenamefont {Bajaj},
  \citenamefont {Liu},\ and\ \citenamefont {Kulik}}]{bajaj2019}%
  \BibitemOpen
  \bibfield  {author} {\bibinfo {author} {\bibfnamefont {A.}~\bibnamefont
  {Bajaj}}, \bibinfo {author} {\bibfnamefont {F.}~\bibnamefont {Liu}},\ and\
  \bibinfo {author} {\bibfnamefont {H.~J.}\ \bibnamefont {Kulik}},\ }\bibfield
  {title} {\bibinfo {title} {Non-empirical, low-cost recovery of exact
  conditions with model-{{Hamiltonian}} inspired expressions in {{jmDFT}}},\
  }\href {https://doi.org/10.1063/1.5091563} {\bibfield  {journal} {\bibinfo
  {journal} {J. Chem. Phys.}\ }\textbf {\bibinfo {volume} {150}},\ \bibinfo
  {pages} {154115} (\bibinfo {year} {2019})}\BibitemShut {NoStop}%
\bibitem [{\citenamefont {Moynihan}(2018)}]{moynihan_thesis}%
  \BibitemOpen
  \bibfield  {author} {\bibinfo {author} {\bibfnamefont {G.}~\bibnamefont
  {Moynihan}},\ }\emph {\bibinfo {title} {A Self-Contained Ground-State
  Approach for the Correction of Self-Interaction Error in Approximate
  Density-Functional Theory}},\ \href@noop {} {\bibinfo {type} {Thesis}},\
  \bibinfo  {school} {Trinity College Dublin. School of Physics. Discipline of
  Physics} (\bibinfo {year} {2018})\BibitemShut {NoStop}%
\bibitem [{\citenamefont {Linscott}\ \emph {et~al.}(2018)\citenamefont
  {Linscott}, \citenamefont {Cole}, \citenamefont {Payne},\ and\ \citenamefont
  {O'Regan}}]{linscott}%
  \BibitemOpen
  \bibfield  {author} {\bibinfo {author} {\bibfnamefont {E.~B.}\ \bibnamefont
  {Linscott}}, \bibinfo {author} {\bibfnamefont {D.~J.}\ \bibnamefont {Cole}},
  \bibinfo {author} {\bibfnamefont {M.~C.}\ \bibnamefont {Payne}},\ and\
  \bibinfo {author} {\bibfnamefont {D.~D.}\ \bibnamefont {O'Regan}},\
  }\bibfield  {title} {\bibinfo {title} {Role of spin in the calculation of
  {{Hubbard}} {$U$} and {{Hund}}'s ${J}$ parameters from first principles},\
  }\href {https://doi.org/10.1103/PhysRevB.98.235157} {\bibfield  {journal}
  {\bibinfo  {journal} {Phys. Rev. B}\ }\textbf {\bibinfo {volume} {98}},\
  \bibinfo {pages} {235157} (\bibinfo {year} {2018})}\BibitemShut {NoStop}%
\bibitem [{\citenamefont {Moynihan}\ \emph {et~al.}(2017)\citenamefont
  {Moynihan}, \citenamefont {Teobaldi},\ and\ \citenamefont
  {O'Regan}}]{moynihan}%
  \BibitemOpen
  \bibfield  {author} {\bibinfo {author} {\bibfnamefont {G.}~\bibnamefont
  {Moynihan}}, \bibinfo {author} {\bibfnamefont {G.}~\bibnamefont {Teobaldi}},\
  and\ \bibinfo {author} {\bibfnamefont {D.~D.}\ \bibnamefont {O'Regan}},\
  }\href {https://doi.org/10.48550/arXiv.1704.08076} {\bibinfo {title} {A
  self-consistent ground-state formulation of the first-principles {{Hubbard
  U}} parameter validated on one-electron self-interaction error}} (\bibinfo
  {year} {2017}),\ \Eprint {https://arxiv.org/abs/1704.08076} {arXiv:1704.08076
  [cond-mat]} \BibitemShut {NoStop}%
\bibitem [{\citenamefont {Ylvisaker}\ \emph {et~al.}(2009)\citenamefont
  {Ylvisaker}, \citenamefont {Pickett},\ and\ \citenamefont
  {Koepernik}}]{kopernik}%
  \BibitemOpen
  \bibfield  {author} {\bibinfo {author} {\bibfnamefont {E.~R.}\ \bibnamefont
  {Ylvisaker}}, \bibinfo {author} {\bibfnamefont {W.~E.}\ \bibnamefont
  {Pickett}},\ and\ \bibinfo {author} {\bibfnamefont {K.}~\bibnamefont
  {Koepernik}},\ }\bibfield  {title} {\bibinfo {title} {Anisotropy and
  magnetism in the {{LSDA+U}} method},\ }\href
  {https://doi.org/10.1103/PhysRevB.79.035103} {\bibfield  {journal} {\bibinfo
  {journal} {Phys. Rev. B}\ }\textbf {\bibinfo {volume} {79}},\ \bibinfo
  {pages} {035103} (\bibinfo {year} {2009})}\BibitemShut {NoStop}%
\bibitem [{\citenamefont {Dudarev}\ \emph {et~al.}(2019)\citenamefont
  {Dudarev}, \citenamefont {Liu}, \citenamefont {Andersson}, \citenamefont
  {Stanek}, \citenamefont {Ozaki},\ and\ \citenamefont
  {Franchini}}]{dudarev_2019}%
  \BibitemOpen
  \bibfield  {author} {\bibinfo {author} {\bibfnamefont {S.~L.}\ \bibnamefont
  {Dudarev}}, \bibinfo {author} {\bibfnamefont {P.}~\bibnamefont {Liu}},
  \bibinfo {author} {\bibfnamefont {D.~A.}\ \bibnamefont {Andersson}}, \bibinfo
  {author} {\bibfnamefont {C.~R.}\ \bibnamefont {Stanek}}, \bibinfo {author}
  {\bibfnamefont {T.}~\bibnamefont {Ozaki}},\ and\ \bibinfo {author}
  {\bibfnamefont {C.}~\bibnamefont {Franchini}},\ }\bibfield  {title} {\bibinfo
  {title} {Parametrization of {{LSDA+U}} for noncollinear magnetic
  configurations: {{Multipolar}} magnetism in f},\ }\href
  {https://doi.org/10.1103/PhysRevMaterials.3.083802} {\bibfield  {journal}
  {\bibinfo  {journal} {Phys. Rev. Materials}\ }\textbf {\bibinfo {volume}
  {3}},\ \bibinfo {pages} {083802} (\bibinfo {year} {2019})}\BibitemShut
  {NoStop}%
\bibitem [{\citenamefont {Shishkin}\ and\ \citenamefont
  {Sato}(2017)}]{shishkin2017}%
  \BibitemOpen
  \bibfield  {author} {\bibinfo {author} {\bibfnamefont {M.}~\bibnamefont
  {Shishkin}}\ and\ \bibinfo {author} {\bibfnamefont {H.}~\bibnamefont
  {Sato}},\ }\bibfield  {title} {\bibinfo {title} {Challenges in computational
  evaluation of redox and magnetic properties of {{Fe-based}} sulfate cathode
  materials of {{Li-}} and {{Na-ion}} batteries},\ }\href
  {https://doi.org/10.1088/1361-648X/aa6667} {\bibfield  {journal} {\bibinfo
  {journal} {J. Phys.: Condens. Matter}\ }\textbf {\bibinfo {volume} {29}},\
  \bibinfo {pages} {215701} (\bibinfo {year} {2017})}\BibitemShut {NoStop}%
\bibitem [{\citenamefont {Shishkin}\ and\ \citenamefont
  {Sato}(2019)}]{shishkin2019}%
  \BibitemOpen
  \bibfield  {author} {\bibinfo {author} {\bibfnamefont {M.}~\bibnamefont
  {Shishkin}}\ and\ \bibinfo {author} {\bibfnamefont {H.}~\bibnamefont
  {Sato}},\ }\bibfield  {title} {\bibinfo {title} {{{DFT}}+{{U}} in
  {{Dudarev}}'s formulation with corrected interactions between the electrons
  with opposite spins: {{The}} form of {{Hamiltonian}}, calculation of forces,
  and bandgap adjustments},\ }\href {https://doi.org/10.1063/1.5090445}
  {\bibfield  {journal} {\bibinfo  {journal} {J. Chem. Phys.}\ }\textbf
  {\bibinfo {volume} {151}},\ \bibinfo {pages} {024102} (\bibinfo {year}
  {2019})}\BibitemShut {NoStop}%
\bibitem [{\citenamefont {Seo}(2007)}]{seo2007}%
  \BibitemOpen
  \bibfield  {author} {\bibinfo {author} {\bibfnamefont {D.-K.}\ \bibnamefont
  {Seo}},\ }\bibfield  {title} {\bibinfo {title} {Self-interaction correction
  in the {{LDA+U}} method},\ }\href
  {https://doi.org/10.1103/PhysRevB.76.033102} {\bibfield  {journal} {\bibinfo
  {journal} {Phys. Rev. B}\ }\textbf {\bibinfo {volume} {76}},\ \bibinfo
  {pages} {033102} (\bibinfo {year} {2007})}\BibitemShut {NoStop}%
\bibitem [{\citenamefont {Cococcioni}\ and\ \citenamefont {{de
  Gironcoli}}(2005)}]{coco-scf-linear-response}%
  \BibitemOpen
  \bibfield  {author} {\bibinfo {author} {\bibfnamefont {M.}~\bibnamefont
  {Cococcioni}}\ and\ \bibinfo {author} {\bibfnamefont {S.}~\bibnamefont {{de
  Gironcoli}}},\ }\bibfield  {title} {\bibinfo {title} {Linear response
  approach to the calculation of the effective interaction parameters in the
  {{LDA+U}} method},\ }\href {https://doi.org/10.1103/PhysRevB.71.035105}
  {\bibfield  {journal} {\bibinfo  {journal} {Phys. Rev. B}\ }\textbf {\bibinfo
  {volume} {71}},\ \bibinfo {pages} {035105} (\bibinfo {year}
  {2005})}\BibitemShut {NoStop}%
\bibitem [{\citenamefont {Prentice}\ \emph {et~al.}(2020)\citenamefont
  {Prentice}, \citenamefont {Aarons}, \citenamefont {Womack}, \citenamefont
  {Allen}, \citenamefont {Andrinopoulos}, \citenamefont {Anton}, \citenamefont
  {Bell}, \citenamefont {Bhandari}, \citenamefont {Bramley}, \citenamefont
  {Charlton}, \citenamefont {Clements}, \citenamefont {Cole}, \citenamefont
  {Constantinescu}, \citenamefont {Corsetti}, \citenamefont {Dubois},
  \citenamefont {Duff}, \citenamefont {Escart{\'i}n}, \citenamefont {Greco},
  \citenamefont {Hill}, \citenamefont {Lee}, \citenamefont {Linscott},
  \citenamefont {O'Regan}, \citenamefont {Phipps}, \citenamefont {Ratcliff},
  \citenamefont {Serrano}, \citenamefont {Tait}, \citenamefont {Teobaldi},
  \citenamefont {Vitale}, \citenamefont {Yeung}, \citenamefont {Zuehlsdorff},
  \citenamefont {Dziedzic}, \citenamefont {Haynes}, \citenamefont {Hine},
  \citenamefont {Mostofi}, \citenamefont {Payne},\ and\ \citenamefont
  {Skylaris}}]{ONETEP}%
  \BibitemOpen
  \bibfield  {author} {\bibinfo {author} {\bibfnamefont {J.~C.~A.}\
  \bibnamefont {Prentice}}, \bibinfo {author} {\bibfnamefont {J.}~\bibnamefont
  {Aarons}}, \bibinfo {author} {\bibfnamefont {J.~C.}\ \bibnamefont {Womack}},
  \bibinfo {author} {\bibfnamefont {A.~E.~A.}\ \bibnamefont {Allen}}, \bibinfo
  {author} {\bibfnamefont {L.}~\bibnamefont {Andrinopoulos}}, \bibinfo {author}
  {\bibfnamefont {L.}~\bibnamefont {Anton}}, \bibinfo {author} {\bibfnamefont
  {R.~A.}\ \bibnamefont {Bell}}, \bibinfo {author} {\bibfnamefont
  {A.}~\bibnamefont {Bhandari}}, \bibinfo {author} {\bibfnamefont {G.~A.}\
  \bibnamefont {Bramley}}, \bibinfo {author} {\bibfnamefont {R.~J.}\
  \bibnamefont {Charlton}}, \bibinfo {author} {\bibfnamefont {R.~J.}\
  \bibnamefont {Clements}}, \bibinfo {author} {\bibfnamefont {D.~J.}\
  \bibnamefont {Cole}}, \bibinfo {author} {\bibfnamefont {G.}~\bibnamefont
  {Constantinescu}}, \bibinfo {author} {\bibfnamefont {F.}~\bibnamefont
  {Corsetti}}, \bibinfo {author} {\bibfnamefont {S.~M.-M.}\ \bibnamefont
  {Dubois}}, \bibinfo {author} {\bibfnamefont {K.~K.~B.}\ \bibnamefont {Duff}},
  \bibinfo {author} {\bibfnamefont {J.~M.}\ \bibnamefont {Escart{\'i}n}},
  \bibinfo {author} {\bibfnamefont {A.}~\bibnamefont {Greco}}, \bibinfo
  {author} {\bibfnamefont {Q.}~\bibnamefont {Hill}}, \bibinfo {author}
  {\bibfnamefont {L.~P.}\ \bibnamefont {Lee}}, \bibinfo {author} {\bibfnamefont
  {E.}~\bibnamefont {Linscott}}, \bibinfo {author} {\bibfnamefont {D.~D.}\
  \bibnamefont {O'Regan}}, \bibinfo {author} {\bibfnamefont {M.~J.~S.}\
  \bibnamefont {Phipps}}, \bibinfo {author} {\bibfnamefont {L.~E.}\
  \bibnamefont {Ratcliff}}, \bibinfo {author} {\bibfnamefont {{\'A}.~R.}\
  \bibnamefont {Serrano}}, \bibinfo {author} {\bibfnamefont {E.~W.}\
  \bibnamefont {Tait}}, \bibinfo {author} {\bibfnamefont {G.}~\bibnamefont
  {Teobaldi}}, \bibinfo {author} {\bibfnamefont {V.}~\bibnamefont {Vitale}},
  \bibinfo {author} {\bibfnamefont {N.}~\bibnamefont {Yeung}}, \bibinfo
  {author} {\bibfnamefont {T.~J.}\ \bibnamefont {Zuehlsdorff}}, \bibinfo
  {author} {\bibfnamefont {J.}~\bibnamefont {Dziedzic}}, \bibinfo {author}
  {\bibfnamefont {P.~D.}\ \bibnamefont {Haynes}}, \bibinfo {author}
  {\bibfnamefont {N.~D.~M.}\ \bibnamefont {Hine}}, \bibinfo {author}
  {\bibfnamefont {A.~A.}\ \bibnamefont {Mostofi}}, \bibinfo {author}
  {\bibfnamefont {M.~C.}\ \bibnamefont {Payne}},\ and\ \bibinfo {author}
  {\bibfnamefont {C.-K.}\ \bibnamefont {Skylaris}},\ }\bibfield  {title}
  {\bibinfo {title} {The {{ONETEP}} linear-scaling density functional theory
  program},\ }\href {https://doi.org/10.1063/5.0004445} {\bibfield  {journal}
  {\bibinfo  {journal} {J. Chem. Phys.}\ }\textbf {\bibinfo {volume} {152}},\
  \bibinfo {pages} {174111} (\bibinfo {year} {2020})}\BibitemShut {NoStop}%
\bibitem [{\citenamefont {Martyna}\ and\ \citenamefont
  {Tuckerman}(1999)}]{martyna}%
  \BibitemOpen
  \bibfield  {author} {\bibinfo {author} {\bibfnamefont {G.~J.}\ \bibnamefont
  {Martyna}}\ and\ \bibinfo {author} {\bibfnamefont {M.~E.}\ \bibnamefont
  {Tuckerman}},\ }\bibfield  {title} {\bibinfo {title} {A reciprocal space
  based method for treating long range interactions in ab initio and
  force-field-based calculations in clusters},\ }\href
  {https://doi.org/10.1063/1.477923} {\bibfield  {journal} {\bibinfo  {journal}
  {J. Chem. Phys.}\ }\textbf {\bibinfo {volume} {110}},\ \bibinfo {pages}
  {2810} (\bibinfo {year} {1999})}\BibitemShut {NoStop}%
\bibitem [{OPI()}]{OPIUM}%
  \BibitemOpen
  \href@noop {} {\bibinfo {title} {{{OPIUM}}: {{The}} optimized pseudopotential
  interface unification module}}\BibitemShut {NoStop}%
\end{thebibliography}%

\clearpage
\renewcommand{\thesection}{SI-\Roman{section}}
\onecolumngrid
\vspace{\columnsep}
\section{Derivation of BLOR}
\label{sec:BLOR-derivation}
\vspace{\columnsep}
\twocolumngrid

To derive the new corrective functional we first consider a single orbital embedded within a material, which can be occupied by up to two electrons of opposite spin. We use the same definition of $U^{\sigma}$ and $J$ as given by equations \ref{eqn:Udefine} and \ref{eqn:Jdefine} in the main text, where the spin resolved Hubbard $U$ parameters is given as:
\begin{equation}
U^{\sigma}=\bigg(\frac{\partial^2 E_{Hxc}^{\rm approx}[\rho_{\rm loc}(\bf{r})]}{\partial (n^{\sigma})^2}\bigg)_{n^{-\sigma}}, 
\end{equation}
and Hund's $J$ parameter is given as:
\begin{equation}
J=-\bigg(\frac{\partial^2 E_{Hxc}^{\rm approx}[\rho_{\rm loc}(\bf{r})]}{\partial (M)^2}\bigg)_{N}, 
\end{equation}
where $n^{\sigma}$, $N$ and $M$ are the spin-resolved subspace occupancy, the subspace electron count and the subspace magnetisation respectively. For the purposes of this work we assume that $U^{\sigma}$ and $J$ are subspace specific constants and leave considerations of higher order partial derivatives to future studies. 

Using these definitions, we can now derive the correction [$E_u(n^{\upharpoonright}=1,n^{\downharpoonright})$] to the Hartree-exchange-correlation energy along the maximally spin up polarised line (along the edge of the diamond), in the upper half plane. The exact Hartree-exchange-correlation energy $E_{Hxc}^{\rm exact}(n^{\upharpoonright},n^{\downharpoonright})$ should follow a linear curve along the maximally spin polarised line:
\begin{equation}
E_{Hxc}^{\rm exact}(n^{\upharpoonright}=1,n^{\downharpoonright})=[E_{Hxc}^{\rm exact}(1,1)]n^{\downharpoonright}.
\end{equation}
While the approximate Hartree-exchange-correlation energy will follow a quadratic curve along the maximally spin polarised line:
\begin{align}
E_{Hxc}^{\rm approx}=&\frac{U^{\downharpoonright}}{2}(n^{\downharpoonright})^2+\bigg[E_{Hxc}^{\rm approx}(1,1)-E_{Hxc}^{\rm approx}(1,0)-\frac{U^{\downharpoonright}}{2}\bigg]n^{\downharpoonright}\nonumber \\ &  +E_{Hxc}^{\rm approx}(1,0).
\end{align}
Now assuming that $E_{Hxc}^{\rm approx}$ yields a close to exact result at integer $n^{\upharpoonright}$ \& $n^{\downharpoonright}$ occupancies we can approximate that:
\begin{equation}
E_{Hxc}^{\rm exact}(1,1)-E_{Hxc}^{\rm approx}(1,1) \approx 0.
\end{equation}
and similarly:
\begin{equation}
E_{Hxc}^{\rm approx}(1,0)\approx E_{Hxc}^{\rm exact}(1,0) = 0.
\end{equation}
Hence the correction to $E_{Hxc}$ along the maximally spin polarised line becomes:
\begin{equation}
E_u(n^{\upharpoonright}=1,n^{\downharpoonright})=E_{Hxc}^{\rm exact}-E_{Hxc}^{\rm approx}=\frac{U^{\downharpoonright}}{2}\bigg[n^{\downharpoonright}-(n^{\downharpoonright})^2 \bigg].
\end{equation}
Along the maximally spin polarised line in the upper half plane, $n^{\upharpoonright}=1$ and $n^{\downharpoonright}=N-1$, where $N$ is the total subspace occupancy. Hence we have that:
\begin{equation}
\label{eqn:correction_along_max_spin_polarised_lines}
E_u(n^{\upharpoonright}=1,n^{\downharpoonright})=\frac{U^{\downharpoonright}}{2}\bigg[(N-1)-(N-1)^2 \bigg].
\end{equation}
We now wish to find an expression for $E_u(n^{\upharpoonright},n^{\downharpoonright})$ at an arbitrary point in the upper half plane. To evaluate this we can define $\Delta_m(n^{\upharpoonright},n^{\downharpoonright})$ as:
\begin{equation}
\Delta_m(n^{\upharpoonright},n^{\downharpoonright})=E_u(n^{\upharpoonright},n^{\downharpoonright})-E_u(1,n^{\upharpoonright}+n^{\downharpoonright}-1).
\end{equation}
Where $\Delta_m$ is the change in the energetic correction on moving from the point $(1,N-1)$ to the point $(n^{\upharpoonright},n^{\downharpoonright})$, where $N=n^{\upharpoonright}+n^{\downharpoonright}$. The exact Hxc energy should be constant as one varies $M$ keeping $N$ fixed. However, $E_{Hxc}^{\rm approx}$  will exhibit a spurious curvature of $-J$. Unlike equation \ref{eqn:correction_along_max_spin_polarised_lines}, there will be no linear term in this case due to the spin symmetry of the system. Hence, the change in the energetic correction is given by:
\begin{equation}
\label{eqn:JminusJmax}
\Delta_m=\frac{J}{2}M^2-\frac{J}{2}M_{\rm max}^2,
\end{equation}
where $M_{\rm max}$ is the subspace magnetisation along the maximally spin polarised line, $M_{\rm max}=2-N$. The constant $\frac{J}{2}M_{\rm max}^2$ term ensures that $\Delta_m=0$ along the maximally spin polarised line. Therefore equation \ref{eqn:JminusJmax} becomes:
\begin{equation}
\label{eqn:deltaM}
\Delta_m=\frac{J}{2}\bigg[M^2-(N-2)^2\bigg].
\end{equation}
We choose to make the further approximation that $U^{\upharpoonright}=U^{\downharpoonright}=U$, in which case we denote the total corrective energy as $E_u^{\rm sym}$. Noting that analogous expressions hold for the corrective functional in the lower half plane, the total corrective functional for a two electron subspace is thus given by:
\begin{widetext}
\begin{align}
E_u^{\rm sym}= \left\{
\begin{array}{*6{>{\displaystyle}c}}
\frac{U}{2}[N-N^2]
&+&
\frac{J}{2}[M^2-N^2]
& , \
N \leq 1
 \\
\frac{U}{2}[(N-1)-(N-1)^2 ]
&+&
\frac{J}{2}[M^2-(N-2)^2]
& , \
N > 1
\end{array}
\right.
\end{align}
\end{widetext}
The previously derived functional assumes that the interaction between the localized electrons and the surrounding environment can be approximated by an effective electric field. However, an effective magnetic field may also be acting on the subspace embedded in the material environment. In this case, the state with one spin up electron is no longer degenerate to the state with one spin down electron. The constancy condition with respect to magnetisation becomes a linearity condition:
\begin{equation}
E[N,M]=\alpha E[N,M_0]+(1-\alpha)E[N,M_0+2] {\rm \hspace{2.5mm}}\forall {\rm \hspace{2.5mm}} 0 \leq \alpha \leq 1, 
\end{equation}
where $M_0$ is an integer and the subspace magnetisation:
\begin{equation}
M=\alpha M_0 +(1-\alpha) (M_0+2).
\end{equation}
The piecewise linearity condition with respect to magnetisation has consequences for the derivation of the corrective functional. In such cases one certainly cannot assume that $U^{\upharpoonright}=U^{\downharpoonright}$. The corrective functional along the maximally spin up polarised line in the upper half plane is still given by equation \ref{eqn:correction_along_max_spin_polarised_lines}. While the corrective functional along the maximally spin down polarised line is given by:
\begin{equation}
E_u(n^{\upharpoonright},n^{\downharpoonright}=1)=\frac{U^{\upharpoonright}}{2}\bigg[(N-1)-(N-1)^2 \bigg].
\end{equation}
Due to the presence of the magnetic field, the spin symmetry of the system is broken, $E[N,M=N-1] \neq E[N,M=1-N]$. Hence $\Delta_m$ in this case will have a linear term in $M$. The linear term in $M$ ensures that $\Delta_m$ continues to give zero contribution to $E_{u}$ along the maximally spin down polarised line, while it gives a non-zero contribution along the maximally spin up polarised line so that the total corrective functional reduces to equation \ref{eqn:correction_along_max_spin_polarised_lines} when $n^{\upharpoonright}=1$. Hence, in the presence of an effective magnetic field equation \ref{eqn:deltaM} becomes:
\begin{widetext}
\onecolumngrid
\begin{equation}
\Delta_m= \frac{J}{2}\bigg[M^2-(N-2)^2\bigg]+\frac{U^{\downharpoonright}-U^{\upharpoonright}}{2}\bigg[(N-1)-(N-1)^2\bigg]\cdot \frac{1}{2}\bigg(1+\frac{M}{2-N}\bigg). 
\end{equation}
Hence the BLOR functional for a single orbital subspace is given as;
\begin{align}
\label{eqn:BLOR_nS_single_orbital_appendix}
E_{\rm BLOR}^{\rm single}= \left\{
\begin{array}{*6{>{\displaystyle}c}}
\frac{U^{\upharpoonright}+U^{\downharpoonright}}{4}[{N}-{N}^2]
&+&
\frac{J}{2}[{M}^2-{N}^2]
&+&
\frac{U^{\upharpoonright}-U^{\downharpoonright}}{4}[{M}-{N}{M}]
& , \
N \leq 1
 \\
\frac{U^{\upharpoonright}+U^{\downharpoonright}}{4}[({N}-1)-({N}-1)^2]
&+&
\frac{J}{2}[{M}^2-({N}-2)^2]
&+&
\frac{U^{\upharpoonright}-U^{\downharpoonright}}{4}[{M}-{N}{M}]
& , \
N > 1
\end{array}
\right.
\end{align}
We now wish to extend this technique to multi-orbital subspaces such as the five d-orbitals at a transition metal site. The naive approach would be to separately apply the corrective functional given by equation \ref{eqn:BLOR_nS_single_orbital_appendix} to each orbital with electron count $N_i$ and magnetisation $M_i$. Assuming the occupancy of each orbital is less than one, the corrective functional would take the following form:
\begin{equation}
E_{\rm BLOR}^{\rm naive}=\sum_i\frac{U^{\upharpoonright}_i+U^{\downharpoonright}_i}{4}\bigg[N_i-N_i^2 \bigg]+\frac{J_i}{2}\bigg[M_i^2-N_i^2\bigg]+\frac{U^{\upharpoonright}_i-U^{\downharpoonright}_i}{4}\bigg[M_i-N_iM_i \bigg].
\end{equation}
However, such a corrective functional would not be rotationally invariant. To counteract this problem we firstly assume that the deviation of each orbital from the flat plane condition can be treated using a subspace averaged Hubbard $U$ and Hund's $J$ parameters. Secondly, we assume that the spin up and spin down subspace occupancy matrices have the same eigenbasis . We can then let the orbitals be equal to the eigenvectors of the susbspace occupancy matrix. This allows us to express the multi-orbital corrective functional in a rotationally invariant form as:
\begin{align}
E_{\rm BLOR}= \left\{
\begin{array}{*6{>{\displaystyle}c}}
\frac{U^{\upharpoonright}+U^{\downharpoonright}}{4}{\rm Tr}[\hat{N}-\hat{N}^2]
&+&
\frac{J}{2}{\rm Tr}[\hat{M}^2-\hat{N}^2]
&+&
\frac{U^{\upharpoonright}-U^{\downharpoonright}}{4}{\rm Tr}[\hat{M}-\hat{N}\hat{M}]
& , \
{\rm Tr}[\hat{N}] \leq {\rm Tr}[\hat{P}]
 \\
\frac{U^{\upharpoonright}+U^{\downharpoonright}}{4}{\rm Tr}[(\hat{N}-\hat{P})-(\hat{N}-\hat{P})^2]
&+&
\frac{J}{2}{\rm Tr}[\hat{M}^2-(\hat{N}-2\hat{P})^2]
&+&
\frac{U^{\upharpoonright}-U^{\downharpoonright}}{4}{\rm Tr}[\hat{M}-\hat{N}\hat{M}]
& , \
{\rm Tr}[\hat{N}] > {\rm Tr}[\hat{P}]
\end{array}
\right.
\end{align}
where $\hat{P}$ is the subspace projection operator. To maintain rotational invariance the above corrective functional has implicitly enforced Hund's First Rule on the multi-orbital subspace. There are of course some multi-orbital subspaces for which Hund's First Rule does not apply, such as low-spin transition metal complexes. An alternative Hubbard type corrective functional will be required for such systems. 
\twocolumngrid
\end{widetext}

\hspace{3cm}

\clearpage

\onecolumngrid
\vspace{\columnsep}
\section{Uniqueness of BLOR}
\label{sec:BLOR-uniqueness}
\vspace{\columnsep}
\twocolumngrid
We now wish to show that for a single orbital subspace BLOR uniquely satisfies conditions (1)-(4). The general expression for the corrective functional $E_{\rm u}$ is given as:
\begin{equation}
\label{eqn:general-eu-expression}
E_{\rm u}=a_0+a_1N+a_2N^2+a_3M+a_4M^2+a_5NM,
\end{equation}
where $\{a_i\}$ are co-efficients yet to be determined and it is assumed that we have a single orbital subspace. The co-efficients of any higher order terms must be equal to zero in order for:
\begin{equation}
\bigg(\frac{\partial^2{ E}_{\rm u}}{\partial M^2}\bigg)_{N}=J {\rm \hspace{3mm}} \& {\rm \hspace{3mm}} \bigg(\frac{\partial^2{ E}_{\rm u}}{\partial (n^{\sigma})^2}\bigg)_{n^{-\sigma}}=-U^{\sigma},
\end{equation}
In order for $E_{\rm u}[N,M]$ to satisfy condition 2 in the lower half plane, we have that:
\begin{align}
\label{eqn:sm1}
&E_{\rm u}[0,0]=a_0=0,\\
\label{eqn:sm2}
&E_{\rm u}[1,1]=a_0+a_1+a_2+a_3+a_4+a_5=0, \\
\label{eqn:sm3}
&E_{\rm u}[1,-1]=a_0+a_1+a_2-a_3+a_4-a_5=0. 
\end{align}
From condition 3 we know that $E_{\rm u}$ should have a curvature of $J$ with respect to $M$:
\begin{equation}
\label{eqn:sm4}
\bigg(\frac{\partial^2{ E}_{\rm u}}{\partial M^2}\bigg)_{N}=2a_4=J.
\end{equation}
One can re-express equation \ref{eqn:general-eu-expression} in terms of $n^{\upharpoonright}$ and $n^{\downharpoonright}$ to allow partial differentiation with respect to $n^{\sigma}$. From condition 4 we then have that:
\begin{align}
\label{eqn:sm5}
\bigg(\frac{\partial^2{ E}_{\rm u}}{\partial (n^{\upharpoonright})^2}\bigg)_{n^\downharpoonright}=&2a_2+2a_4+2a_5=-U^{\upharpoonright} \\
\label{eqn:sm6}
\bigg(\frac{\partial^2{ E}_{\rm u}}{\partial (n^{\upharpoonright})^2}\bigg)_{n^\downharpoonright}=&2a_2+2a_4-2a_5=-U^{\downharpoonright}.
\end{align}
Solving the simultaneous equations \ref{eqn:sm1} to \ref{eqn:sm6} and substituting back into equation \ref{eqn:general-eu-expression} yields:
\begin{widetext}
\begin{equation}
 E_{\rm u}=0+\bigg(\frac{U^{\upharpoonright}+U^{\downharpoonright}}{4}\bigg)N-\bigg(\frac{U^{\upharpoonright}+U^{\downharpoonright}+2J}{4}\bigg)N^2+\bigg(\frac{U^{\upharpoonright}-U^{\downharpoonright}}{4}\bigg)M+\bigg(\frac{J}{2}\bigg)M^2-\bigg(\frac{U^{\upharpoonright}-U^{\downharpoonright}}{4}\bigg)NM.
\end{equation}
\end{widetext}
This is nothing more than a re-arrangement of the BLOR functional for a single orbital subspace in the lower half-plane.

Repeating this procedure for the upper-half plane, we start again with the general expression for the corrective functional ($E_{\rm u}$) as given by equation \ref{eqn:general-eu-expression}. Condition (2) in this case yields:
\begin{align}
\label{eqn:sm1upper}
&E_{\rm u}[1,1]=a_0+a_1+a_2+a_3+a_4+a_5=0, \\
\label{eqn:sm2upper}
&E_{\rm u}[1,-1]=a_0+a_1+a_2-a_3+a_4-a_5=0, \\
\label{eqn:sm3upper}
&E_{\rm u}[2,0]=a_0+2a_1+4a_2=0.
\end{align}
Conditions (3) and (4) again in this case yield:
\begin{align}
\label{eqn:sm4upper}
&\bigg(\frac{\partial^2{ E}_{\rm u}}{\partial M^2}\bigg)_{N}=2a_4=J, \\
\label{eqn:sm5upper}
&\bigg(\frac{\partial^2{ E}_{\rm u}}{\partial (n^{\upharpoonright})^2}\bigg)_{n^\downharpoonright}=2a_2+2a_4+2a_5=-U^{\upharpoonright}, \\
\label{eqn:sm6upper}
&\bigg(\frac{\partial^2{ E}_{\rm u}}{\partial (n^{\upharpoonright})^2}\bigg)_{n^\downharpoonright}=2a_2+2a_4-2a_5=-U^{\downharpoonright}.
\end{align}

Solving the simultaneous equations \ref{eqn:sm1upper} to \ref{eqn:sm6upper} and substituting back into equation \ref{eqn:general-eu-expression} yields:
\begin{widetext}
\begin{equation}
 E_{\rm u}=-2\bigg(\frac{U^{\upharpoonright}+U^{\downharpoonright}+4J}{4}\bigg)+\bigg(\frac{3U^{\upharpoonright}+3U^{\downharpoonright}+8J}{4}\bigg)N-\bigg(\frac{U^{\upharpoonright}+U^{\downharpoonright}+2J}{4}\bigg)N^2+\bigg(\frac{U^{\upharpoonright}-U^{\downharpoonright}}{4}\bigg)M+\bigg(\frac{J}{2}\bigg)M^2-\bigg(\frac{U^{\upharpoonright}-U^{\downharpoonright}}{4}\bigg)NM.
\end{equation}
This is just a re-arrangement of the BLOR functional for a single orbital subspace in the upper half-plane. 
\end{widetext}

\hspace{1mm}

\onecolumngrid
\vspace{\columnsep}
\section{Use of BLOR functional with current DFT coding packages}
\label{sec:BLOR-with-current-DFT-code}
\vspace{\columnsep}
From the dissocaited molecular test cases BLOR was found to yield an incorrect KS orbital ordering. For future practical use we propose BLOR is applied non-self consistently  as follows:
\begin{enumerate}
    \item compute a Hubbard $U$ and Hund’s $J$ using either Cococcioni et al's self-consistent field linear response methodology \cite{coco-scf-linear-response} or Linscott et al's scalar 2x2 method \cite{linscott}. 
    \item evaluate $U_{\rm eff}=U_{\rm calc}-J_{\rm calc}$ and run a DFT$+U$ calculation using Dudarev's 1998 functional
    \item use the BLOR functional to apply a non-self consistent correction to the total DFT$+U$ energy as follows:
\end{enumerate}
\begin{align}
\Delta_{\rm NSCF}= \left\{
\begin{array}{*6{>{\displaystyle}c}}
\sum_{\sigma m m'}(0)n^{\sigma}_{mm'}\delta_{mm'}
&-&
\bigg(\frac{U_{\rm calc}+J_{\rm calc}}{2}\bigg)n^{\sigma}_{mm'}n^{\bar{\sigma}}_{m'm}
&\bf{}&
\bf{}
& , \
N \leq 2l+1
\\
\sum_{\sigma m m'}(U_{\rm calc}+J_{\rm calc})n^{\sigma}_{mm'}\delta_{mm'}
&-&
\bigg(\frac{U_{\rm calc}+J_{\rm calc}}{2}\bigg) n^{\sigma}_{mm'}n^{\bar{\sigma}}_{m'm}
&-&
\frac{U_{\rm calc}+J_{\rm calc}}{2(2l+1)}
& , \
N > 2l+1
\end{array}
\right.
\end{align}
As well as providing an erroneous KS potential, self-consistent application of BLOR is inhibited by a current lack of functionality in many common DFT software packages, typically it is not possible to assign spin resolved Hubbard $U$ parameters. The BLOR (sym) functional offers a close emulation of the full BLOR functional without the use of spin resolved Hubbard $U$ parameters. For non spin-polarised systems $U^{\upharpoonright}=U^{\downharpoonright} =U$ and BLOR (see equation \ref{eqn:BLOR_sum_version}) simplifies to:
\begin{align}
{\rm E}_{\rm BLOR}^{\rm sym}= \left\{
\begin{array}{*6{>{\displaystyle}c}}
\sum_{\sigma m m'}\frac{U}{2}n^{\sigma}_{mm'}\delta_{mm'}-\frac{U}{2}n^{\sigma}_{mm'}n^{\sigma}_{m'm}
&-&
\frac{U+2J}{2}n^{\sigma}_{mm'}n^{\bar{\sigma}}_{m'm}
&\bf{}&
\bf{}
& , \
N \leq 2l+1
\vspace{2mm}
\\
\sum_{\sigma m m'}\frac{3U+4J}{2}n^{\sigma}_{mm'}\delta_{mm'}-\frac{U}{2}n^{\sigma}_{mm'}n^{\sigma}_{m'm}
&-&
\frac{U+2J}{2}n^{\sigma}_{mm'}n^{\bar{\sigma}}_{m'm}
&-&
\frac{(U+2J)}{2(2l+1)}
& , \
N > 2l+1
\end{array}
\right.
\end{align}
For practical use of BLOR (sym) in both spin polarised and non-spin polarised systems, we let $U=f^{\sigma \sigma}$ where $\sigma$ is the majority spin channel in the lower half plane and the minority spin channel in the upper half plane. With this prescription BLOR (sym) is equivalent to BLOR for atomic subspaces which are located along the edge or fold of the diamond as well as non-spin polarised systems. However, the authors recommend the non-self consistent scheme over the self consistent application of the BLOR (sym) for reasons already outlined. 

\vspace{8cm}

\pagebreak
\newpage
\clearpage

\clearpage
\renewcommand{\thesection}{SI-\Roman{section}}
\onecolumngrid
\vspace{\columnsep}
\section{Dissociated Molecular Test Systems}
\label{sec:dimer-bar-charts}
\vspace{\columnsep}
\twocolumngrid

\subsection*{Dissociated H$_2$}
\begin{figure}[H]
\includegraphics[scale=0.20]{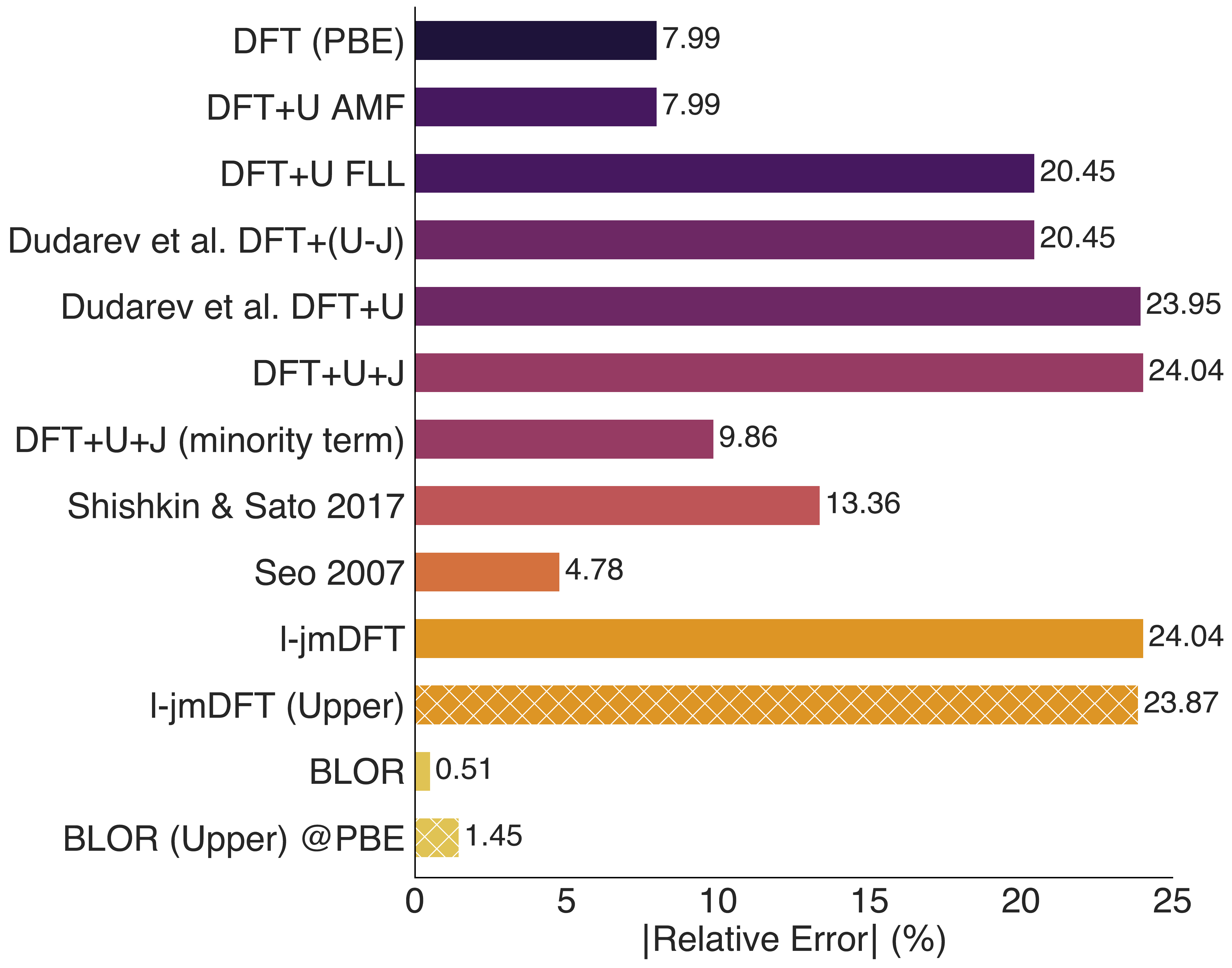}
\caption{  Bar chart of the relative errors in the total energies of H$_2$ at a bond length of 9 bohr radii using different corrective functionals \cite{PBE,kopernik,dudarev,dudarev_2019,moynihan_thesis,coco,shishkin2017,seo2007,bajaj2017,bajaj2019}. Note that BLOR and BLOR$_{\rm nS}$ are equivalent for non-spin polarised systems.}
\end{figure}

\subsection*{Dissociated He$_2^+$}
\begin{figure}[H]
\includegraphics[scale=0.2]{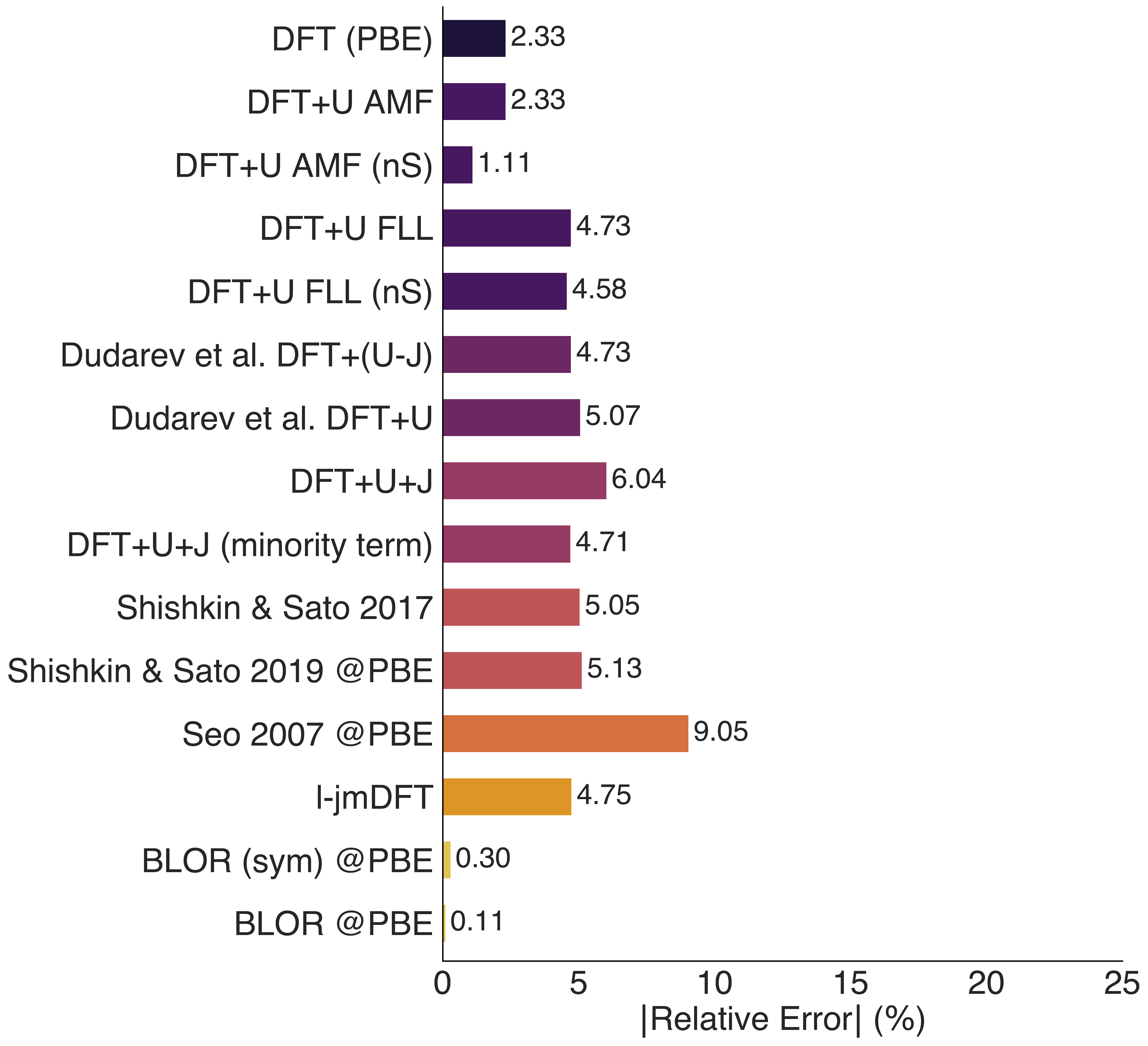}
\caption{ Bar chart of the relative errors in the total energies of He$_2^+$ at a bond length of 5 bohr radii using different corrective functionals \cite{PBE,kopernik,dudarev,dudarev_2019,moynihan_thesis,coco,shishkin2017,shishkin2019,seo2007,bajaj2017,bajaj2019}.}
\label{fig:he2+_complete_bar_chart}
\end{figure}

\subsection*{Dissociated Li$_2$}
\begin{figure}[H]
\includegraphics[scale=0.2]{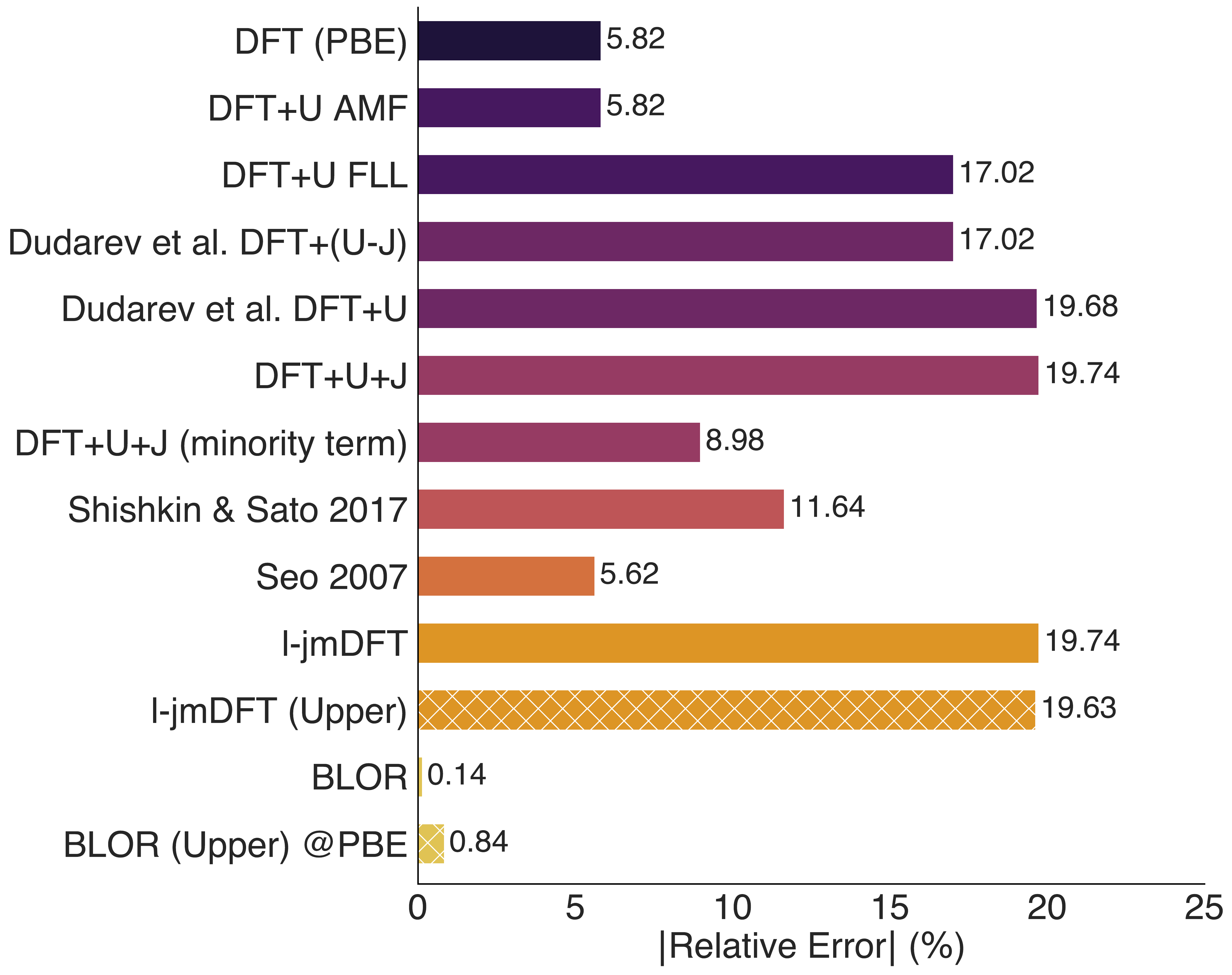}
\caption{ Bar chart of the relative errors in the total energies of Li$_2$ at a bond length of 15 bohr radii using different corrective functionals \cite{PBE,kopernik,dudarev,dudarev_2019,moynihan_thesis,coco,shishkin2017,seo2007,bajaj2017,bajaj2019}. Note that BLOR and BLOR$_{\rm nS}$ are equivalent for non-spin polarised systems.}
\end{figure}

\subsection*{Dissociated Be$_2^+$}
\begin{figure}[H]
\includegraphics[scale=0.2]{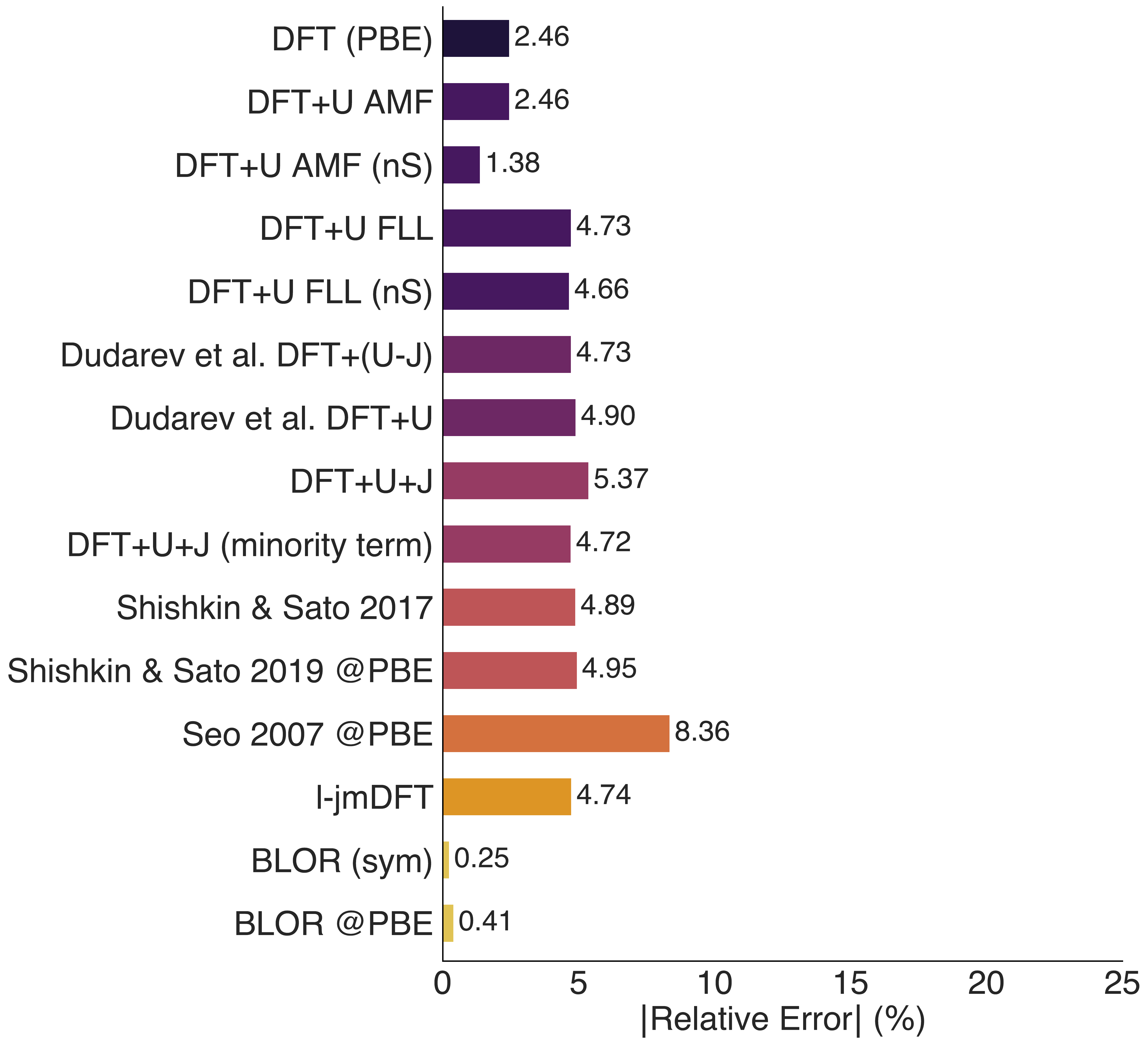}
\caption{ Bar chart of the relative errors in the total energies of Be$_2^+$ at a bond length of 10 bohr radii using different corrective functionals \cite{PBE,kopernik,dudarev,dudarev_2019,moynihan_thesis,coco,shishkin2017,shishkin2019,seo2007,bajaj2017,bajaj2019}.}
\end{figure}

\subsection*{Dissociated H$_5^+$ @PBE}
\begin{figure}[H]
\includegraphics[scale=0.2]{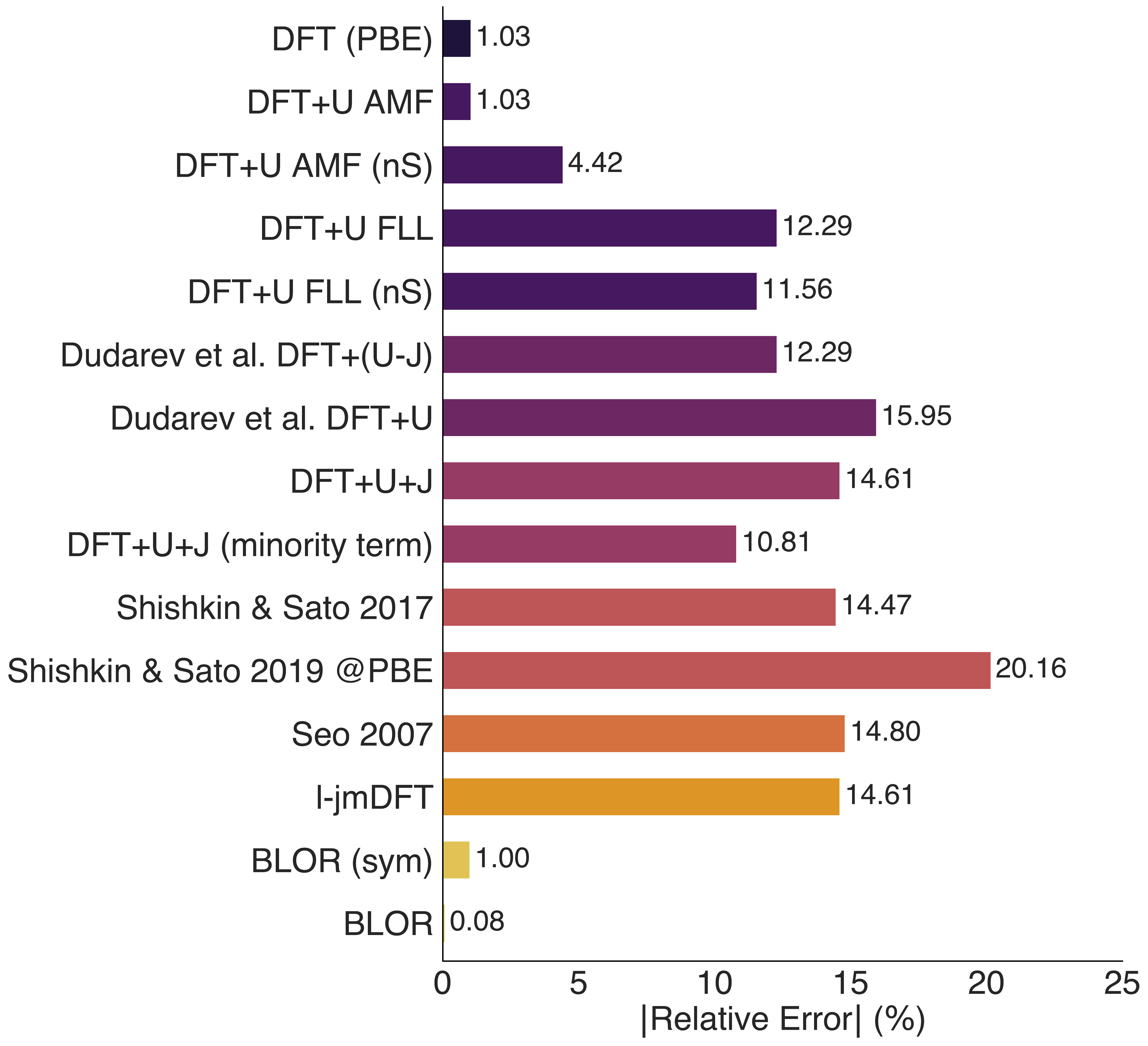}
\caption{Bar chart of the relative errors in the total energy of H$_5^+$ at an internuclear separation of 8 bohr radii using different corrective functionals \cite{PBE,kopernik,dudarev,dudarev_2019,moynihan_thesis,coco,shishkin2017,shishkin2019,seo2007,bajaj2017,bajaj2019}, which have been applied non-self consistently on the PBE density.}
\label{fig:h5+_complete_bar_chart}
\end{figure}

\hspace{1mm}

\newpage

\vspace{10cm}

\clearpage
\renewcommand{\thesection}{SI-\Roman{section}}
\onecolumngrid
\vspace{\columnsep}
\section{Alternative Methods to Evaluate ${\bf{U}}$ \& ${\bf{J}}$}
\label{sec:hubbard-param-tests}

\renewcommand{\thesection}{SI-\Roman{subsection}}
\vspace{\columnsep}
\subsection{Simple 2x2 Method @PBE}

In the main text, the simple $2 \times 2$ method (as given by equations \ref{eqn:Usimple}$\&$ \ref{eqn:Jsimple}), was used to compute the $U$ and $J$ parameters for each corrective functional with the exception of the BLOR functional. In this subsection for completeness, the $U$ and $J$ parameters for all corrective functionals including BLOR were evaluated using the simple $2 \times 2$ method. BLOR has spin resolved Hubbard parameters $U^{\sigma}$ and hence it is necessary to set:
\begin{equation}
U^{\upharpoonright}=U^{\downharpoonright}=\frac{1}{4}(f^{\upharpoonright \upharpoonright}+f^{\upharpoonright \downharpoonright}+f^{\downharpoonright \upharpoonright}+f^{\upharpoonright \downharpoonright}).
\end{equation}
By setting $U^{\upharpoonright}=U^{\downharpoonright}$ the results for BLOR will be equal to BLOR (sym). Evaluating the Hubbard $U$ parameter for BLOR using the simple $2 \times 2$ method for He$_2^+$ (as shown in figure \ref{fig:he2+_fdd}) yields a significant relative error of $4.92\%$. Evaluating the spin resolved Hubbard parameters $U^{\sigma}$, via this method is thus not a suitable choice for the BLOR functional. 

\begin{figure}[H]
\includegraphics[scale=0.2]{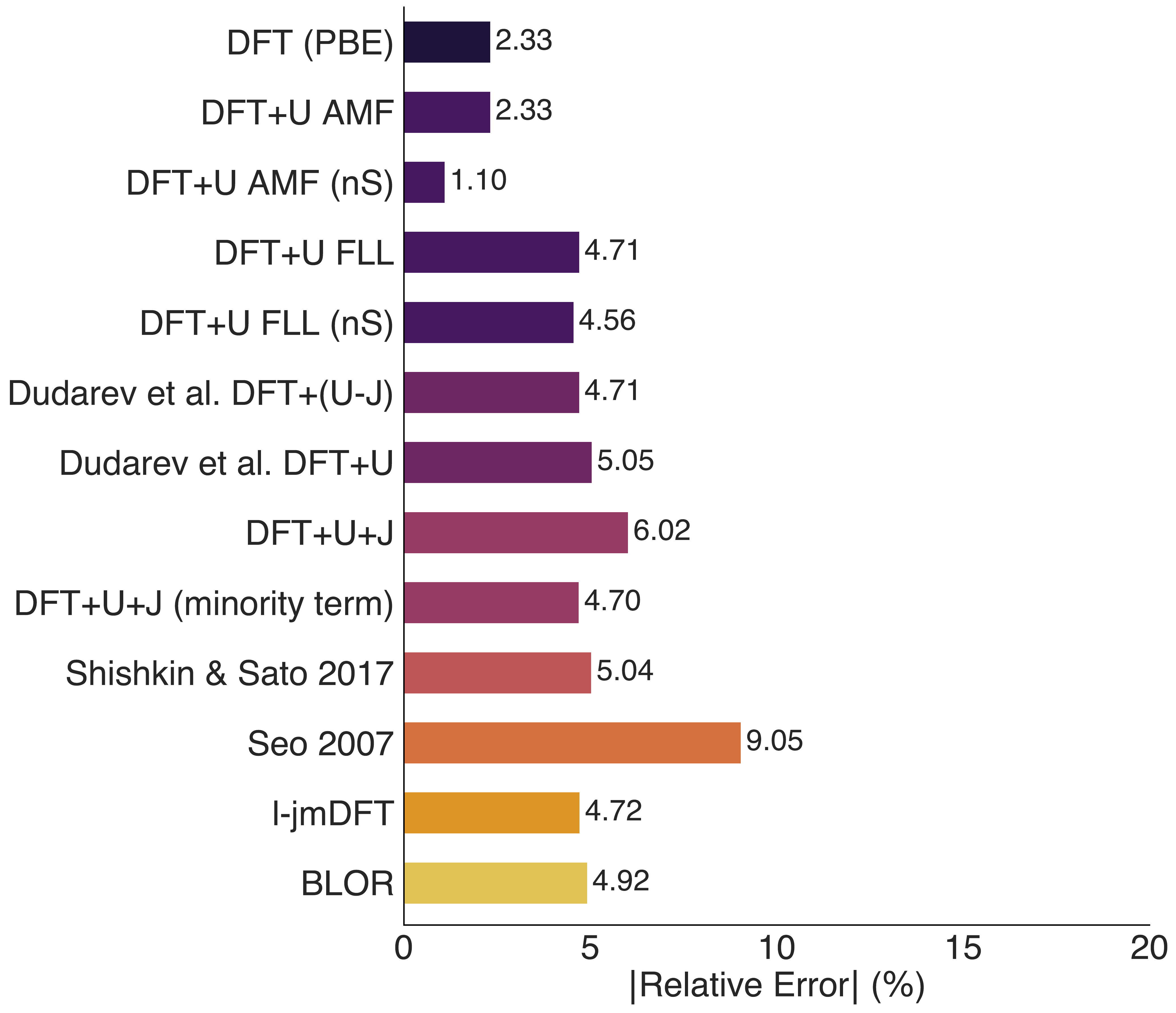}
\caption{Bar chart of the relative errors in the total energies of He$_2^+$ at a bond length of 5 bohr radii using various corrective functionals \cite{PBE,kopernik,dudarev,dudarev_2019,moynihan_thesis,coco,shishkin2017,shishkin2019,seo2007,bajaj2017,bajaj2019}, which have been applied non-self consistently on the PBE density. Where the Hubbard $U$ and Hund's $J$ parameter have both been computed by the simple 2x2 method \cite{linscott,moynihan}.}
\label{fig:he2+_usimple}
\end{figure}

\begin{figure}[H]
\includegraphics[scale=0.2]{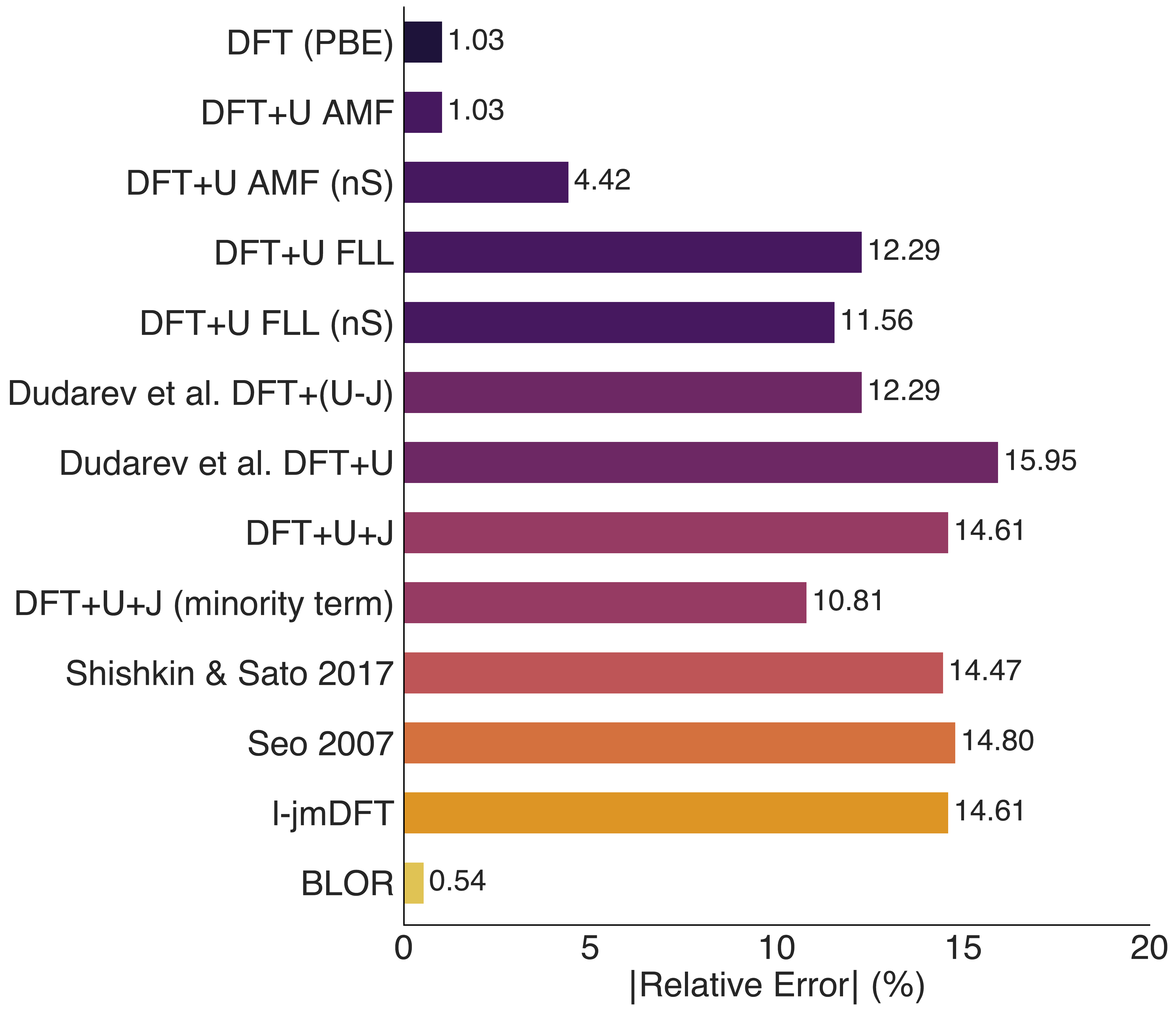}
\caption{Bar chart of the relative errors in the total energies of H$_5^+$ at an internuclear separation of 8 bohr radii using different corrective functionals \cite{PBE,kopernik,dudarev,dudarev_2019,moynihan_thesis,coco,shishkin2017,shishkin2019,seo2007,bajaj2017,bajaj2019}, which have been applied non-self consistently on the PBE density. Where the Hubbard $U$ and Hund's $J$ parameter have both been computed by the simple 2x2 method \cite{linscott,moynihan}.}
\label{fig:h5+_usimple}
\end{figure}

With the exception of the BLOR functional, the results for He$_2^+$ and H$_5^+$ in figures \ref{fig:he2+_usimple} $\&$ \ref{fig:h5+_usimple} below are equivalent to the results given in figures \ref{fig:he2+_complete_bar_chart} $\&$ \ref{fig:h5+_complete_bar_chart}. The minor numerical differences between the He$_2^+$ bar charts of figures \ref{fig:he2+_complete_bar_chart} $\&$ \ref{fig:he2+_usimple} are due to density self-consistency effects. The total energies in this section have been evaluated at the PBE density while in figure \ref{fig:he2+_complete_bar_chart} the total energies were evaluated self-consistently. The density self consistency effects for these dissociated molecular test systems are clearly negligible and will not be considered for the remainder of \ref{sec:hubbard-param-tests}.

\vspace{10mm}

\twocolumngrid

\newpage

\renewcommand{\thesection}{SI-\Roman{subsection}}
\onecolumngrid
\vspace{\columnsep}
\subsection{Simple 2x2 method for J and set U=$f^{\sigma \sigma}$ @PBE}
For the BLOR functional in the main text, the simple $2 \times 2$ method was used to evaluate the Hund's $J$ parameter and $U^{\sigma}$ is set as $f^{\sigma \sigma}$. In this subsection for completeness, the spin agnostic Hubbard $U$ parameter associated with all other corrective functionals was set as $f^{\sigma \sigma}$ where $\sigma$ is the majority spin channel in the lower half plane and the minority spin channel in the upper half plane. Similarly, for the BLOR functional we let $U^{\upharpoonright}=U^{\downharpoonright}=f^{\sigma \sigma}$, so that BLOR simplifies to BLOR (sym). 

\begin{figure}[H]
\includegraphics[scale=0.2]{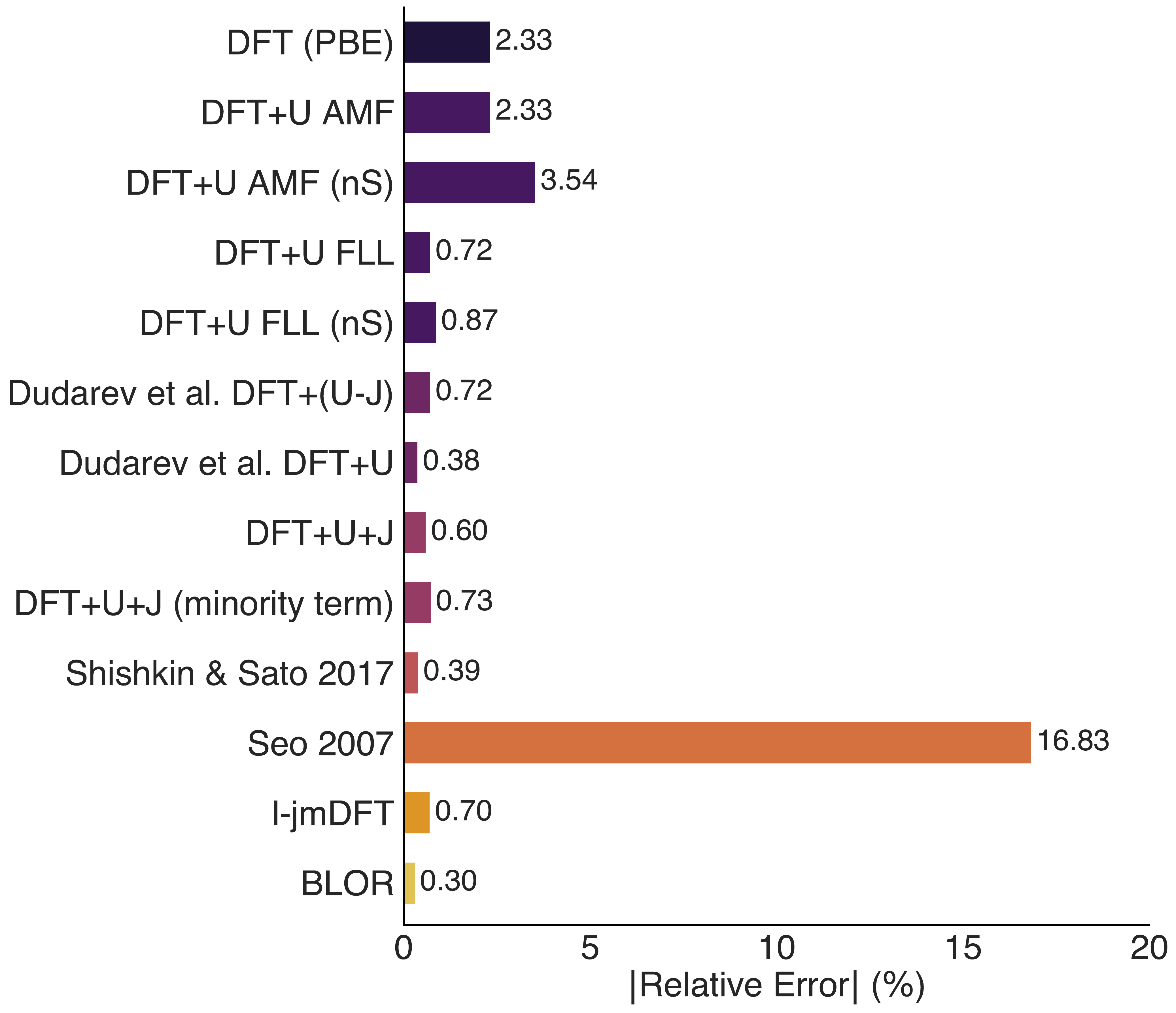}
\caption{Bar chart of the relative errors in the total energies of He$_2^+$ at a bond length of 5 bohr radii using various corrective functionals \cite{PBE,kopernik,dudarev,dudarev_2019,moynihan_thesis,coco,shishkin2017,shishkin2019,seo2007,bajaj2017,bajaj2019}, which have been applied non-self consistently on the PBE density. Where we evaluate the Hubbard $U$ parameter as $U=f^{\downharpoonright \downharpoonright}$ and evaluate the Hund's $J$ parameter via the simple 2x2 method \cite{linscott,moynihan}.}
\label{fig:he2+_fdd}
\end{figure}

Using this prescription, many corrective functionals yield extremely low relative errors for He$_2^+$ as shown in figure \ref{fig:he2+_fdd}. The atomic subspaces of He$_2^+$ are approximately located along the edges of the diamond. For an atomic subspace located perfectly along the edge of the diamond the BLOR functional for a single orbital subspace, simplifies to:
\begin{equation}
E_{\rm BLOR}=\frac{U^{\sigma}}{2}(n^{\sigma}-n^{\sigma}n^{\sigma}),
\end{equation}
where $\sigma$ is the majority spin channel in the lower half plane and the minority spin channel in the upper half plane. Therefore, if we set the effective Hubbard parameter $U_{\rm eff}$ in Dudarev et al's funcitonal \cite{dudarev} to $f^{\sigma \sigma}$, it will perfectly emulate the BLOR functional for systems with atomic susbpaces located along the edges of the diamond. Therefore, the extremely low realtive error associated with Dudarev et al's DFT+$U$ functional of $0.38 \%$ is unsurprising. 

\begin{figure}[H]
\includegraphics[scale=0.2]{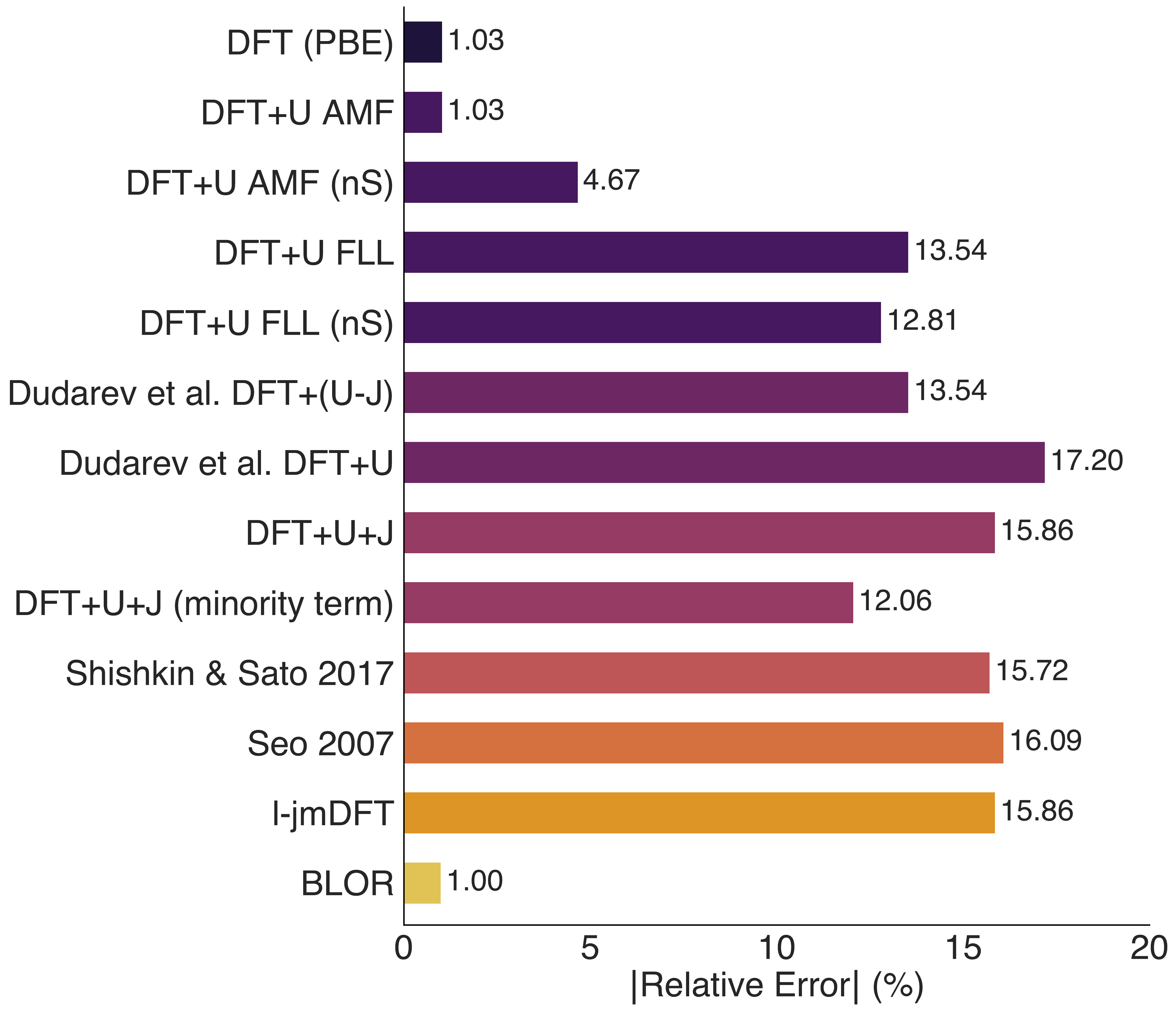}
\caption{Bar chart of the relative errors in the total energies of H$_5^+$ at an internuclear separation of 8 bohr radii using different corrective functionals \cite{PBE,kopernik,dudarev,dudarev_2019,moynihan_thesis,coco,shishkin2017,shishkin2019,seo2007,bajaj2017,bajaj2019}, which have been applied non-self consistently on the PBE density. Where we evaluate the Hubbard $U$ parameter as $U=f^{\upharpoonright \upharpoonright}$ and evaluate the Hund's $J$ parameter via the simple 2x2 method \cite{linscott,moynihan}.}
\label{fig:h5+_fuu}
\end{figure}

For systems with atomic susbpaces not located along the edge of the diamond (such as H$_5^+$) setting $U=f^{\sigma \sigma}$ does not allow Dudarev et al's DFT+$U$ functional to perfectly emulate the BLOR functional. The corrective functionals whihc performed well for He$_2^+$ using this method for evaluating $U$ and $J$ parameters clearly fails for H$_5^+$ as shown in figure \ref{fig:h5+_fuu}.   
\vspace{10mm}

\twocolumngrid

\vspace{40mm}

\newpage

\renewcommand{\thesection}{SI-\Roman{subsection}}
\onecolumngrid
\vspace{\columnsep}
\subsection{Scaled 2x2 Method @PBE}
In this subsection the $U$ and $J$ parameters were computed using Linscott et al's scaled $2 \times 2$ formulae \cite{linscott}. In the case of BLOR both $U^{\upharpoonright}$ and $U^{\downharpoonright}$ were set equal to the scaled Hubbard $U$ parameter. With $U^{\upharpoonright}=U^{\downharpoonright}$ the results for BLOR will be equal to the results for BLOR (sym). 

\begin{figure}[H]
\includegraphics[scale=0.2]{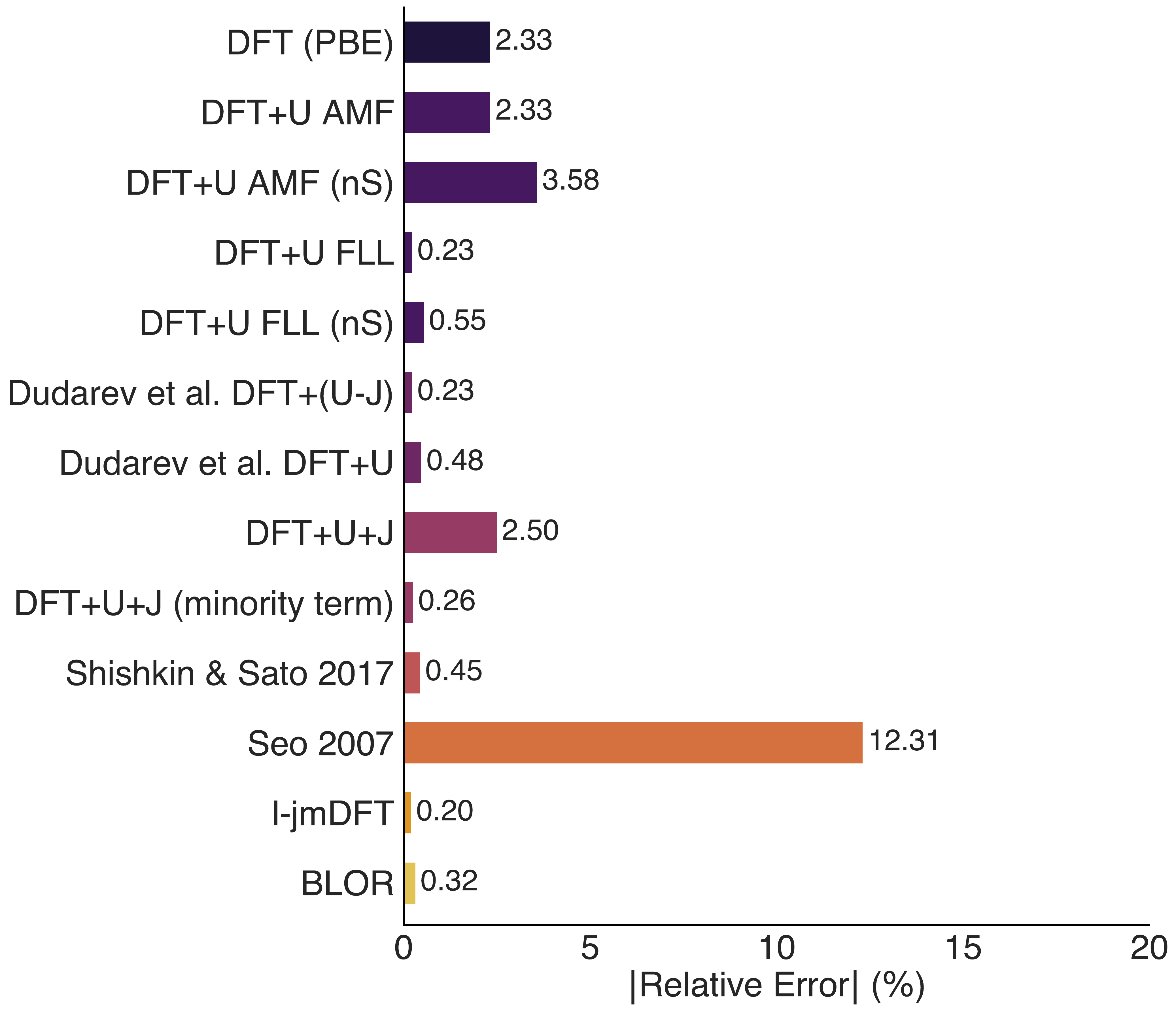}
\caption{Bar chart of the relative errors in the total energies of He$_2^+$ at a bond length of 5 bohr radii using various corrective functionals \cite{PBE,kopernik,dudarev,dudarev_2019,moynihan_thesis,coco,shishkin2017,shishkin2019,seo2007,bajaj2017,bajaj2019}, which have been applied non-self consistently on the PBE density. Where the Hubbard $U$ and Hund's $J$ parameter have both been computed by the scaled 2x2 method \cite{linscott,moynihan}.}
\label{fig:he2+_scaled_U}
\end{figure}

\begin{figure}[H]
\includegraphics[scale=0.2]{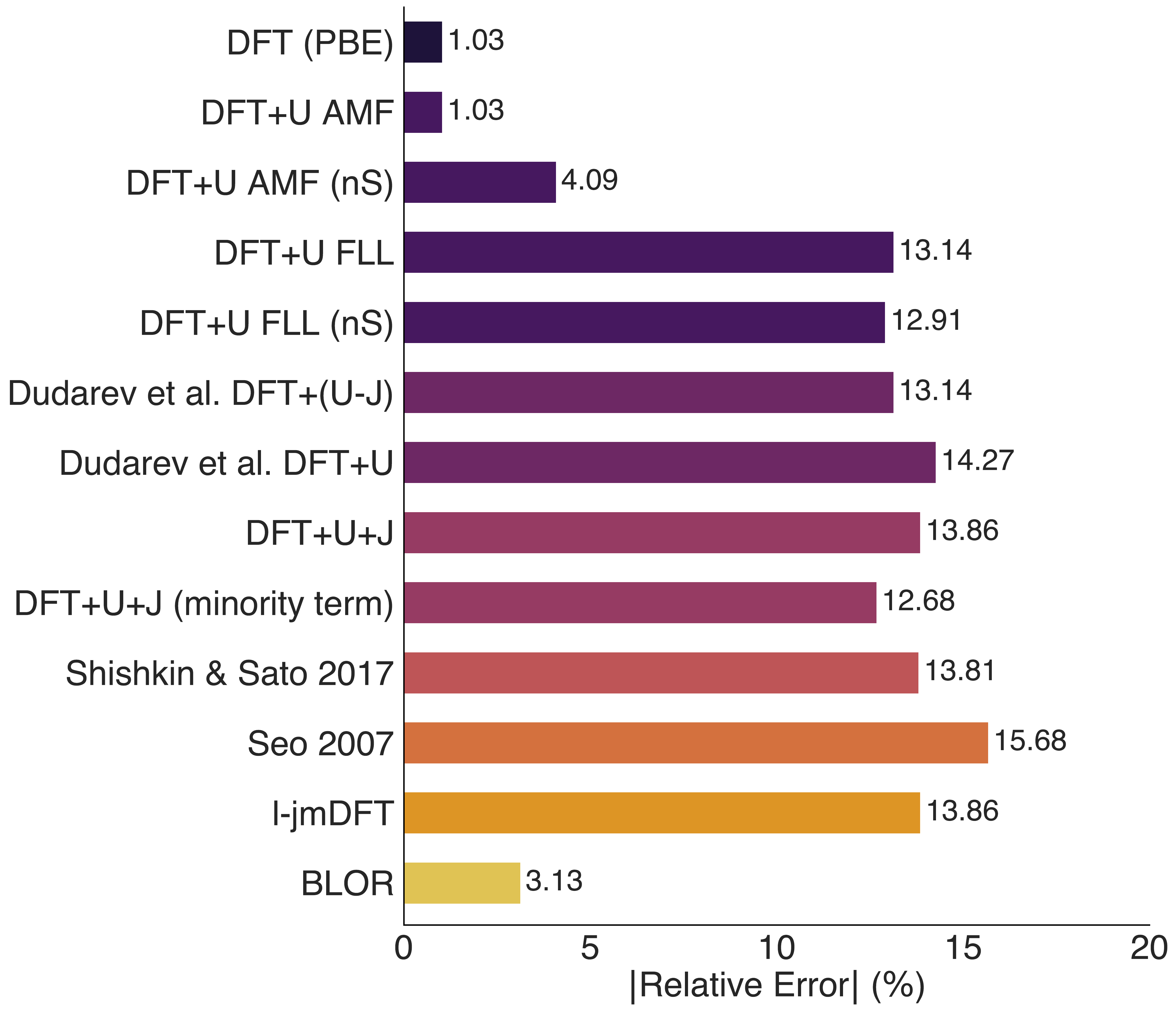}
\caption{Bar chart of the relative errors in the total energies of H$_5^+$ at an internuclear separation of 8 bohr radii using different functionals \cite{PBE,kopernik,dudarev,dudarev_2019,moynihan_thesis,coco,shishkin2017,shishkin2019,seo2007,bajaj2017,bajaj2019}, which have been applied non-self consistently on the PBE density. Where the Hubbard $U$ and Hund's $J$ parameter have both been computed by the scaled 2x2 method \cite{linscott,moynihan}. }
\label{fig:h5++_scaled_U}
\end{figure}

Many of the corrective functionals perform excellently for He$_2^+$ as shown in figre \ref{fig:he2+_scaled_U}. However, the atomic subspaces of He$_2^+$ is approximately located along the edge of the diamond and hence the system is dominated by local-MSIE. These corrective functionals, which yield low relative errors for He$_2^+$, perform poorly when significant portions of both local-MSIE and local-SCE are present. This is shown in figure \ref{fig:h5++_scaled_U} for H$_5^+$ where all corrective functionals except BLOR significantly worsen the PBE result. Thus, low relative errors can be achieved for systems dominated by local-MSIE when the scaled $2 \times 2$ method is used to evaluate $U$ and $J$ parameters for use in any of a wide variety of corrective functionals. 

As shown in figure \ref{fig:h5++_scaled_U}, the scaled $2 \times 2$ method with the BLOR functional, yields a significantly larger relative error for H$_5^+$ of $3.13 \%$ compared to the $0.08 \%$ achieved from using the simple $2 \times 2$ formula for $J$ and letting $U^{\sigma}=f^{\sigma \sigma}$.

\vspace{10mm}

\twocolumngrid

{\vspace{20cm}}

\clearpage
\renewcommand{\thesection}{SI-\Roman{section}}
\onecolumngrid
\vspace{\columnsep}
\section{Computational Details}
\label{sec:comp-details}
\vspace{\columnsep}
\twocolumngrid
All calculations were completed using the ONETEP (Order-N Electronic Total Energy Package) DFT code \cite{ONETEP}. The ONETEP code constructs the density matrix $\rho({\bf r},{\bf r}')$ from a set of Non-orthogonal Generalized Wannier Functions (NGWFs) $\{ \phi_{\alpha} \}$, as follows:
\begin{equation}
\rho({\bf r},{\bf r}')=\sum_{\alpha, \beta}\phi_{\alpha}({\bf r})K^{\alpha \beta}\phi_{\beta}({\bf r}'),
\end{equation}
where $K^{\alpha \beta}$ is the density kernel. The total energy of the system is minimised by optimizing both $K^{\alpha \beta}$ and $\{ \phi_{\alpha} \}$.

All calculations were completed using the PBE \cite{PBE} exchange-correlation functional at a high cutoff energy of no lower than $2,300$ eV. The dissociated molecular test systems were located in a large simulation cell, no smaller than $80 \times 60 \times 60$ $a_0^3$, with a Martyna-Tuckerman periodic boundary correction cutoff of $7.0$ $a_0$ \cite{martyna}.

For a system with $N_{\sigma}$ spin $\sigma$ KS particles, the occupancy of the lowest $N_{\sigma}$ KS particles was set equal to one, and otherwise set equal to zero. The convergence threshold of the root-mean-square gradient of the density kernel and the NGWFs was set at $1\times 10^{-6}$ and $1\times 10^{-7}$ respectively, and the electronic energy tolerance was set at $1\times 10^{-6}$ eV. The kerfix parameter was set equal to 1, but the occ-mix parameter was varied between test systems.

Four NGWFs were assigned per atom using the split-valence approach, with $15\%$ of the norm set to be beyond the matching radius $r_m$. The NGWF cutoff was set to $14$ $a_0$. A bespoke set of hard, norm-conserving pseudopotentials were made using the OPIUM code \cite{OPIUM}. 

For the spin polarised systems (He$_2^+$, Be$_2^+$ \& H$_5^+$) the elements of the Hxc kernel $f^{\sigma \sigma'}$ for the subspace were computed from a series of spin resolved linear response calculations on the atomic subspace. A spin up perturbation can be achieved by setting $\alpha=\beta$ (see equation \ref{eqn:dft_code}) and a spin down perturbation can be achieved by setting $\alpha=-\beta$. From each perturbative calculation the subspace averaged spin $\sigma$ KS potential is calculated:
\begin{equation}
V_{\rm KS}^{\sigma}=\frac{{\rm Tr}[\hat{P}^I \hat{V}_{\rm KS}^{\sigma}]}{{\rm Tr}[\hat{P}^I]}.
\end{equation}
The spin resolved subspace occupancy $N^{\sigma}$ and spin resolved perturbation strength $d V_{\rm ext}^{\sigma}$ is also recorded. The slopes of the plots of $V_{\rm KS}^{\sigma}[d V_{\rm ext}^{\sigma'} ]$ and $N^{\sigma}[d V_{\rm ext}^{\sigma'} ]$ were then used to evaluate $f^{\sigma \sigma'}$ as detailed by Linscott et al \cite{linscott}:
\begin{equation}
f^{\sigma \sigma'}=\bigg[\bigg(\frac{dV_{\rm KS}}{dV_{\rm ext}}-1\bigg)\bigg(\frac{dn}{V_{\rm ext}}\bigg)^{-1}\bigg]^{\sigma \sigma'}.
\end{equation}
$f^{\sigma \sigma'}$ can also be evaluated using the subspace averaged Hxc kernel $(V_{\rm Hxc})$ as follows:
\begin{equation}
f^{\sigma \sigma'}=\bigg[\bigg(\frac{dV_{\rm Hxc}}{dV_{\rm ext}}\bigg)\bigg(\frac{dn}{V_{\rm ext}}\bigg)^{-1}\bigg]^{\sigma \sigma'}.
\end{equation}
For H$_5^+$ $f^{\sigma \sigma'}$ was evaluated with a stabilising potential of the form:
\begin{equation}
\label{eqn:stabilising_potential}
\hat{v}^{\sigma}=G\hat{n}^{-\sigma}
\end{equation}
applied to the atomic subspaces. The Hubbard parameters were evaluated at a series of values of $G$ and these results were extrapolated to $G=0$ (thus following the same procedure as the bare PBE result for this system). 

For the non-spin polarised systems (H$_2$ \& Li$_2$) full spin polarised perturbations were not required because of the spin symmetry of the system. From a series of alpha perturbations the slope of $V_{\rm Hxc}[N]$ can be evaluated. It can be shown that this is equal to twice the value of the Hubbard $U$ parameter (from the simple 2x2 method). Similarly, from a series of beta perturbations, the slope of $[V_{\rm Hxc}^{\upharpoonright}-V_{\rm Hxc}^{\downharpoonright}][M]$, can be ascertained, where $M$ is the subspace magnetisation. Similarly, it can be shown that the slope of this curve is equal to minus two times the value of the Hund's $J$ parameter (from the simple 2x2 method). Finally, $f^{\upharpoonright \upharpoonright}$=U-J, for non-spin polarised systems. 

The non-spin polarised PBE solution for stretched $H_2$ \& Li$_2$ are at a point of unstable equilibirum in the PBE energy landscape. Applying a perturbation to the atomic subspace can cause the calculation to converge to the lower energy spin polarised solution. A stabilising potential of the form given by equation \ref{eqn:stabilising_potential} was used to stabilise the non-spin polarised solution. The Hund's $J$ parameter was evaluated at non-zero values of $G$ and extrapolated to $G=0$. This allows one to evaluate the Hund's $J$ parameter for the non-spin polarised systems as opposed to the spin polarised system. This technique was also required for the evaluation of the Hubbard $U$ parameter for Li$_2$. 

The Hund's $J$ parameter was evaluated using the minimum tracking linear response method, which defines the Hund's $J$ parameter as:
\begin{equation}
J=-\frac{1}{2}\frac{dV_{\rm Hxc}^{\upharpoonright}-dV_{\rm Hxc}^{\downharpoonright}}{d(n^{\upharpoonright}-n^{\downharpoonright})},
\end{equation}
one can approximate this as:
\begin{equation}
\label{eqn:hundsj-define}
J\approx-\frac{1}{2}\bigg(\frac{f^{\upharpoonright \upharpoonright}\delta n^{\upharpoonright}+f^{\upharpoonright \downharpoonright}\delta n^{\downharpoonright}-f^{\downharpoonright \upharpoonright}\delta n^{\upharpoonright}-f^{\downharpoonright \downharpoonright}\delta n^{\downharpoonright}}{\delta(n^{\upharpoonright}-n^{\downharpoonright})}\bigg),
\end{equation}
in the case of the BLOR functional, the Hund’s J parameter should be computed from varying the subspace magnetisation, keeping the total subspace occupancy constant, i.e. $\delta n^{\upharpoonright}=-\delta n^{\downharpoonright}$. Hence, equation \ref{eqn:hundsj-define} reduces to Linscott et al’s \cite{linscott} simple $2\times 2$ method:
\begin{equation}
J=-\frac{1}{4}(f^{\upharpoonright \upharpoonright}-f^{\upharpoonright \downharpoonright}-f^{\downharpoonright \upharpoonright}+f^{\upharpoonright \downharpoonright}). 
\end{equation}
for the evaluation of $J$. $J$ was evaluated by the simple $2\times 2$ method for use in all corrective functionals in the main text.

\clearpage

\clearpage

\clearpage
\renewcommand{\thesection}{SI-\Roman{section}}
\onecolumngrid
\vspace{\columnsep}
\section{Implementation of Corrective Functionals}
\label{sec:param-table}
\vspace{\columnsep}
\twocolumngrid

\onecolumngrid
\vspace{\columnsep}
Given any DFT code with the following DFT$+U$ functionality:
\begin{equation}
\label{eqn:dft_code}
\frac{U-J}{2}\sum_{\sigma m m'} \bigg[n^{\sigma}_{mm'}\delta_{mm'}-n^{\sigma}_{mm'}n^{\sigma}_{m'm}\bigg] +\frac{J}{2}\sum_{\sigma m m'}n^{\sigma}_{mm'}n^{\bar{\sigma}}_{m'm}+\alpha \sum_{\sigma m m'} n^{\sigma}_{mm'}\delta_{mm'}+\beta \sum_{m m'}\bigg[ n^{\upharpoonright}_{mm'}\delta_{mm'}-n^{\downharpoonright}_{mm'}\delta_{mm'}\bigg]+C,
\end{equation}
where $C$ is a constant and $\alpha$ \& $\beta$ are free parameters, typically used to apply perturbations to the subspace. Careful choice of the value of $U_{\rm input}$, $J_{\rm input}$, $\alpha_{\rm input}$, $\beta_{\rm input}$ and $C$ specified in the DFT input file allows one to use the DFT code to simulate the corrective functionals listed in the proceeding table. The proceeding table specifies the required input parameters for spin polarised systems. Please note that the BLOR corrective functional also requires separate $U$ parameters for the two spin channels.

\vspace{\columnsep}
\twocolumngrid

\noindent
\begin{table}[H]
\hspace{-0.8cm}
\begin{tabular}{ | m{5.0cm}| m{3.5cm}|m{3cm}|m{2.5cm}|m{1cm}|m{3.5cm}|}
 \hline
 {\vspace{2mm}} Hubbard Functional &$U_{\rm input}$ &$J_{\rm input}$&$\alpha_{\rm input}$ &$\beta_{\rm input}$&$C$\\ [2.5ex] 
 \hline\hline
 {\vspace{2mm}} DFT+U+J $(U_{\rm eff}=U)$ NSCF  & $U+J$ &$J$& $0$ & $0$  &0\\ [2.5ex] 
 \hline
  {\vspace{2mm}}DFT+U+J $(U_{\rm eff}=U)$ SCF & $-2J$&$-U-J$& $0$ & $0$ &0\\[2.5ex] 
 \hline
{\vspace{2mm}}DFT+U+J no minority spin term & $U$&$J$&  $0$ & $0$ &0\\[2.5ex] 
 \hline
 {\vspace{2mm}}DFT+U+J with minority spin term & $U$&$J$&  $-J/2$ & $J/2$&0\\[2.5ex] 
 \hline
  {\vspace{2mm}} DFT+J  $(U_{\rm eff}=0)$  & $-2J$&$-2J$& $0$  & $0$ &0\\[2.5ex]
 \hline 
  {\vspace{2mm}}Shiskin \& Sato (2017) &$U+J$&$J$&  $-J/2$ &  $J/2$&0\\[2.5ex] 
 \hline
{\vspace{2mm}}Shiskin \& Sato  (2019) & $U+U_{\uparrow \downarrow}$&$U_{\uparrow \downarrow}$&  $-U_{\uparrow \downarrow}/2$ & $U_{\uparrow \downarrow}/2$&0\\[2.5ex] 
\hline
{\vspace{2mm}}Dudarev  et al (1998) & $U-J$&$0$&  $0$ & $0$&0\\[2.5ex] 
 \hline
 {\vspace{2mm}}Dudarev  et al (2019) & $U$&$0$&  $0$ & $0$&0\\[2.5ex] 
 \hline
{\vspace{2mm}}Bajaj, Kulik et al (lower)  & $U$&$J$&  $0$ & $0$&0\\[2.5ex] 
 \hline
   {\vspace{2mm}}Bajaj, Kulik et al (upper)   &$U$&$J'$&  $-J'$ &  0&$J(2l+1)$\\[2.5ex] 
 \hline
  {\vspace{2mm}} BLOR$_{\rm nS}$ (lower) & $-2J$&$-U-2J$&  $0$& $0$ &0\\[2.5ex] 
 \hline
{\vspace{2mm}}  BLOR$_{\rm nS}$ (upper) & $-2J$&$-U-2J$&  $U+2J$& $0$ &$(-U-2J)(2l+1)$\\[2.5ex] 
 \hline
 {\vspace{2mm}}BLOR (lower)  &$U^{\sigma}_{input}=$&  $-\frac{1}{2}(U^{\upharpoonright}+U^{\downharpoonright}+4J)$ &0  &0&0\\[1.5ex]   & $U^{\sigma}-\frac{1}{2}(U^{\upharpoonright}+U^{\downharpoonright}+4J)$&&&&\\[2.5ex] 
 \hline
  {\vspace{2mm}}BLOR (upper) &$U^{\sigma}_{input}=$&  $-\frac{1}{2}(U^{\upharpoonright}+U^{\downharpoonright}+4J)$&  $\frac{1}{2}(U^{\upharpoonright}+U^{\downharpoonright}+4J)$ &  0&$-\frac{1}{2}(U^{\upharpoonright}+U^{\downharpoonright} +4J)(2l+1)$\\[1.5ex]   &$U^{\sigma}-\frac{1}{2}(U^{\upharpoonright}+U^{\downharpoonright}+4J)$ &&&&\\[2.5ex] 
 \hline
 \hline
\end{tabular}
\end{table}

\clearpage

\onecolumngrid
\vspace{\columnsep}
\section{Evaluation of Hubbard Parameters}
\subsection{Dissociated H$_2$}
\vspace{\columnsep}
\twocolumngrid

\begin{figure}[H]
\includegraphics[scale=0.9]{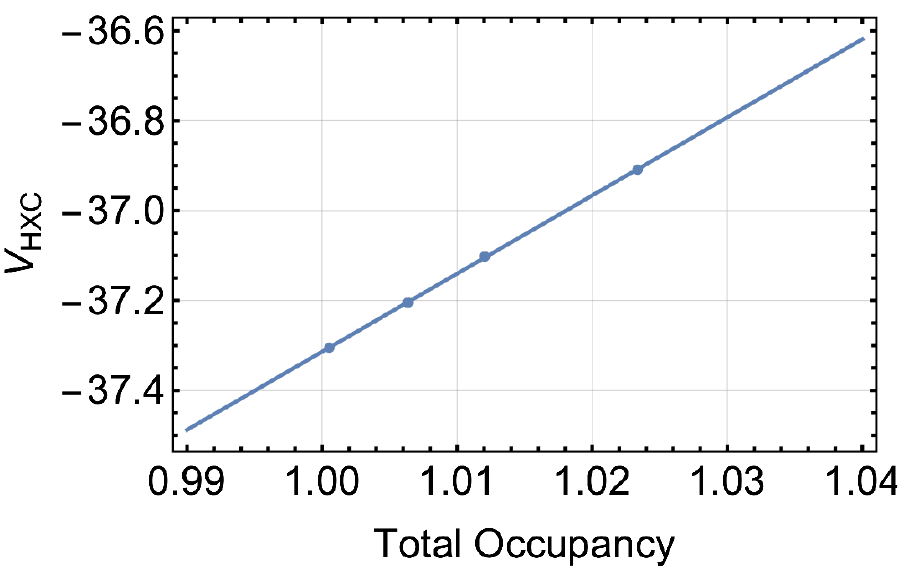}
\caption{Linear response curve of $V_{\rm Hxc}$ versus total occupancy for the H$_2$ molecule at $9$ $a_0$. The slope of the bestfit line is used to compute the Hubbard $U$ parameter.}
\end{figure}

\begin{figure}[H]
\includegraphics[scale=0.95]{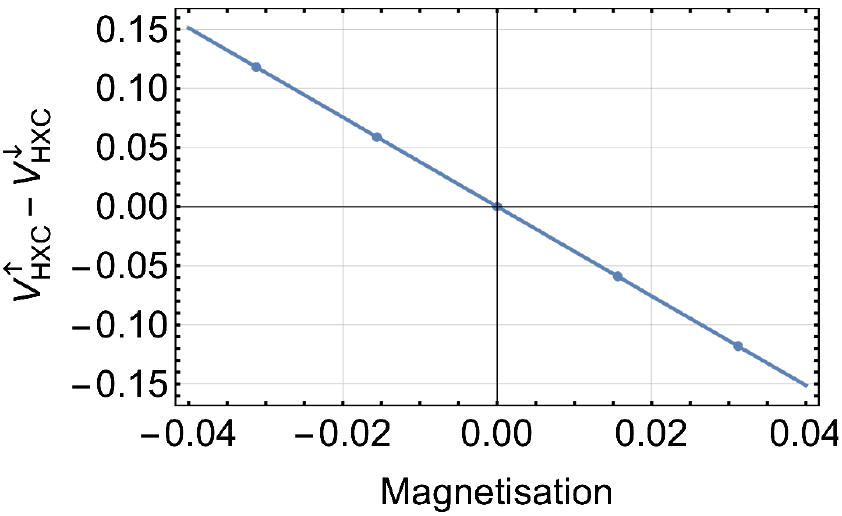}
\caption{Linear response curve of $V_{\rm Hxc}^{\upharpoonright}-V_{\rm Hxc}^{\downharpoonright}$ versus subspace magnetisation for the H$_2$ molecule at $9$ $a_0$ with a stabilising potential of $G=-10$ eV (see equation \ref{eqn:stabilising_potential2}). The slope of the bestfit line is used to compute the Hund's $J$ parameter at $G=-10$ eV.}
\end{figure}

\vspace{5cm}
\hphantom{x}

\begin{figure}[H]
\includegraphics[scale=0.865]{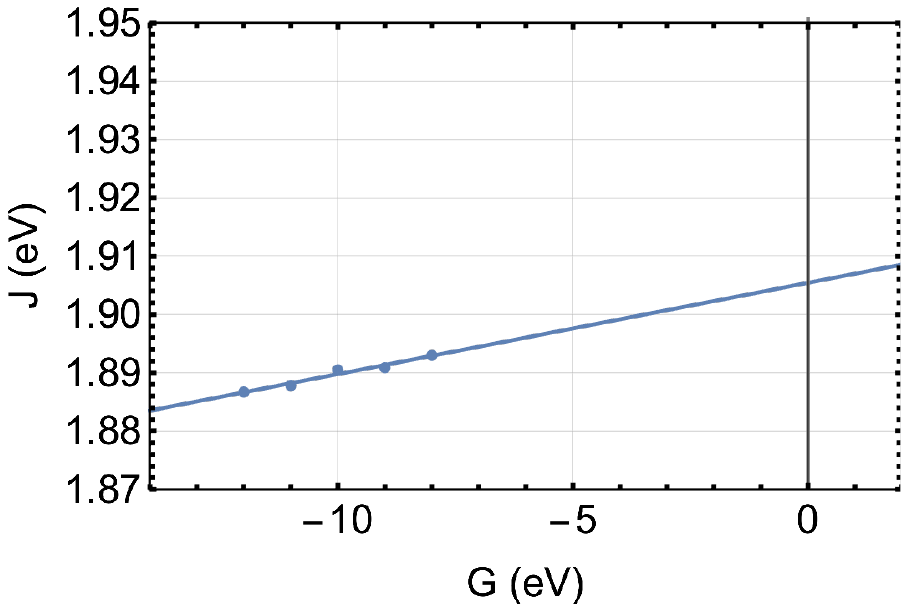}
\caption{Plot of the Hund's $J$ parameter versus the stabilising potential parameter $G$ (see equation \ref{eqn:stabilising_potential2}). Extrapolation of the bestfit line to $G=0$ eV, yields the Hund's $J$ parameter for use in the corrective functionals.}
\end{figure}

\vspace{10cm}
\bf{}

\clearpage

\onecolumngrid
\vspace{\columnsep}
\subsection{Dissociated H$_5^+$}
\vspace{\columnsep}
\twocolumngrid

\begin{figure}
\includegraphics[scale=0.7]{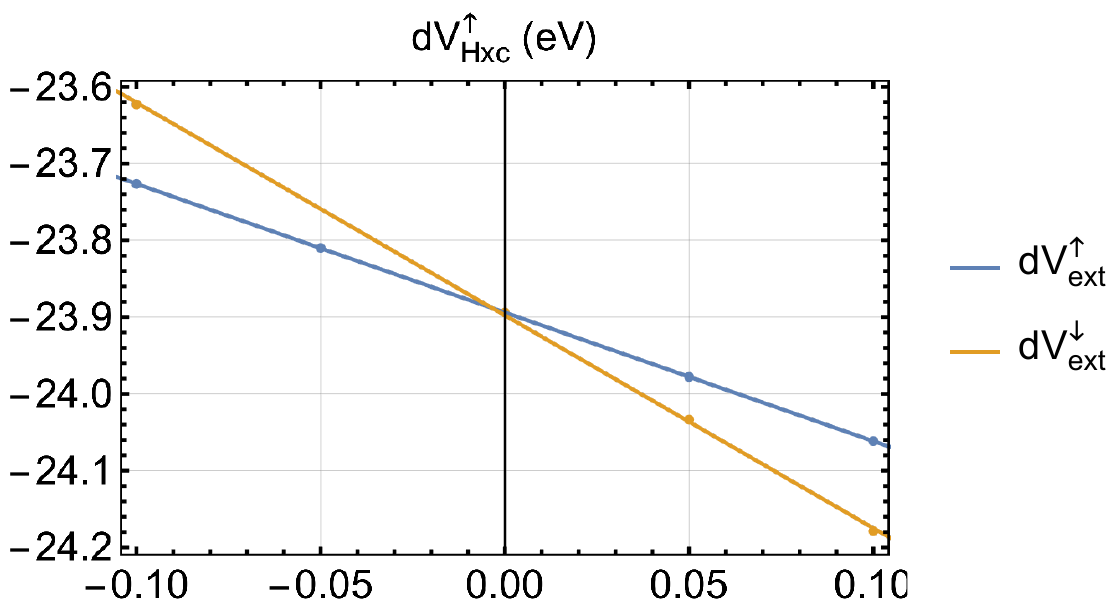}
\caption{Linear response graph of the variation in $V_{\rm Hxc}^{\upharpoonright}$ with respect to spin up and spin down perturbations for H$_5^+$ at 8 $a_0$ with a stabilising potential of $G=-14$ eV (see equation \ref{eqn:stabilising_potential2}).}
\end{figure}

\begin{figure}[H]
\includegraphics[scale=0.7]{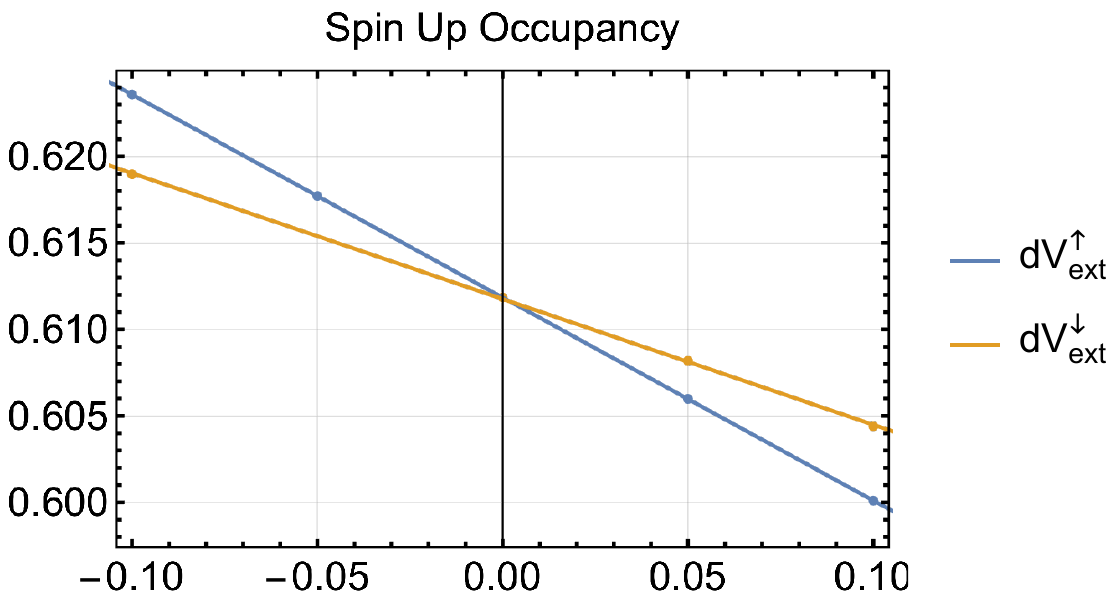}
\caption{Linear response graph of the variation in the spin up subspace occupancy with respect to spin up and spin down perturbations for H$_5^+$ at 8 $a_0$ with a stabilising potential of $G=-14$ eV (see equation \ref{eqn:stabilising_potential2}).}
\end{figure}

\begin{figure}[H]
\includegraphics[scale=0.72]{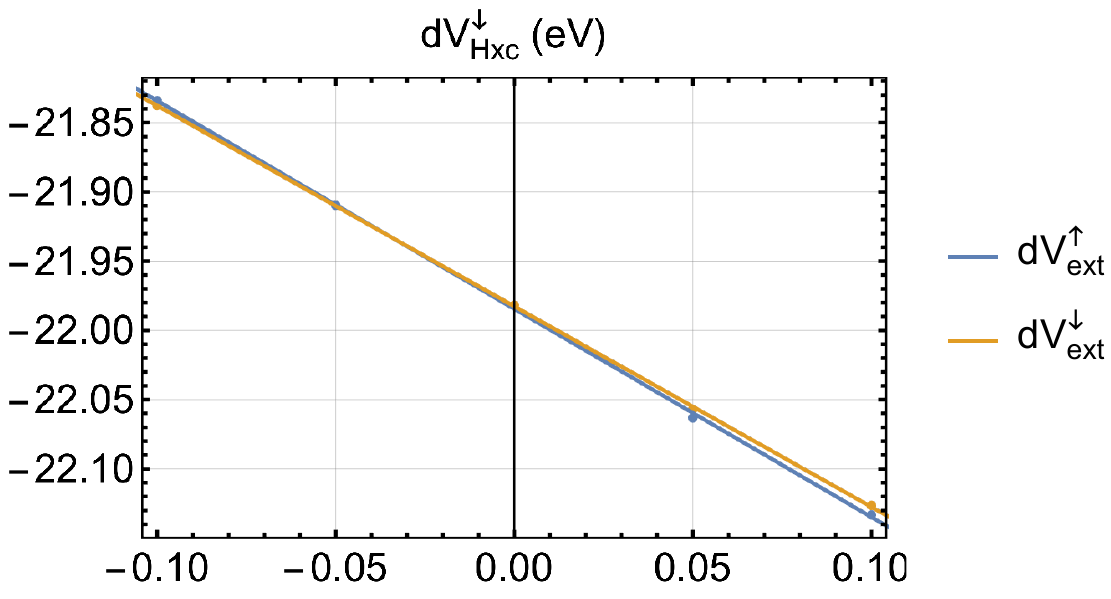}
\caption{Linear response graph of the variation in $V_{\rm Hxc}^{\downharpoonright}$ with respect to spin up and spin down perturbations for H$_5^+$ at 8 $a_0$ with a stabilising potential of $G=-14$ eV (see equation \ref{eqn:stabilising_potential2}).}
\end{figure}

\begin{figure}[H]
\includegraphics[scale=0.7]{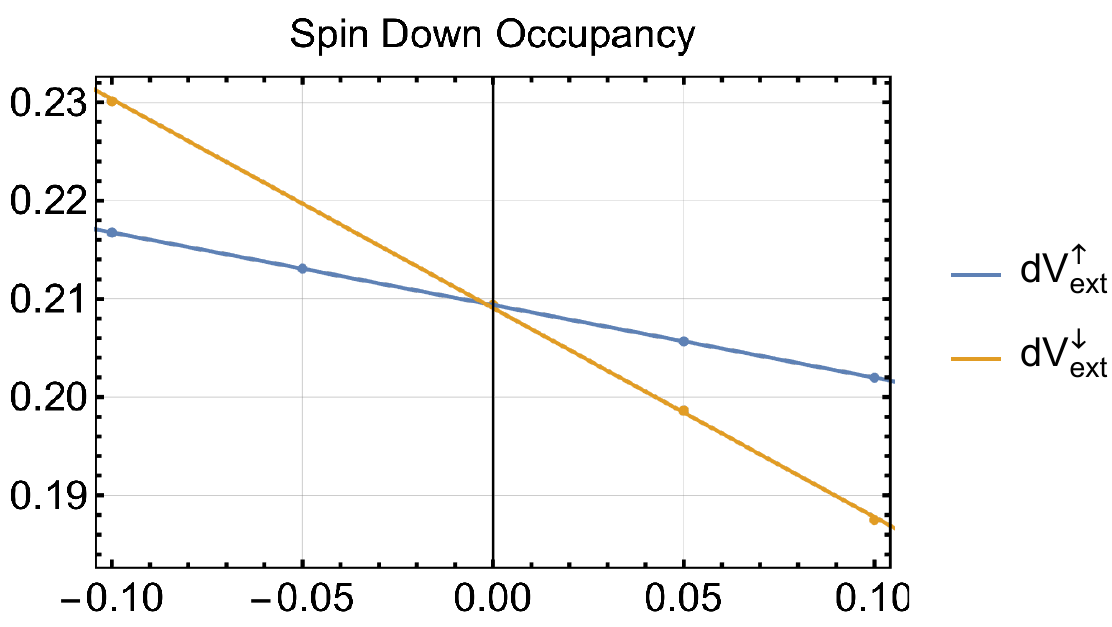}
\caption{Linear response graph of the variation in the spin down subspace occupancy with respect to spin up and spin down perturbations for H$_5^+$ at 8 $a_0$ with a stabilising potential of $G=-14$ eV (see equation \ref{eqn:stabilising_potential2}).}
\end{figure}

\onecolumngrid

\begin{figure}[H]
\centering
\includegraphics[scale=0.6]{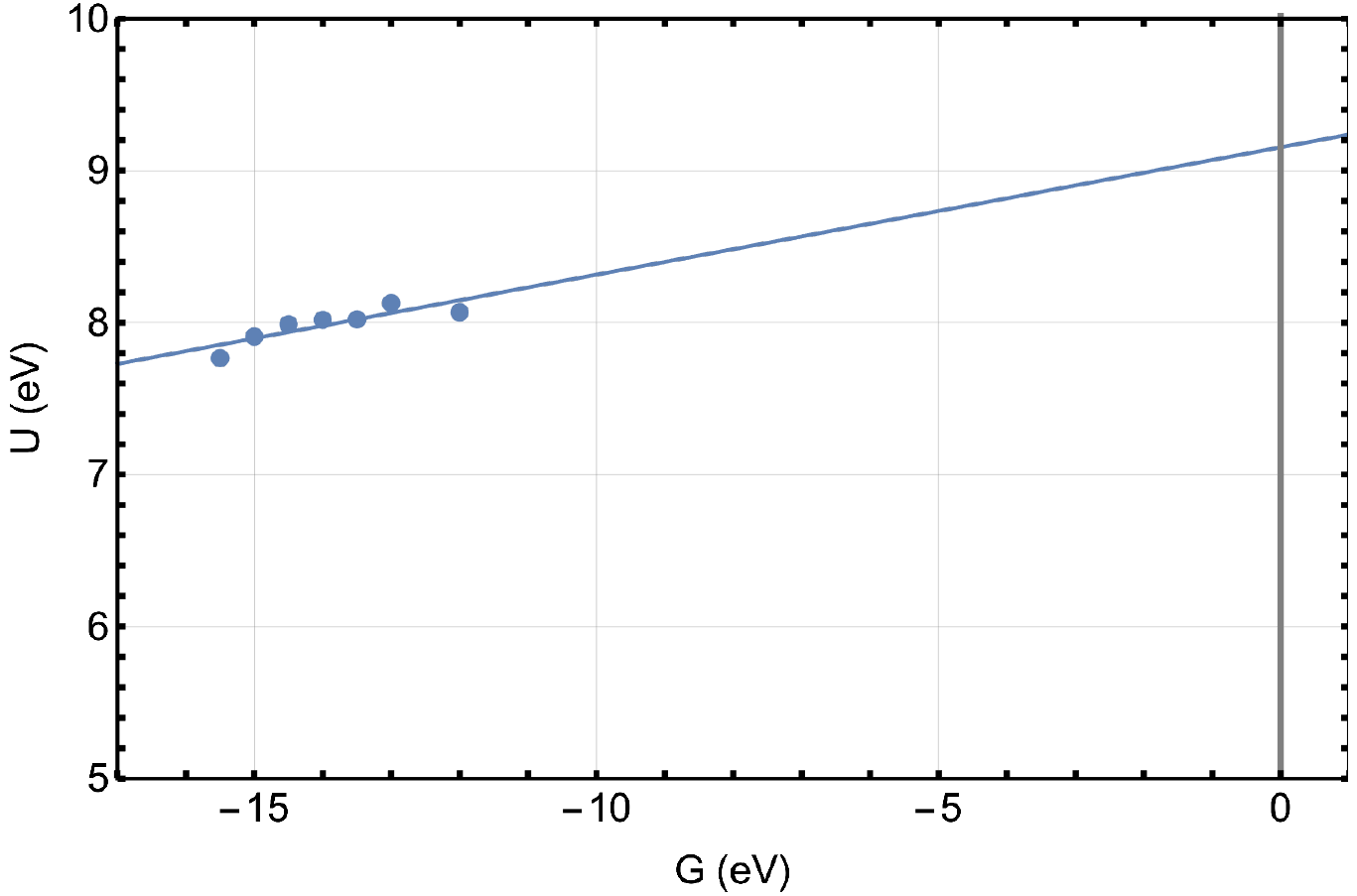}
\caption{Plot of the Hubbard $U$ parameter versus the stabilising potential parameter $G$ (see equation \ref{eqn:stabilising_potential2}). Extrapolation of the bestfit line to $G=0$ eV, yields the Hubbard $U$ parameter for use in the corrective functionals.}
\end{figure}

\end{document}